\documentclass[preprint,prd,aps,showpacs,preprintnumbers,amsmath,amssymb]{revtex4-1}
\usepackage{graphics,epsfig,subfigure}
\usepackage{diagbox}
\usepackage[usenames]{color}
\usepackage{color}
\usepackage{graphicx}
\usepackage{amsfonts}
\usepackage{indentfirst}
\usepackage{booktabs}
\usepackage{hyperref}
\hypersetup{hypertex=ture,backref=true,colorlinks=ture,linkcolor=blue,anchorcolor=blue,citecolor=blue}
\usepackage{float}
\usepackage{multirow}

\begin{document}
\renewcommand{\baselinestretch}{1.3}
\newcommand\beq{\begin{equation}}
\newcommand\eeq{\end{equation}}
\newcommand\beqn{\begin{eqnarray}}
\newcommand\eeqn{\end{eqnarray}}
\newcommand\nn{\nonumber}
\newcommand\fc{\frac}
\newcommand\lt{\left}
\newcommand\rt{\right}
\newcommand\pt{\partial}

\title{\Large \bf Dirac-boson stars}
\author{Chen Liang\footnote{liangch2020@lzu.edu.cn}, Ji-Rong Ren\footnote{renjr@lzu.edu.cn}, Shi-Xian Sun\footnote{sunshx20@lzu.edu.cn},  and Yong-Qiang Wang\footnote{yqwang@lzu.edu.cn, corresponding author
}
}

\affiliation{ $^{1}$Lanzhou Center for Theoretical Physics, Key Laboratory of Theoretical Physics of Gansu Province,
	School of Physical Science and Technology, Lanzhou University, Lanzhou 730000, China\\
	$^{2}$Institute of Theoretical Physics $\&$ Research Center of Gravitation, Lanzhou University, Lanzhou 730000, China}

\begin{abstract}
In this paper, we construct \textit{Dirac-boson stars} (DBSs) model composed of a scalar field and two Dirac fields. The scalar field and both Dirac fields are in the ground state. We consider the solution families of the DBSs for the synchronized frequency $\tilde{\omega}$ and the nonsynchronized frequency $\tilde{\omega}_D$ cases, respectively. We find several different solutions when the Dirac mass $\tilde{\mu}_D$ and scalar field frequency $\tilde{\omega}_S$ are taken in some particular ranges. In contrast, no similar case has been found in previous studies of multistate boson stars. Moreover, we discuss the characteristics of each type of solution family of the DBSs and present the relationship between the ADM mass $M$ of the DBSs and the synchronized frequency $\tilde{\omega}$ or the nonsynchronized frequency $\tilde{\omega}_D$. Finally, we calculate the binding energy $E_B$ of the DBSs and investigate the relationship of $E_B$ with the synchronized frequency $\tilde{\omega}$ or the nonsynchronized frequency $\tilde{\omega}_D$.
\end{abstract}

\maketitle

\section{Introduction}\label{Sec1}

In the 1950s, J. A. Wheeler studied the classical fields of electromagnetism coupled to Einstein's gravity, and obtained an unstable solution, so-called \textit{geons}~\cite{Wheeler:1955zz,Power:1957zz}. In 1968, D.~J.~Kaup constructed the theory of complex scalar field coupled to Einstein gravity~\cite{Kaup:1968zz}, called the Einstein-Klein-Gordon theory. He found a stable solution of this configuration. Related work was subsequently done by R. Ruffini and S. Bonazzola~\cite{Ruffini:1969qy}. Solitons formed by scalar fields under their gravity are called \textit{Boson stars} (BSs). In astrophysics, BSs are thought to be a possible component of dark matter~\cite{Sahni:1999qe,Matos:2000ng,Hu:2000ke,Suarez:2013iw,Hui:2016ltb,Padilla:2019fju}. Furthermore, BSs have been used to mimic the black holes~\cite{Cardoso:2019rvt,Herdeiro:2021lwl}. More recently, BSs have been used to analyze gravitational-wave signal~\cite{Bustillo:2020syj,Bezares:2022obu}.

BSs with a single complex scalar field of matter has been extensively studied. The BSs with self-interaction have been studied in Refs.~\cite{Colpi:1986ye,Mielke:1980sa,Kling:2017hjm,Herdeiro:2020jzx}. In addition, coupling the complex scalar field with the electromagnetic field can be considered to obtain the charged boson stars~\cite{Jetzer:1989av,Jetzer:1992tog,Kleihaus:2009kr,Kumar:2014kna}. Solving Einstein's Klein-Gordon equation in the Newtonian limit yields the Newtonian boson stars~\cite{harrison2002numerical}. In addition, rotating bosons with angular momentum are studied in Refs.~\cite{Schunck:1996he,Yoshida:1997qf,Kleihaus:2005me,Hartmann:2010pm,Kling:2020xjj}. Einstein's gravity can also be coupled to fields with non-zero spin. In 2015, Brito \textit{et al}.~\cite{Brito:2015pxa} proposed the Einstein-Proca theory. They studied the static solutions of the system in which the Proca field (spin-1) coupled to Einstein gravity, called \textit{Proca stars}. Shortly afterward, I.~Salazar Landea and F.~Garc\'\i{}a proposed the charged Proca stars~\cite{SalazarLandea:2016bys}. In addition, Finster \textit{et al}.~\cite{Finster:1998ws} constructed a spherically symmetric system with two spinor fields (spin-1/2) and Einstein gravity coupling, which is the first study of the \textit{Dirac stars}. Moreover, the coupling of the spinor field with the electromagnetic field can constitute the charged Dirac stars~\cite{Finster:1998ux,Bohun:1999zdr}. In Ref.~\cite{Dzhunushaliev:2018jhj}, V.~Dzhunushaliev and V.~Folomeev studied the Dirac stars with self-interaction. More recently, the rotating Dirac stars were first proposed by Brito \textit{et al}.~\cite{Herdeiro:2019mbz}. There are also other studies on the coupling of the spinor field with gravity~\cite{Daka:2019iix,Blazquez-Salcedo:2019qrz,Blazquez-Salcedo:2019uqq,Minamitsuji:2020hpl,Leith:2021urf}.

Also interesting are systems in which Einstein's gravity is coupled to multiple matter fields. In Ref.~\cite{Bernal:2009zy}, Bernal \textit{et al}. constructed a system consisting of two complex scalar fields in the ground state and the first excited state, respectively, called \textit{Multi-state Boson Stars} (MSBS). Later, the rotating multistate boson stars were studied in Refs.~\cite{Li:2019mlk,Li:2020ffy}. The matter fields of the spherically symmetric boson stars can be extended to an odd number $N$ of complex scalar fields~\cite{Alcubierre:2018ahf,Alcubierre:2019qnh}. Systems in which the axion field and the complex scalar field are coupled are considered in Refs.~\cite{Guerra:2019srj,Delgado:2020udb} and are called \text{axion boson stars} (ABSs). Multi-field models that include the Proca field have also been studied in Ref.~\cite{Delgado:2020hwr}. This work aims to solve Einstein-Dirac-Klein-Gordon equations numerically, construct the spherical Dirac-boson stars (DBSs) consisting of two spinor fields and a complex scalar field, and investigate the characteristics of various families of the solutions.

The paper is organized as follows. In Sec.~\ref{sec2}, we present the model four-dimensional
Einstein gravity coupled to a complex massive scalar and two Dirac fields. In Sec.~\ref{sec3}, the boundary conditions of the Dirac boson stars are studied. We show the numerical results of the equations of motion and exhibit the properties of the multistate for two different cases in Sec.~\ref{sec4}. We conclude in Sec.~\ref{sec5} with a summary and illustrate the range for future work.

\section{The model setup}\label{sec2}

we consider the system of two massive Dirac fields and a massive complex scalar field, which is minimally coupled to ($3 + 1$)-dimensional Einstein gravity. The action is given by
\begin{equation}
 S=\int \sqrt{-g} d^4 x \left( \frac{R}{16\pi G} + {\cal L}_{S} + {\cal L}_{D}\right)\,,
\end{equation}
\noindent where $G$ is the gravitational constant, $R$ is the Ricci scalar, and ${\cal L}_{S}$ and ${\cal L}_{D}$ denote the Lagrangians of scalar and Dirac fields, respectively,
\begin{equation}
 {\cal L}_{S}=-g^{\alpha\beta}\partial_\alpha \psi^*\partial_\beta\psi - \mu_S^2\psi^*\psi\,,
\end{equation}
\begin{equation}
 {\cal L}_{D}=-i\sum\limits_{k=1}^2 \left[ \frac{1}{2}\left(\hat{D}_\mu\overline{\Psi}^{(k)}\gamma^\mu\Psi^{(k)} - \overline{\Psi}^{(k)}\gamma^\mu\hat{D}_\mu\Psi^{(k)}\right) + \mu_D\overline{\Psi}^{(k)}\Psi^{(k)}\right]\,,
\end{equation}
 \noindent where $\mu_S$ and $\mu_D$ are the scalar field mass and the common mass of the Dirac fields, respectively. Here $\psi$ is a complex scalar field, $\Psi^{(k)}~(k = 1, 2)$ are the Dirac fields (to construct a spherically symmetric configuration, we need at least two spinors). $\overline{\Psi}^{(k)} = \Psi^{(k)\dagger}\zeta$ are the Dirac conjugate, with $\Psi^{(k)\dagger}$ stands for the Hermitian conjugate. For the Hermitizing matrix $\zeta$, we choose $\zeta = \gamma^0$~\cite{Dolan:2015eua}. $\gamma^\mu$ are the gamma matrices in curved spacetime, $\hat{D}_\mu = \partial_\mu + \Gamma_\mu$ is the spinor covariant derivative, where $\Gamma_\mu$ are the spinor connection matrices. We show the details of the Dirac fields in the appendix. The field equations are given by
\begin{equation}\label{equ4}
 R_{\alpha\beta} - \frac{1}{2}g_{\alpha\beta}R = 8\pi G\left(T^S_{\alpha\beta} + T^D_{\alpha\beta}\right)\,,
\end{equation}
\begin{equation}\label{equ5}
 \nabla^2\psi - \mu_S^2\psi = 0\,,
\end{equation}
\begin{equation}\label{equ6}
 \gamma^\mu\hat{D}_\mu\Psi^{(k)} - \mu_D\Psi^{(k)} = 0\,,
\end{equation}
 \noindent where $T^S_{\alpha\beta}$ and $T^D_{\alpha\beta}$ are the energy-momentum tensors of the scalar and Dirac fields, respectively,
\begin{equation}
 T^S_{\alpha\beta} = \partial_\alpha\psi^*\partial_\beta\psi + \partial_\beta\psi^*\partial_\alpha\psi - g_{\alpha\beta}\left[\frac{1}{2}g^{\mu\nu}\left(\partial_\mu\psi^*\partial_\nu\psi + \partial_\nu\psi^*\partial_\mu\psi\right) + \mu_S^2\psi^*\psi\right]\,,
\end{equation}
\begin{equation}
 T^D_{\alpha\beta} = \sum\limits_{k=1}^2 -\frac{i}{2}\left(\overline{\Psi}^{(k)}\gamma_\alpha\hat{D}_\beta\Psi^{(k)} + \overline{\Psi}^{(k)}\gamma_\beta\hat{D}_\alpha\Psi^{(k)} - \hat{D}_\alpha\overline{\Psi}^{(k)}\gamma_\beta\Psi^{(k)} - \hat{D}_\beta\overline{\Psi}^{(k)}\gamma_\alpha\Psi^{(k)}\right)\,.
\end{equation}

The action of the matter fields are invariant under the $U(1)$ transformation $\psi\rightarrow e^{i\alpha}\psi$, $\Psi^{(k)}\rightarrow e^{i\alpha}\Psi^{(k)}$ with a constant $\alpha$. According to Noether's theorem, there are conserved currents corresponding to these two matter fields:
\begin{equation}\label{equ9}
 J_S^{\mu} = -i\left(\psi^*\partial^\mu\psi - \psi\partial^\mu\psi^*\right),\qquad J_D^{\mu} = \overline{\Psi}\gamma^\mu\Psi\,.
\end{equation}
Integrating the timelike component of these conserved currents on a spacelike hypersurface $\cal{S}$, then we can obtain the Noether charges:
\begin{equation}\label{equ10}
 Q_S = \int_{\cal S}J_S^t\,,\qquad Q_D = \int_{\cal S}J_D^t\,.
\end{equation}

We will construct static spherically symmetric Dirac-boson stars, for which we should choose the following form of the spacetime metric:
\begin{equation}
 ds^2 = -N(r)\sigma^2(r)dt^2 + \frac{dr^2}{N(r)} + r^2\left(d\theta^2 + \sin^2\theta d\varphi^2\right)\,,
\end{equation}
where $N(r) = 1 - {2m(r)}/{r}$, and the mass function $m(r)$ and $\sigma(r)$ depend only on the radial distance $r$. In addition, for the static spherically symmetric system, we use the following ansatz of the scalar and Dirac fields~\cite{Herdeiro:2017fhv}:
\begin{equation}
 \psi = \phi(r)e^{-i\omega_St}\,,
\end{equation}
\begin{equation}
 \Psi^{(1)} = \begin{pmatrix}\cos(\frac{\theta}{2})[(1 + i)f(r) - (1 - i)g(r)]\\ i\sin(\frac{\theta}{2})[(1 - i)f(r) - (1 + i)g(r)]\\-i\cos(\frac{\theta}{2})[(1 - i)f(r) - (1 + i)g(r)]\\ -\sin(\frac{\theta}{2})[(1 + i)f(r) - (1 - i)g(r)] \end{pmatrix}e^{i\frac{\varphi}{2} - i\omega_Dt}\,,
\end{equation}
\begin{equation}
 \Psi^{(2)} = \begin{pmatrix}i\sin(\frac{\theta}{2})[(1 + i)f(r) - (1 - i)g(r)]\\ \cos(\frac{\theta}{2})[(1 - i)f(r) - (1 + i)g(r)]\\ \sin(\frac{\theta}{2})[(1 - i)f(r) - (1 + i)g(r)]\\ i\cos(\frac{\theta}{2})[(1 + i)f(r) - (1 - i)g(r)] \end{pmatrix}e^{-i\frac{\varphi}{2} - i\omega_Dt}\,,
\end{equation}
where $\phi(r)$, $f(r)$ and $g(r)$ are real functions. Besides, the constants $\omega_S$ and $\omega_D$ are the frequency of the scalar and Dirac fields, respectively. When  $\omega_S$ and $\omega_D$ meet $\omega_S = \omega_D = \omega$, we call $\omega$ as the synchronized frequency, and when $\omega_S\ne\omega_D$, these two frequencies are called nonsynchronized frequencies.

Substituting the above ansatz into the field equations~(\ref{equ4}--\ref{equ6}), we can get the following equations for $\phi(r)$, $f(r)$, $g(r)$, $m(r)$ and $\sigma(r)$:
\begin{equation}\label{equ15}
 \phi^{\prime\prime}+\left(\frac{2}{r} + \frac{N^\prime}{N} + \frac{\sigma^\prime}{\sigma}\right)\phi^\prime + \left(\frac{\omega_S^2}{N\sigma^2} - \mu_S^2\right)\frac{\phi}{N} = 0\,,
\end{equation}
\begin{equation}\label{equ16}
 f^\prime + \left(\frac{N^\prime}{4N} + \frac{\sigma^\prime}{2\sigma} + \frac{1}{r\sqrt{N}} + \frac{1}{r}\right)f + \left(\frac{\mu_D}{\sqrt{N}} - \frac{\omega_D}{N\sigma}\right)g = 0 \,,
\end{equation}
\begin{equation}\label{equ17}
 g^\prime + \left(\frac{N^\prime}{4N} + \frac{\sigma^\prime}{2\sigma} - \frac{1}{r\sqrt{N}} + \frac{1}{r}\right)g + \left(\frac{\mu_D}{\sqrt{N}} + \frac{\omega_D}{N\sigma}\right)f = 0\, ,
\end{equation}
\begin{equation}
 m^\prime = r^2N\phi_n^{\prime2} + \left(\mu_S^2 + \frac{\omega_S^2}{N\sigma}\right)r^2\phi^2 + \frac{8r^2\omega_D(f^2 + g^2)}{\sqrt{N}\sigma}\,,
\end{equation}
\begin{equation}
 \frac{\sigma^\prime}{\sigma} = 2r\left(\phi^{\prime2} + \frac{\omega^2\phi^2}{N^2\sigma^2}\right) + \frac{8r}{\sqrt{N}}\left[gf^\prime - fg^\prime + \frac{\omega_D(f^2 + g^2)}{N\sigma}\right]\,.
\end{equation}
Also, the specific forms of the Noether charges obtained from Eq.~(\ref{equ9}) and Eq.~(\ref{equ10}) are
\begin{equation}
 Q_S = 8\pi\int_0^\infty r^2\frac{\omega_S\phi^2}{N\sigma}dr\,,\qquad Q_D = 16\pi\int_0^\infty r^2\frac{f^2 + g^2}{\sqrt{N}}dr\,.
\end{equation}

\section{Boundary conditions}\label{sec3}
We need to impose appropriate boundary conditions to solve the system of ordinary differential equations obtained in the previous section. For asymptotically flat solutions, the boundary conditions satisfied by the metric functions are
\begin{equation}
m(0) = 0,\qquad \sigma(0) = \sigma_0,\qquad m(\infty) = M,\qquad \sigma(\infty) = 1,
\end{equation}
where $M$ and $\sigma_0$ are currently unknown, the values of these two quantities can be obtained after finding the solution of the differential equation system.
For the matter field functions, at infinity we require
\begin{equation}
\phi(\infty) = 0,\qquad f(\infty) = 0,\qquad g(\infty) = 0.
\end{equation}
Additionally, by considering the form of the field equation~(\ref{equ15}--\ref{equ17}) at the origin, we can obtain the following boundary conditions satisfied by the field functions:
\begin{equation}
\left.\frac{d\phi(r)}{dr}\right|_{r = 0} = 0,\qquad f(0) = 0,\qquad \left.\frac{dg(r)}{dr}\right|_{r = 0} = 0.
\end{equation}

\section{Numerical results}\label{sec4}
To facilitate numerical calculations, we use dimensionless quantities:
\begin{equation}
\begin{split}
\tilde{r} = r/\rho,\quad \tilde{\phi} = \frac{\sqrt{4\pi}}{M_{Pl}}\phi,\quad \tilde{\omega}_S = \omega_S\rho,\quad \tilde{\mu}_S = \mu_S\rho,\quad\\ \tilde{f} = \frac{\sqrt{4\pi\rho}}{M_{Pl}}f,\quad \tilde{g} = \frac{\sqrt{4\pi\rho}}{M_{Pl}}g,\quad \tilde{\omega}_D = \omega_D\rho,\quad \tilde{\mu}_D = \mu_D\rho,
\end{split}
\end{equation}
where $M_{Pl} = 1/\sqrt{G}$ is the Planck mass, $\rho$ is a positive constant whose dimension is length, we let the constant $\rho$ be $1/\mu_S$. Additionally, we introduce a new radial variable
\begin{equation}
x = \frac{\tilde{r}}{1+\tilde{r}},
\end{equation}
where the radial coordinate $\tilde{r}\in[0,\infty)$, so $x\in[0,1]$. We numerically solve the system of differential equations based on the finite element method. The number of grid points in the integration region $0\le x\le 1$ is 1000. The iterative method we use is the Newton-Raphson method. To ensure that the calculation results are correct, we require the relative error to be less than $10^{-5}$.

For the ground state of the Dirac field, the field functions $f$ and $g$ do not have radial nodes, so we denote the ground state of the Dirac field as $D_0$, where the superscript is the number of radial nodes of the matter fields. Similarly, we denote the complex scalar field $\phi$ in the ground state as $S_0$. Therefore, We denote the coexisting state consisting of a scalar field and two Dirac fields in the ground state as $D_0S_0$. Next, we will discuss the coexisting state $D_0S_0$ of the DBSs in different cases.

\subsection{ Synchronized frequency }

Based on the characteristics of the solutions of DBSs, we divide the families of synchronized frequency solutions of DBSs into two categories: the \textit{one-branch} solution family and the \textit{multi-branch} solution family (multi-branch means that a given set of parameters can correspond to more than one solution). When $0.9051 \le \tilde{\mu}_D \le 0.9982$, the solution family of DBSs is a \textit{one-branch} solution family; when $0.559 \le \tilde{\mu}_D \le 0.905$, the solution family of DBSs is a \textit{multi-branch} solution family. We will next study the characteristics of these two solution families.

\begin{figure}[!htbp]
\begin{center}
\includegraphics[height=.16\textheight]{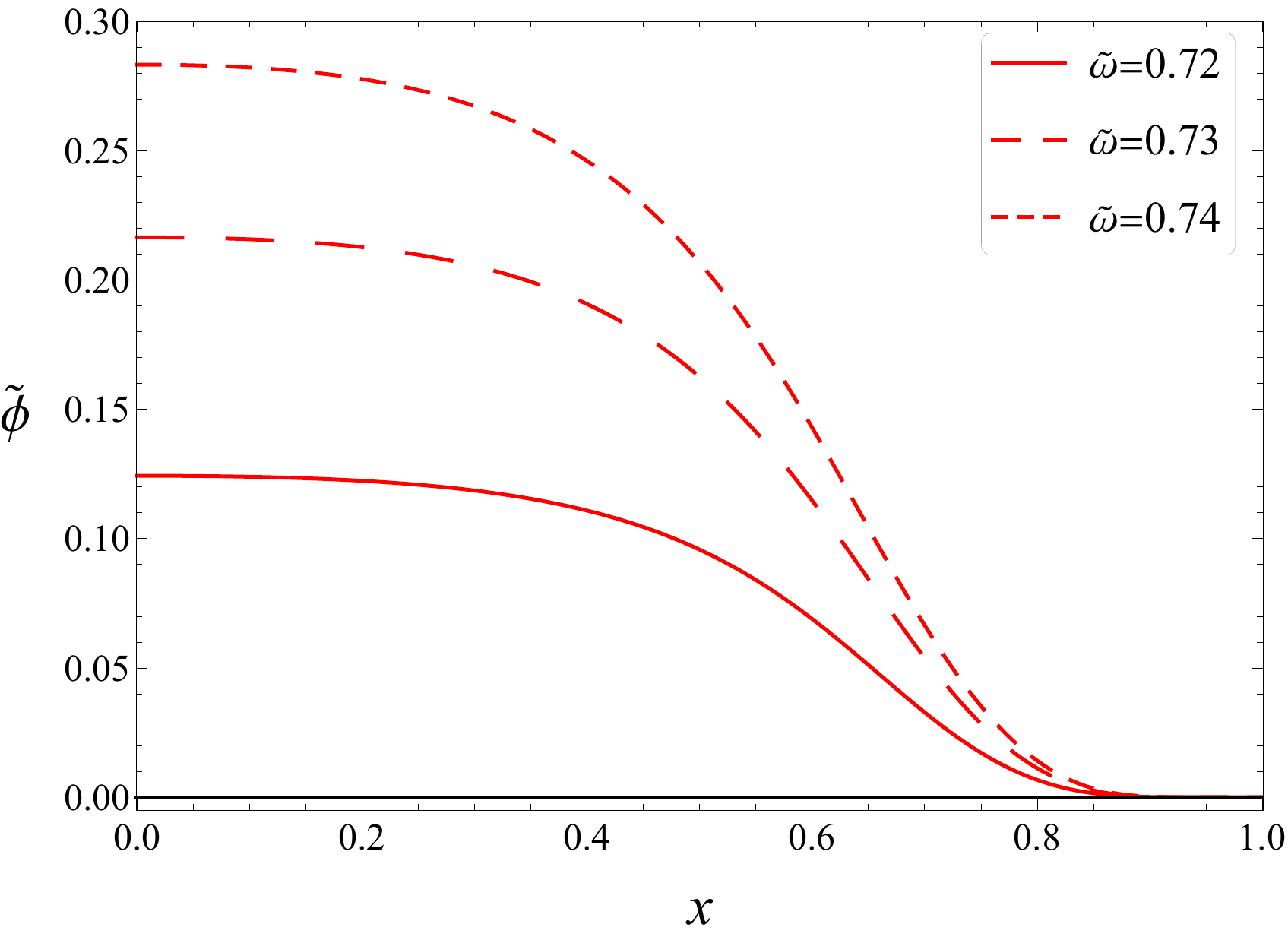}
\includegraphics[height=.16\textheight]{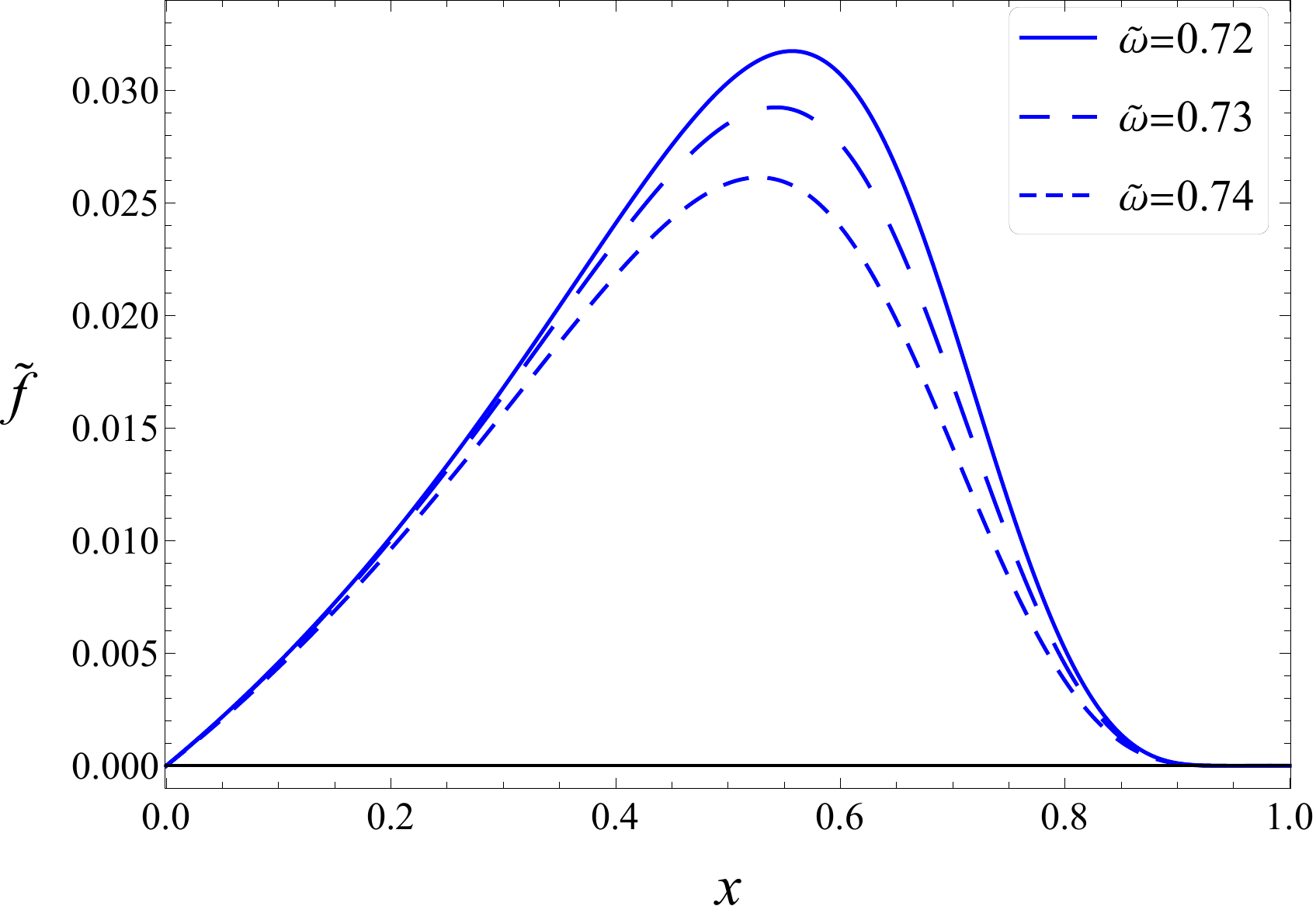}
\includegraphics[height=.16\textheight]{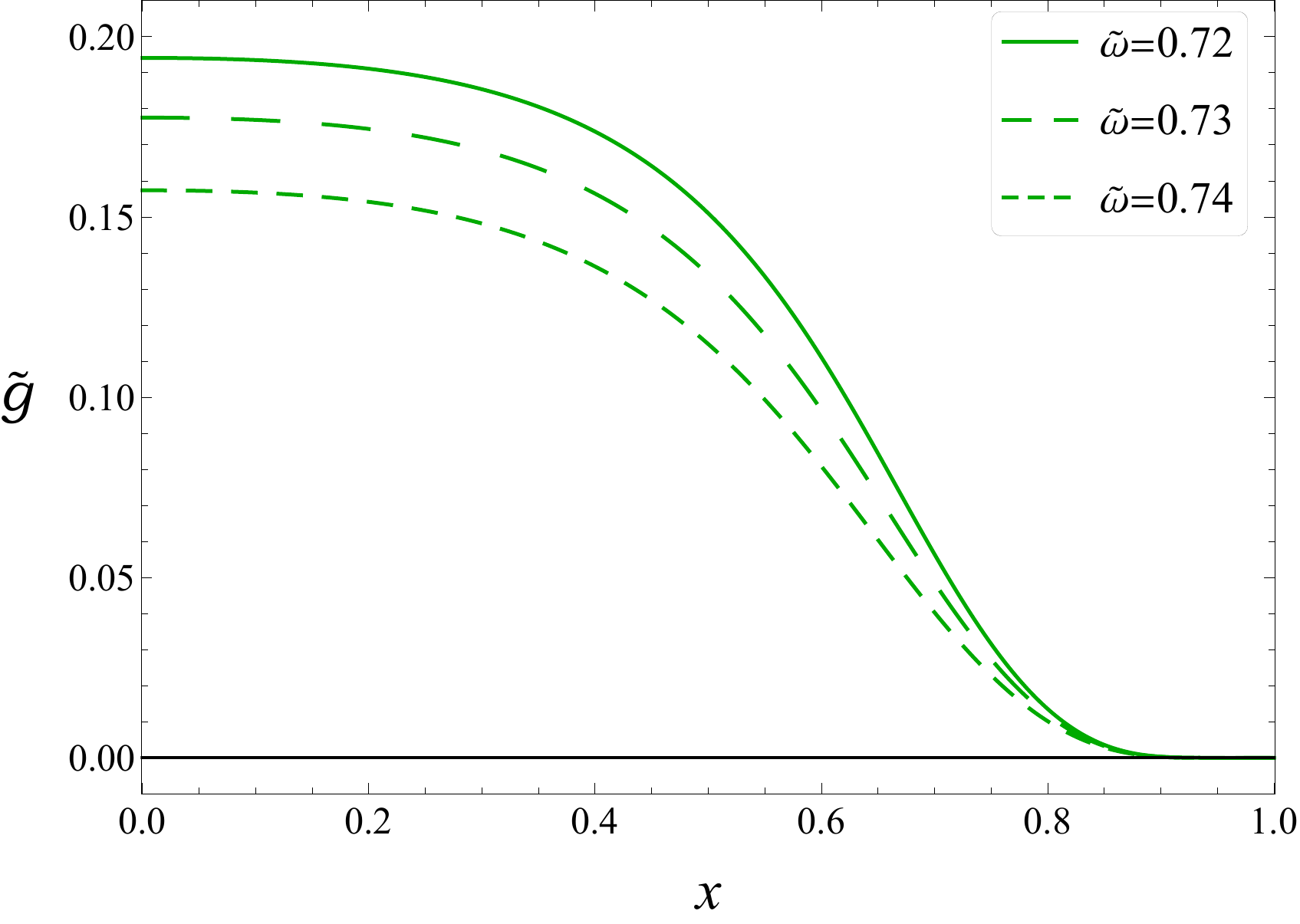}
\end{center}
\caption{Scalar field function $\tilde{\phi}$(left panel) and Dirac field functions $\tilde{f}$(middle panel) and $\tilde{g}$(right panel) as functions of $x$ for $\tilde{\omega} = \tilde{\omega}_S = \tilde{\omega}_D = 0.72, 0.73, 0.74$. All solutions have $\tilde{\mu}_D= 0.95$ and $\tilde{\mu}_S=1$.}
\label{psi_f_g}
\end{figure}

\subsubsection{One-Branch}

The profiles of the field functions $\tilde{\phi}$, $\tilde{f}$ and $\tilde{g}$ with several values of synchronized frequency $\tilde{\omega} = \tilde{\omega}_D =  \tilde{\omega}_S$ in the solutions of the $D_0S_0$ state are shown in Fig.~\ref{psi_f_g}. For the scalar field, as the frequency $\tilde{\omega}$ increases, $\tilde{\phi}_{max}$ (the maximum value of the field function $\tilde{\phi}$) increases. However, for the Dirac field, as the frequency $\tilde{\omega}$ increases, $\tilde{f}_{max}$ and $\tilde{g}_{max}$ (the maximum values of the field functions $\tilde{f}$ and $\tilde{g}$) decrease. Furthermore, we would like to study DBSs in the $D_0S_0$ state, where both the scalar and Dirac fields are in the ground state. It can be seen that there are no nodes in the matter functions in Fig.~\ref{psi_f_g}, which means that the scalar and Dirac fields are in the ground state.

\begin{figure}[!htbp]
\begin{center}
\includegraphics[height=.24\textheight]{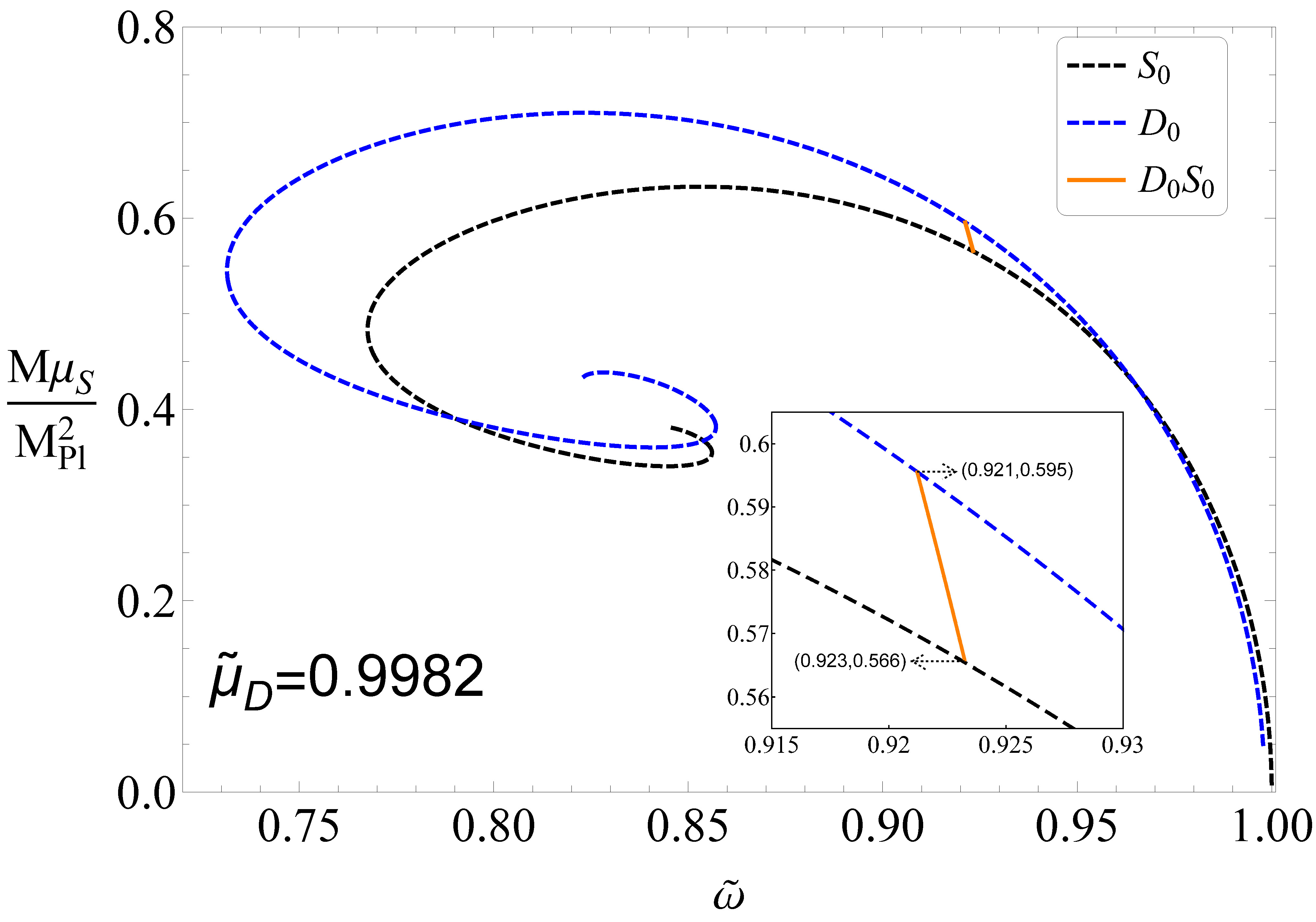}
\includegraphics[height=.24\textheight]{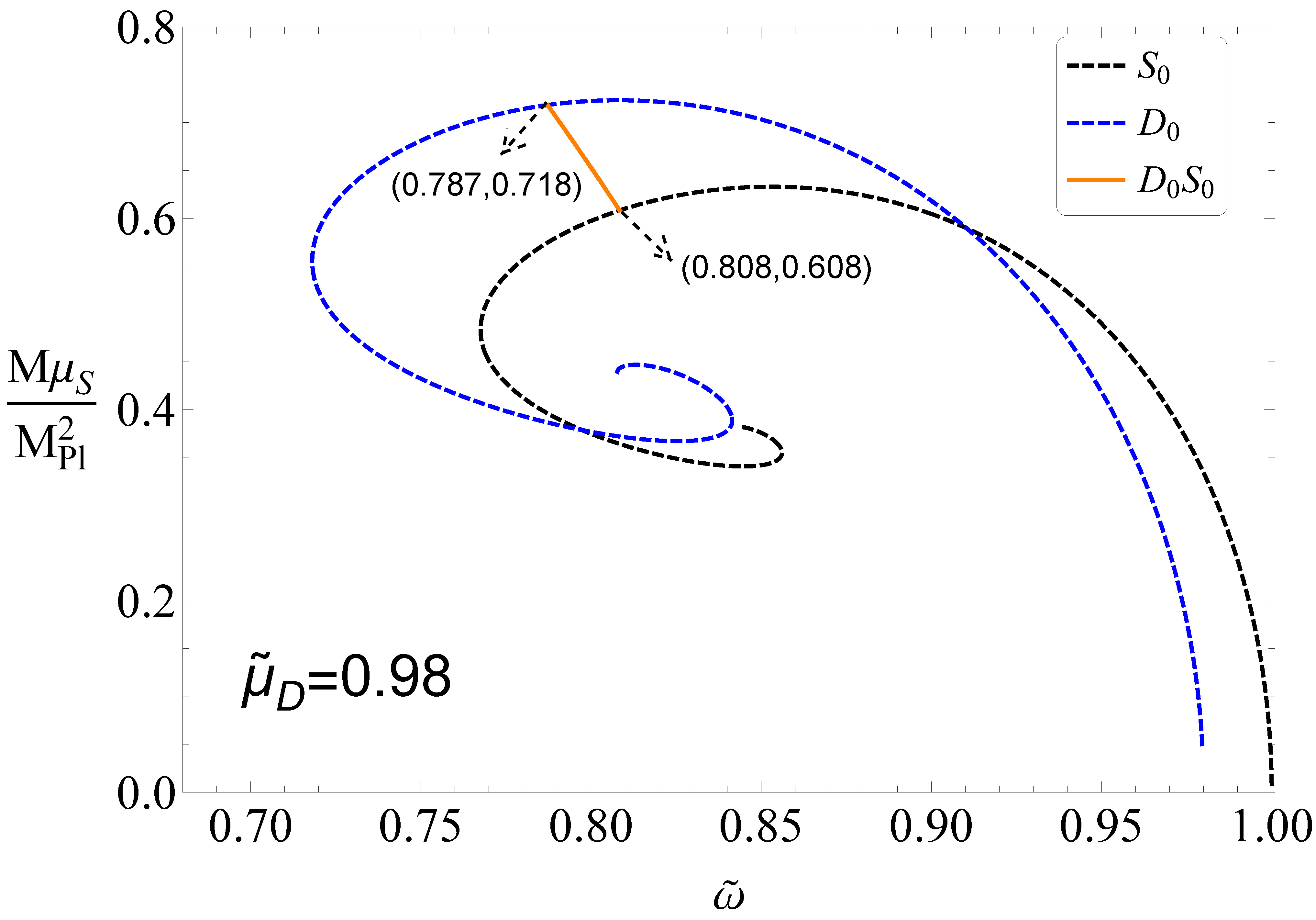}
\includegraphics[height=.24\textheight]{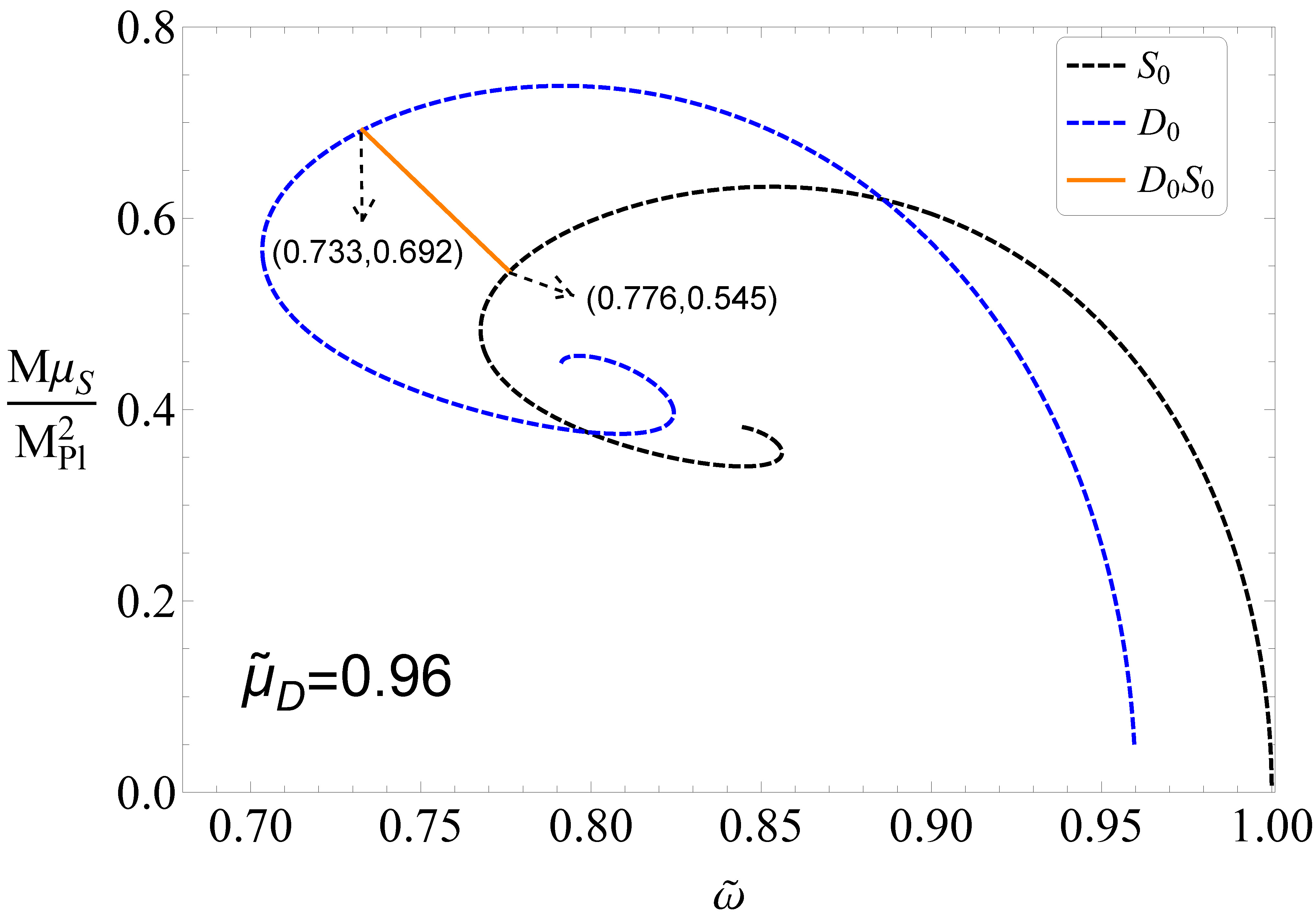}
\includegraphics[height=.24\textheight]{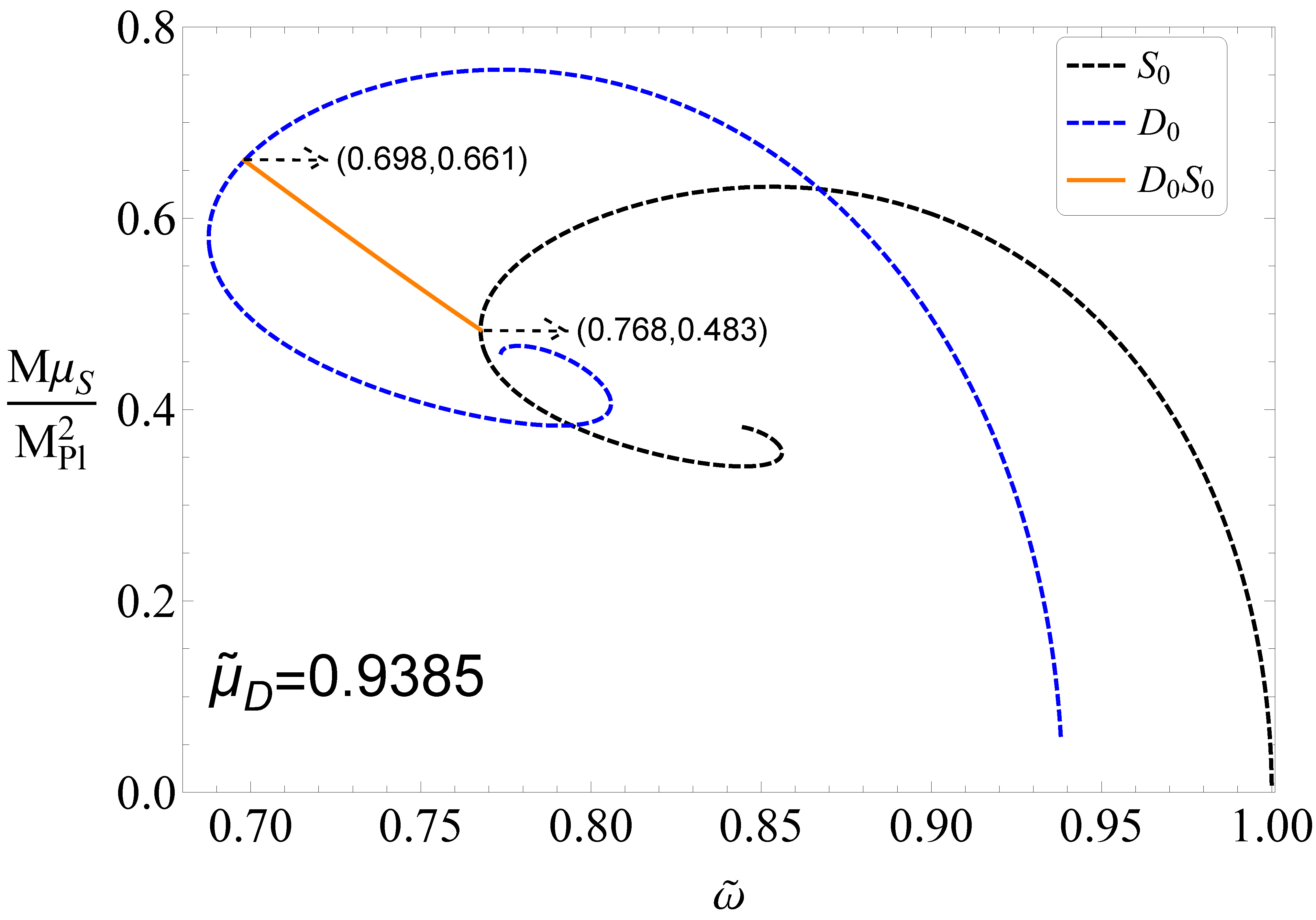}
\includegraphics[height=.24\textheight]{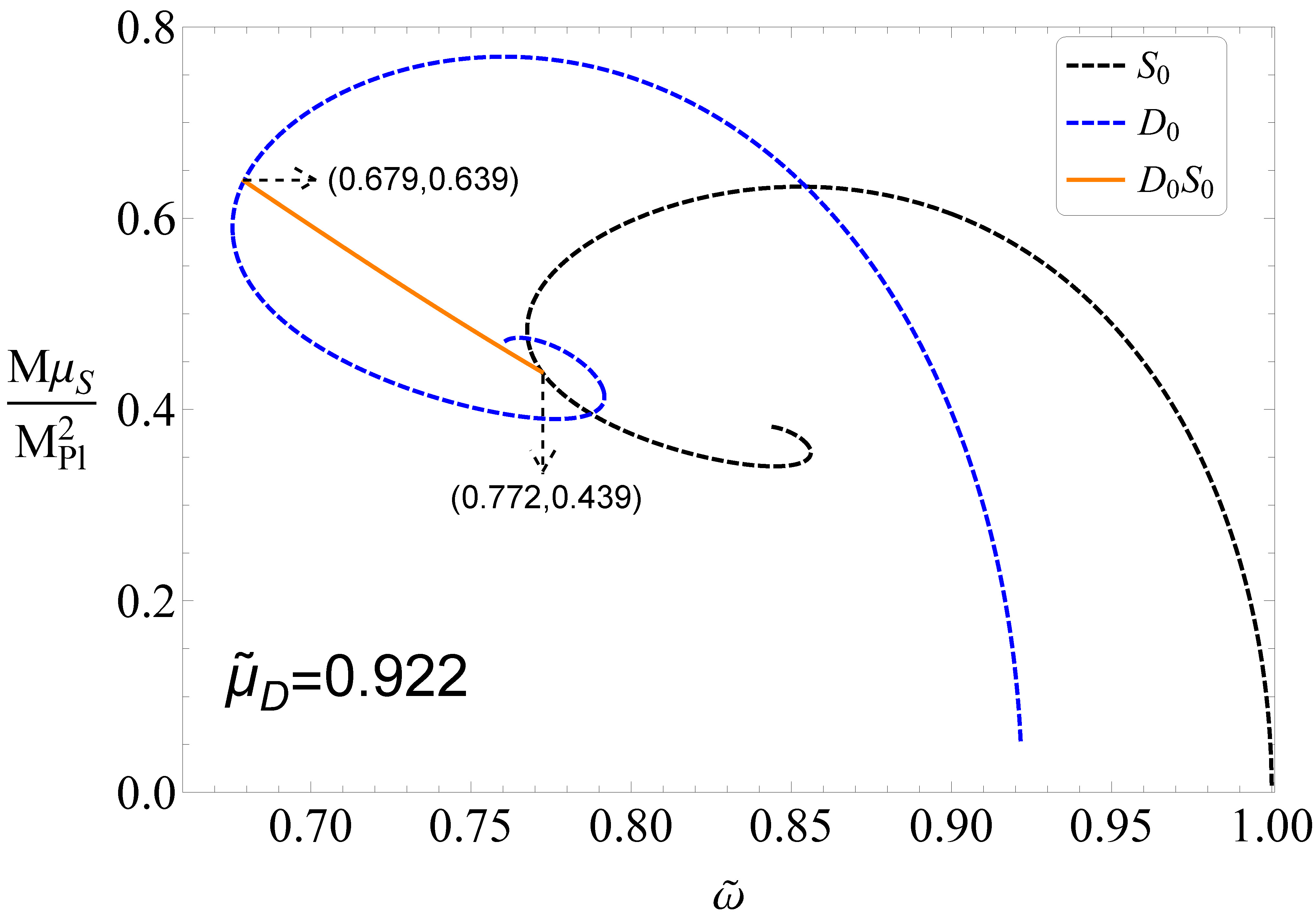}
\includegraphics[height=.24\textheight]{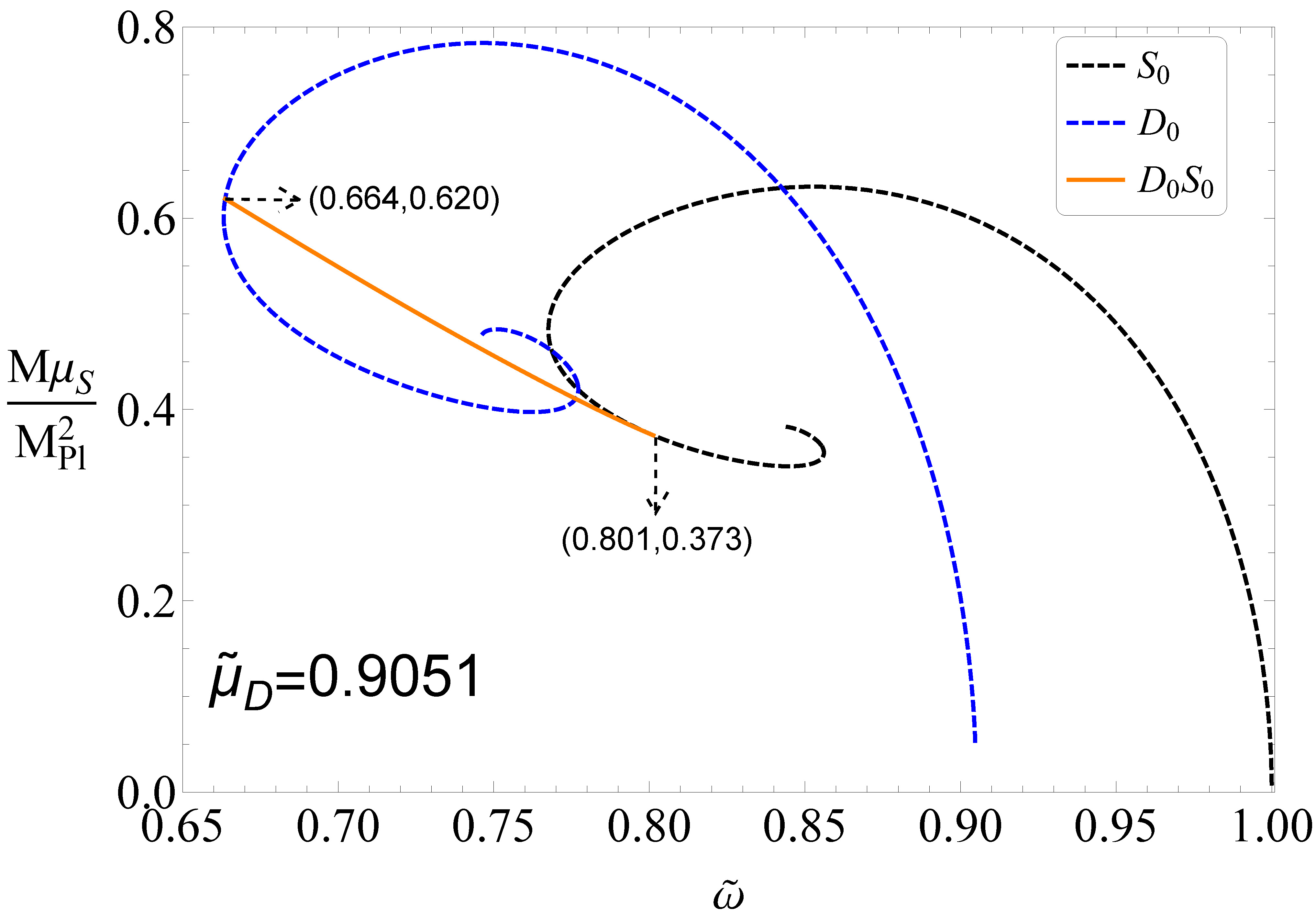}
\end{center}
\caption{The ADM mass $M$ of the DBSs as a function of the synchronized frequency $\tilde{\omega}$ for $\tilde{\mu}_D=0.9982, 0.98, 0.96, 0.9385, 0.922, 0.9051$. The black dashed line represents the $S_0$ state solutions with $\tilde{\mu}_S = 1$, the blue dashed line represents the $D_0$ state solutions, and the orange line denote the coexisting state $D_0S_0$. All solutions have $\tilde{\mu}_S=1$.}
\label{mu_k-adm}
\end{figure}

The ADM mass $M$ of the DBSs versus the synchronized frequency $\tilde\omega$ for several values of the mass $\tilde{\mu}_D$ are presented in Fig.~\ref{mu_k-adm}. The black dashed line represents the $S_0$ state solutions of the boson stars with $\tilde{\mu}_S = 1$, the blue dashed line represents the $D_0$ state solutions of the Dirac stars, and the orange line denotes the coexisting state $D_0S_0$ of the DBSs. The relationship between the ADM mass of the DBSs and the synchronized frequency is similar to the case of the $^{1}S^{2}S$ or $^{1}S^{2}P$ of the RMSBSs in Ref.~\cite{Li:2019mlk}, the solutions of DBSs have only one branch. It can be seen that the mass $M$ of the DBSs decrease with increasing synchronized frequency $\tilde\omega$. The two ends of the orange line intersect with the blue and black dashed lines, respectively. We can see from Fig.~\ref{psi_f_g} that the amplitude of the scalar field function $\tilde{\phi}$ increases with the increase of the synchronization frequency, and the amplitude of the Dirac field functions $\tilde{f}$ and $\tilde{g}$ decrease with the increase of the synchronization frequency. When the ADM mass $M$ of the DBSs reaches a maximum, the scalar field disappears, and the DBSs transform into Dirac stars, while when the ADM mass $M$ of the DBSs comes to a minimum, the Dirac fields disappear, and the DBSs transform into BSs.

In addition, when $\tilde{\mu}_D > 0.9385$, the intersection of the orange line and the black dashed line is at the first branch of the black dashed line; when $\tilde{\mu}_D = 0.9385$, the intersection is at the inflection point of the black dashed line; when $\tilde{\mu}_D < 0.9385$, the intersection is at the black dashed line the second branch. And for $0.9051 \le \tilde{\mu}_D \le 0.9982$, the intersection of the orange line and the blue dashed line is always in the first branch of the blue dashed line. It can be seen from the coordinates at both ends of the orange line marked in Fig.~\ref{mu_k-adm} that as the mass of the Dirac field decreases, the existence domain of the synchronized frequency gradually increases.

\begin{figure}[!htbp]
\begin{center}
\includegraphics[height=.16\textheight]{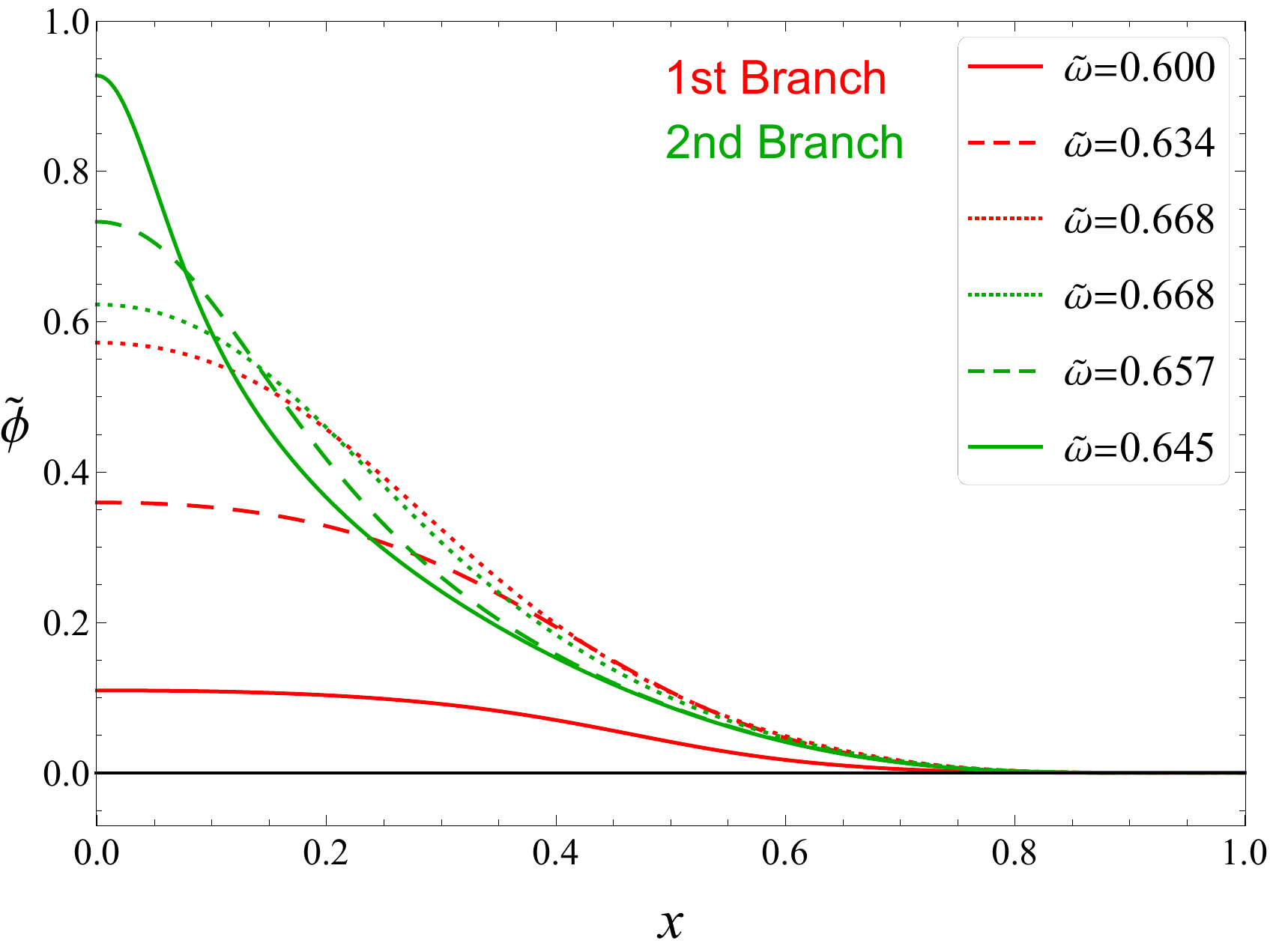}
\includegraphics[height=.16\textheight]{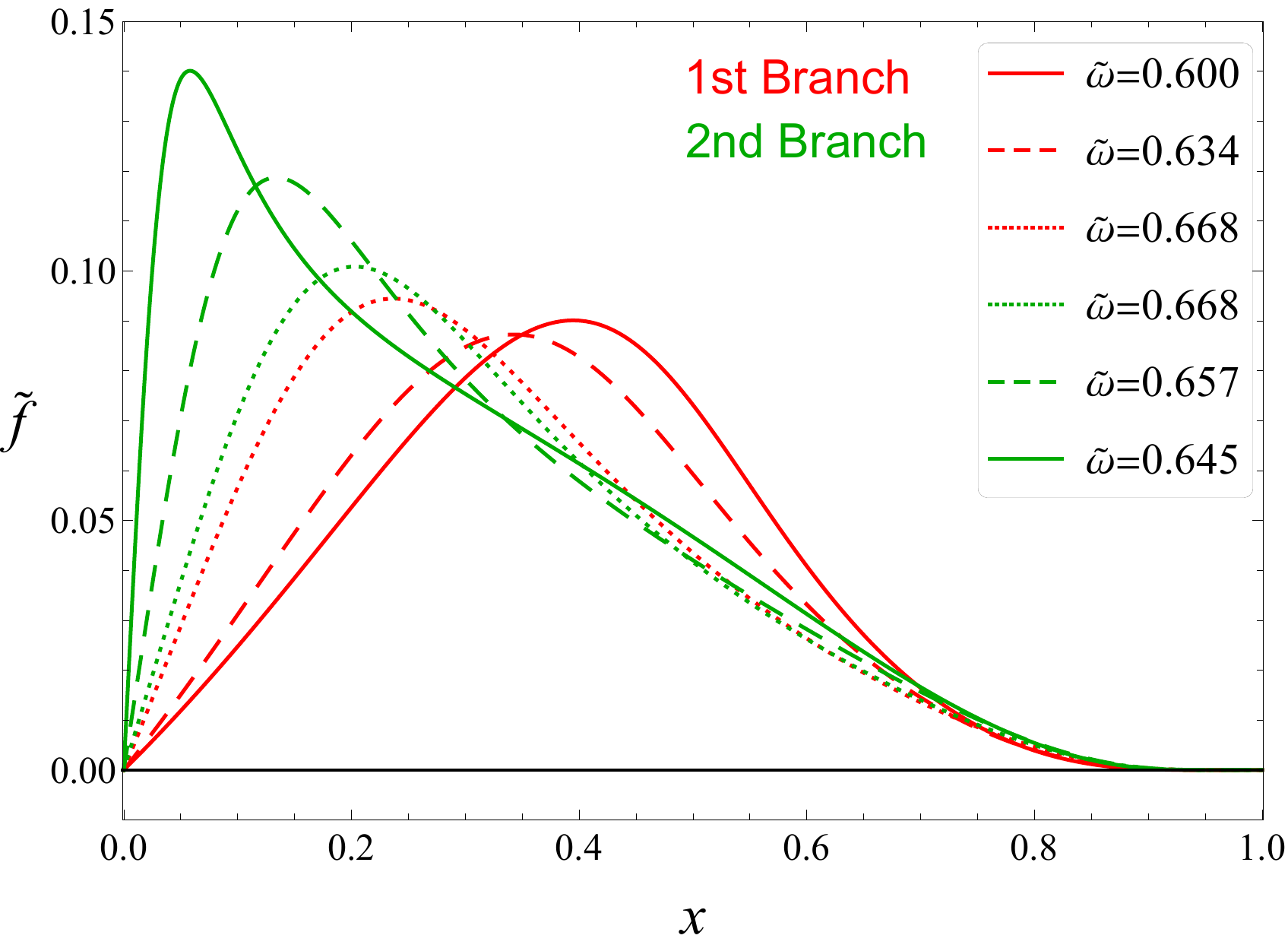}
\includegraphics[height=.16\textheight]{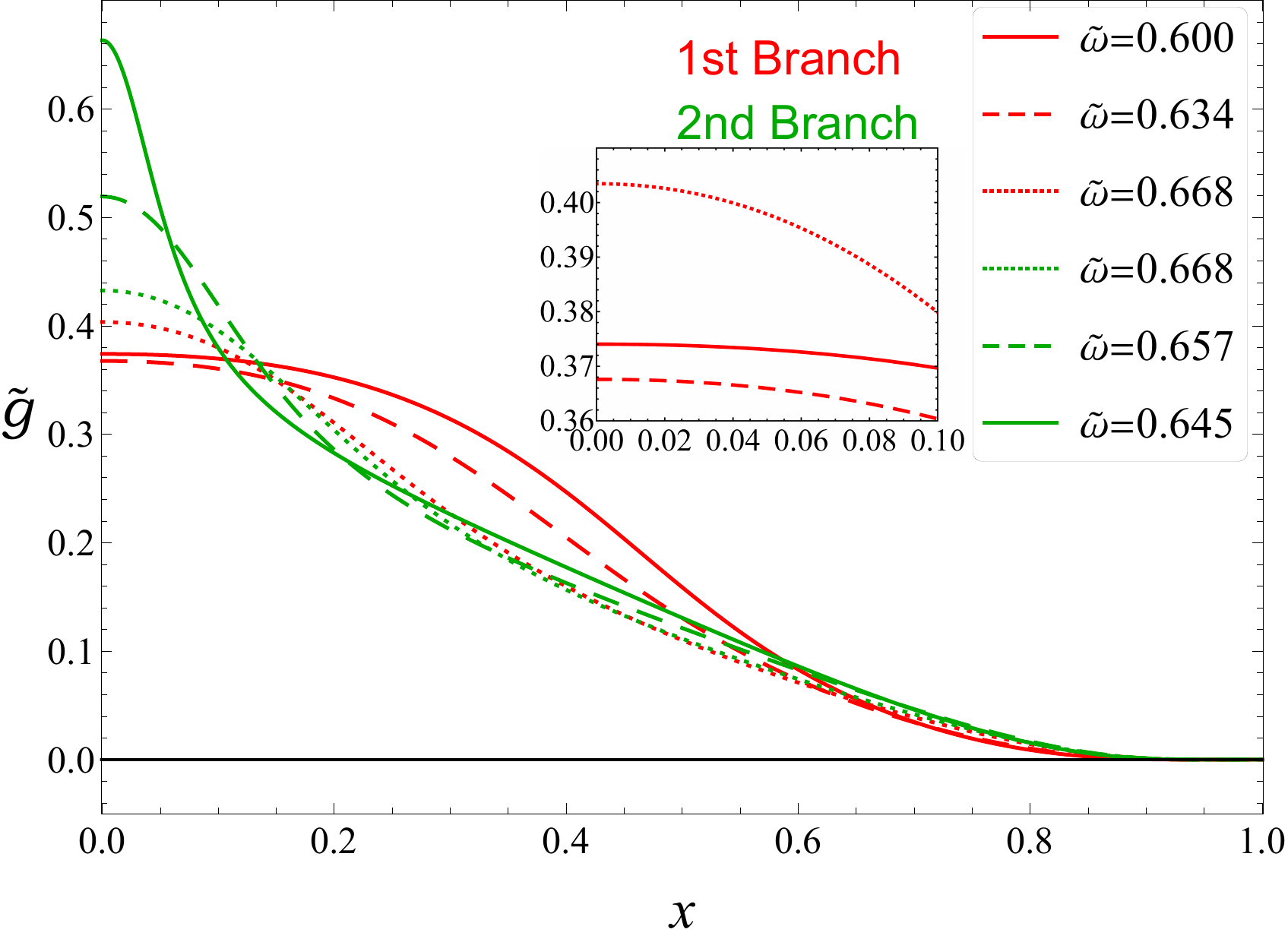}
\includegraphics[height=.16\textheight]{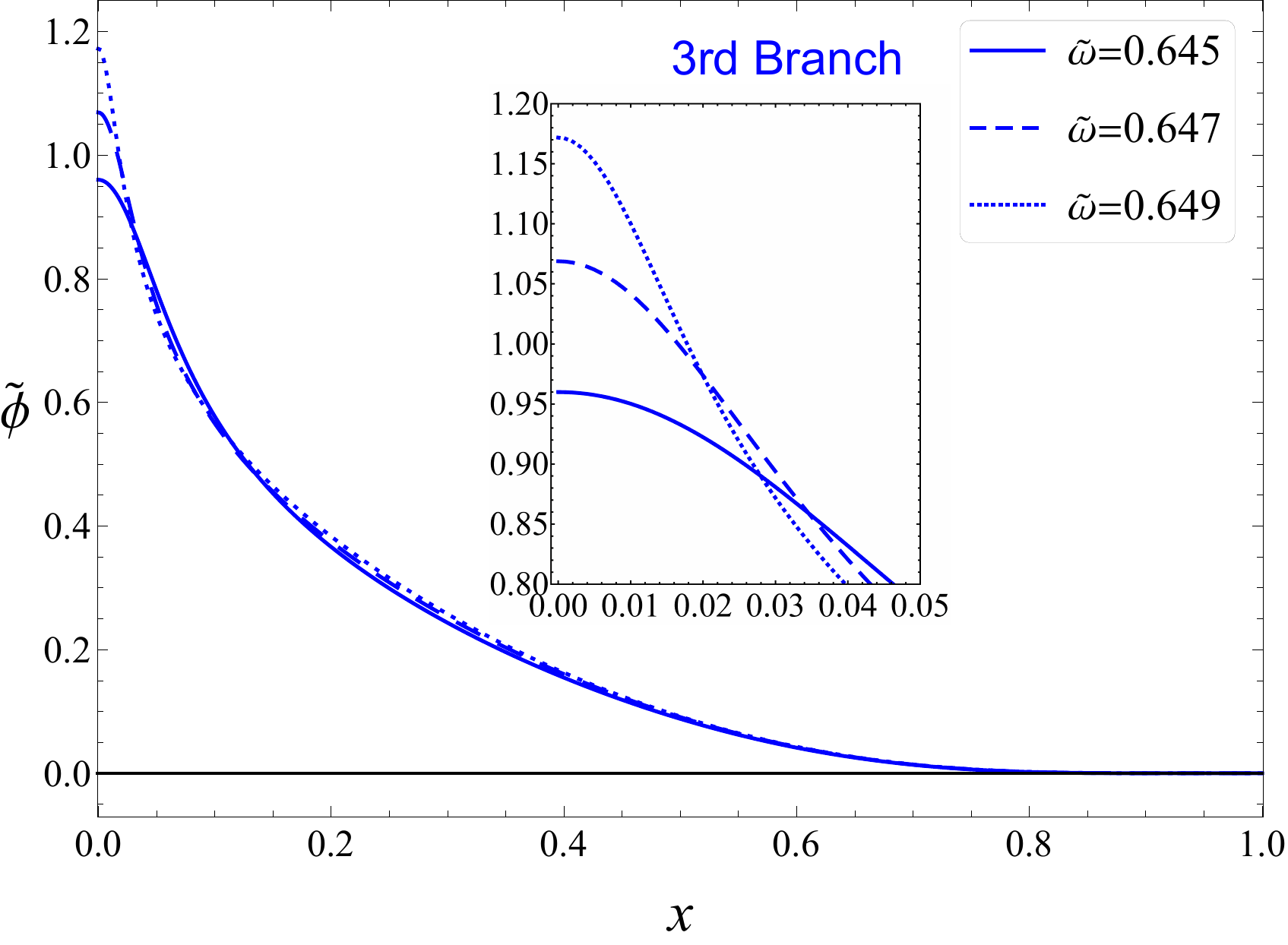}
\includegraphics[height=.16\textheight]{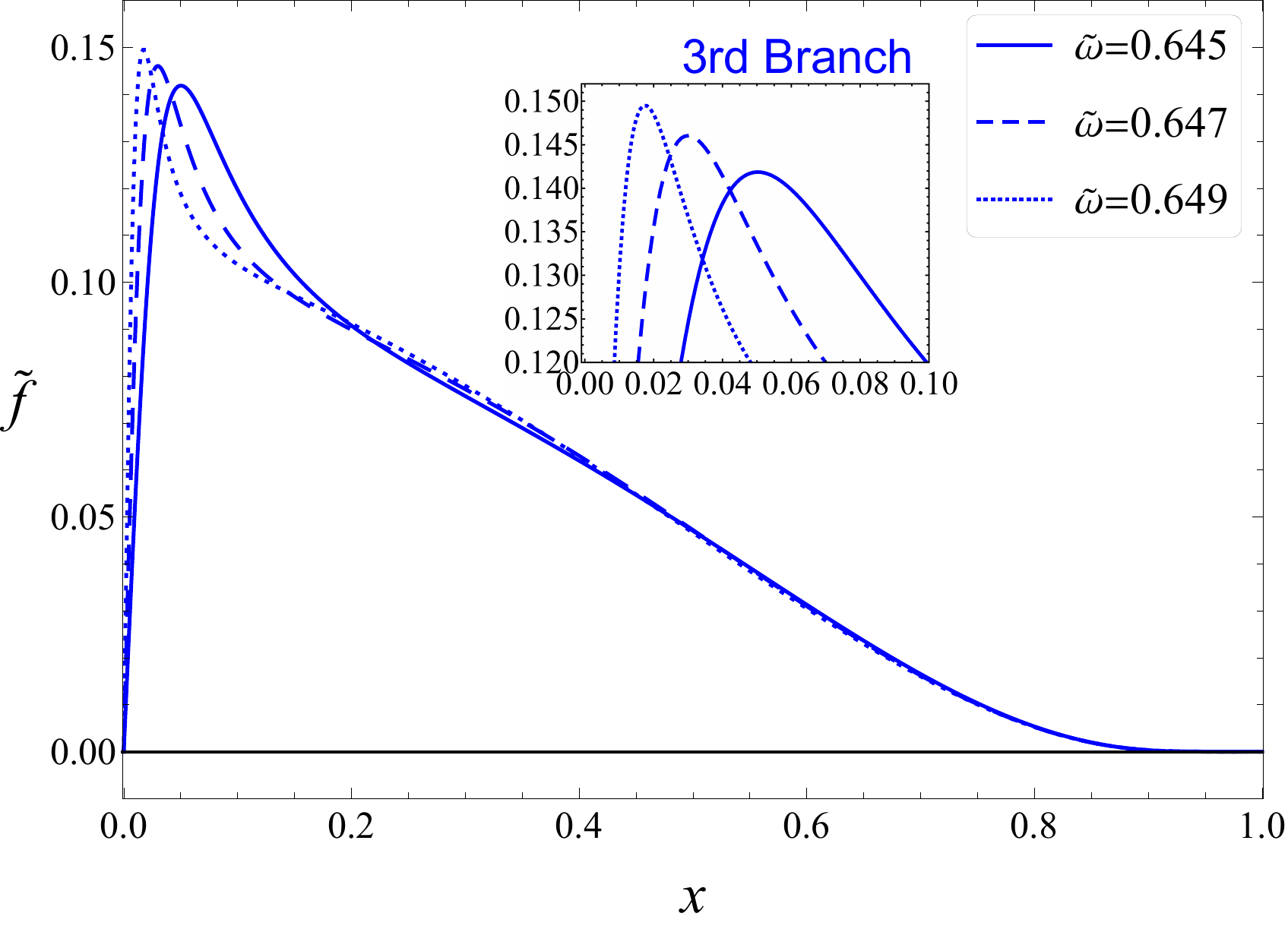}
\includegraphics[height=.16\textheight]{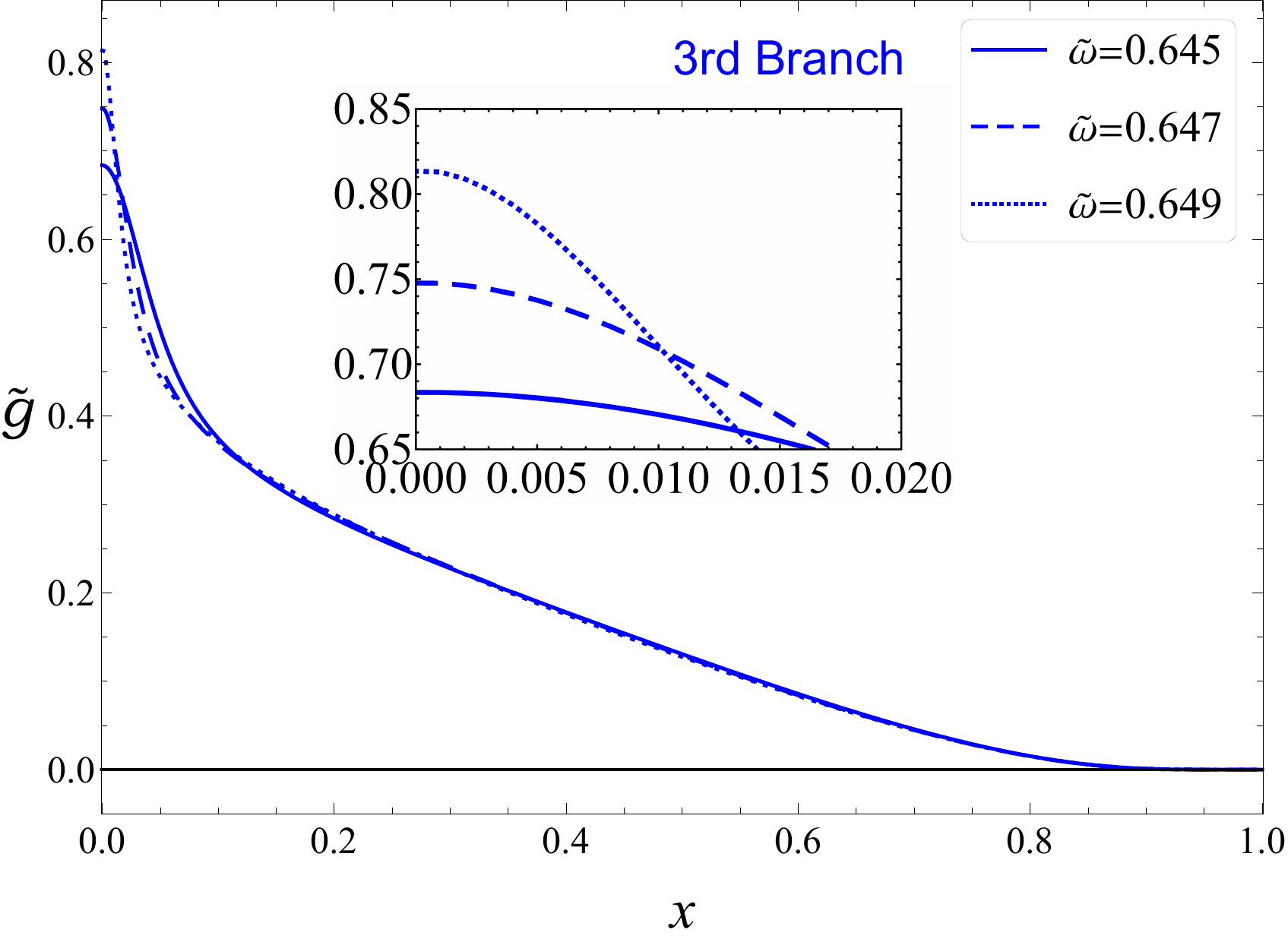}
\end{center}
\caption{Scalar field function $\tilde{\phi}$(left panel) and Dirac field functions $\tilde{f}$(middle panel) and $\tilde{g}$(right panel) as functions of $x$ with several values of synchronized frequency $\tilde{\omega}$, where the field functions on the first, second and third branches are represented by the red, orange and blue lines. All solutions have $\tilde{\mu}_D= 0.781$ and $\tilde{\mu}_S=1$.}
\label{psi_f_g_mb}
\end{figure}

\subsubsection{Multi-Branch}

For the multi-branch solution family, the amplitudes of the field functions $\tilde{\psi}$, $\tilde{f}$ and $\tilde{g}$ on each branch with the synchronized frequency is shown in Fig.~\ref{psi_f_g_mb}. The field functions on the first, second and third branches are represented by the red, orange and blue lines, respectively. For the first branch solutions, $\tilde{\psi}_{max}$ gradually increases as the synchronized frequency $\tilde{\omega}$ increases, while $\tilde{f}_{max}$ and $\tilde{g}_{max}$ first decrease and then increase. For the second branch solutions, $\tilde{\psi}_{max}$, $\tilde{f}_{max}$ and $\tilde{g}_{max}$ increase as the synchronized frequency $\tilde{\omega}$ decreases. For the third branch solutions, $\tilde{\psi}_{max}$, $\tilde{f}_{max}$ and $\tilde{g}_{max}$ decrease with increasing synchronized frequency.

\begin{figure}[!htbp]
\begin{center}
\includegraphics[height=.24\textheight]{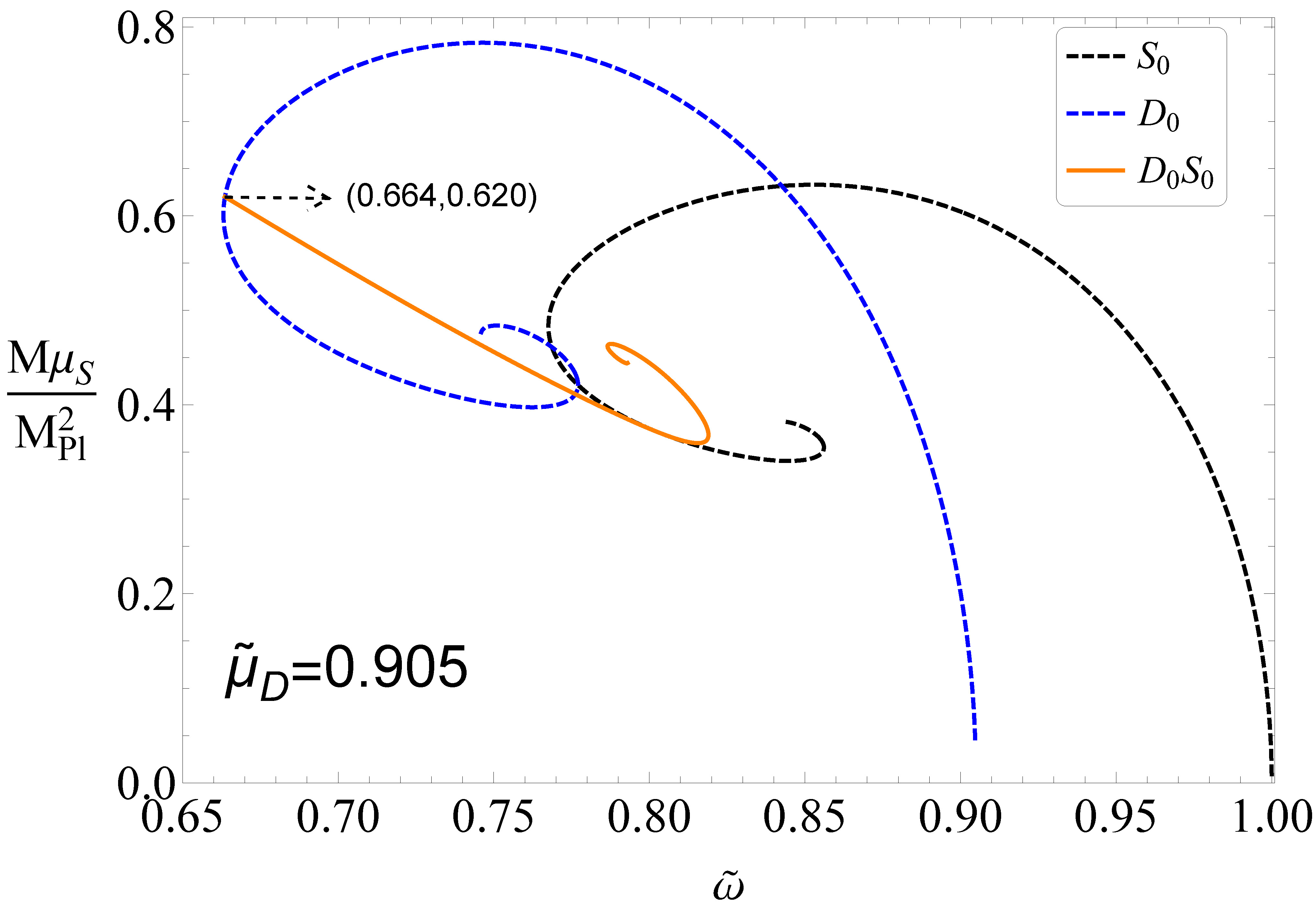}
\includegraphics[height=.24\textheight]{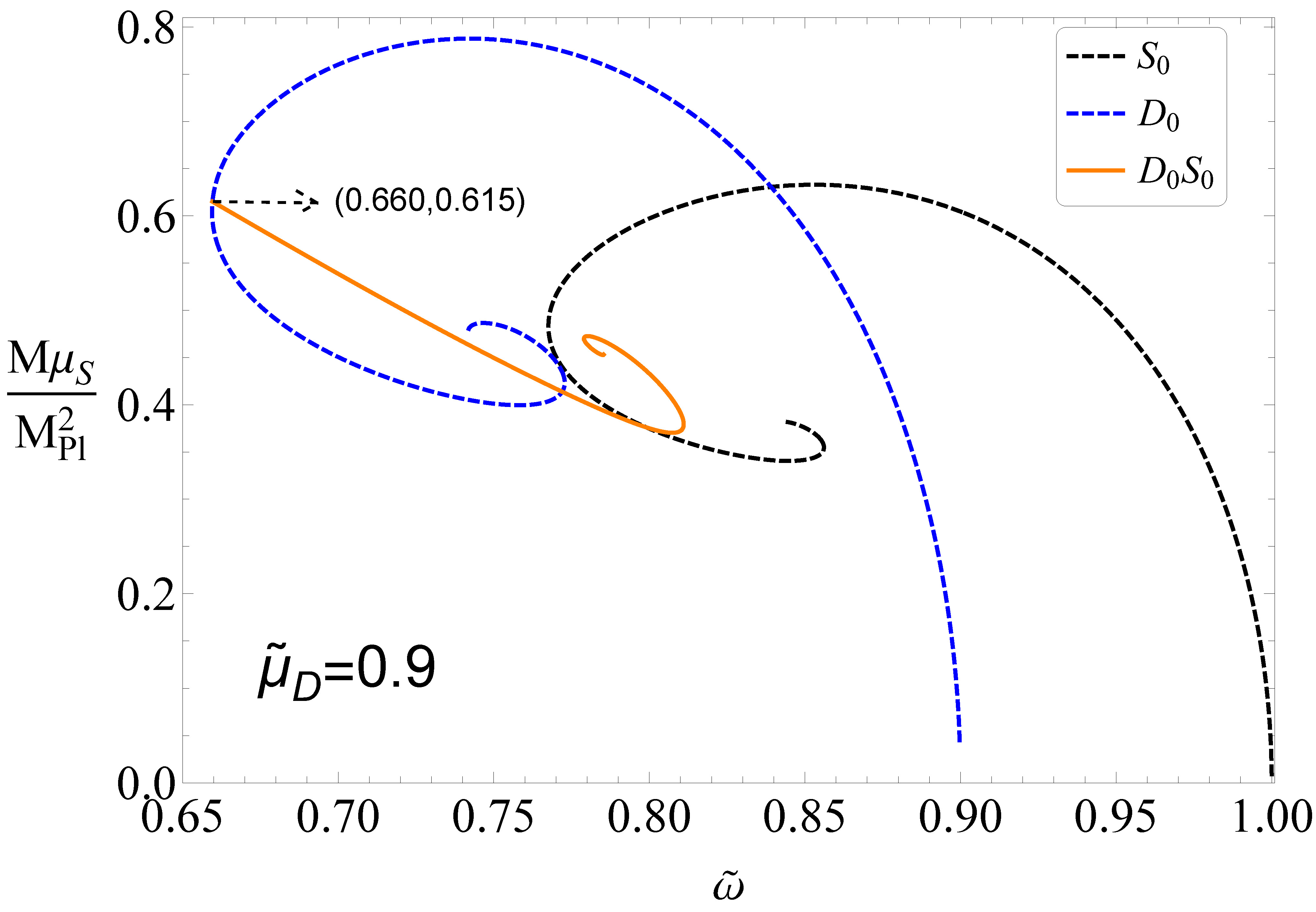}
\includegraphics[height=.24\textheight]{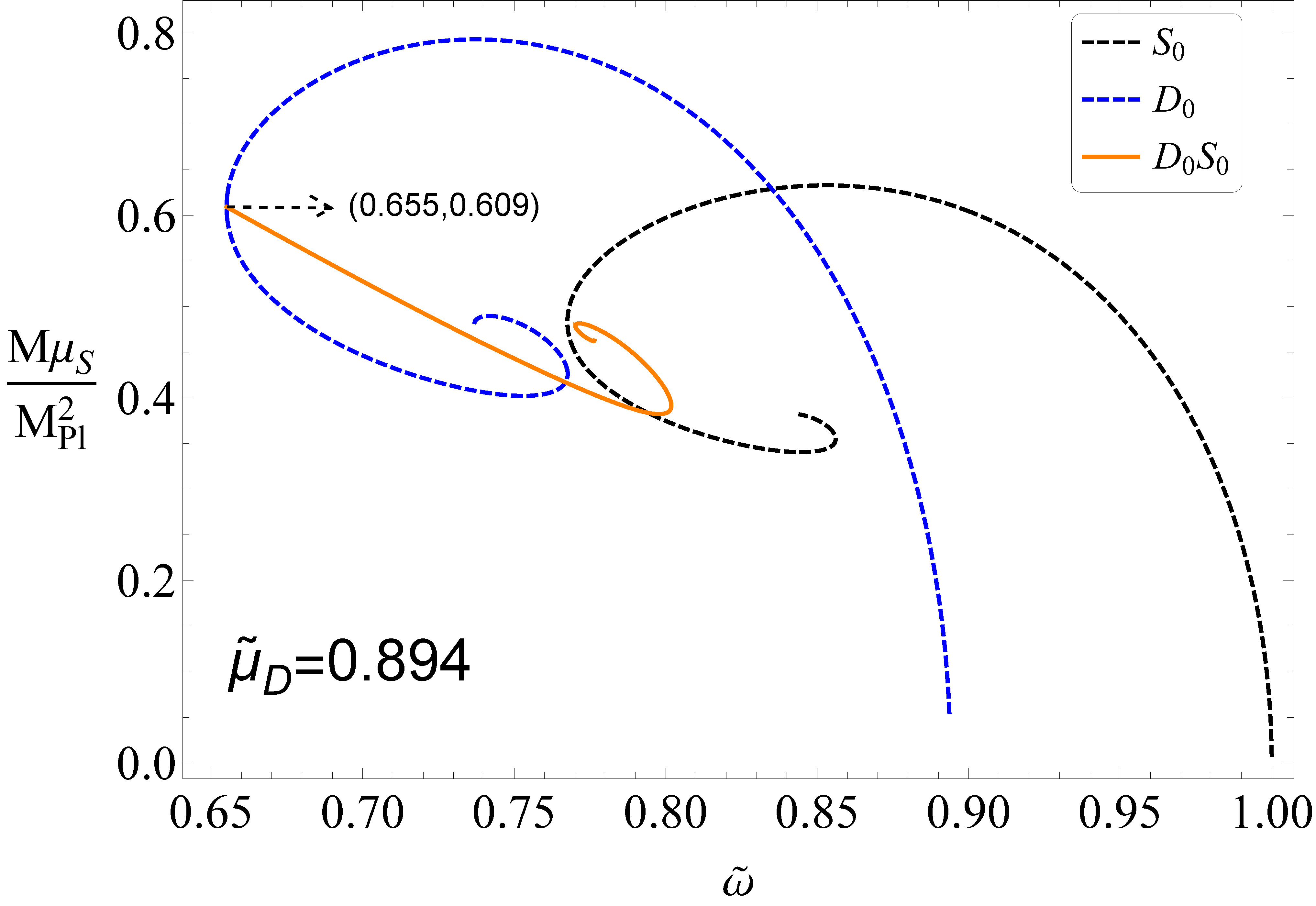}
\includegraphics[height=.24\textheight]{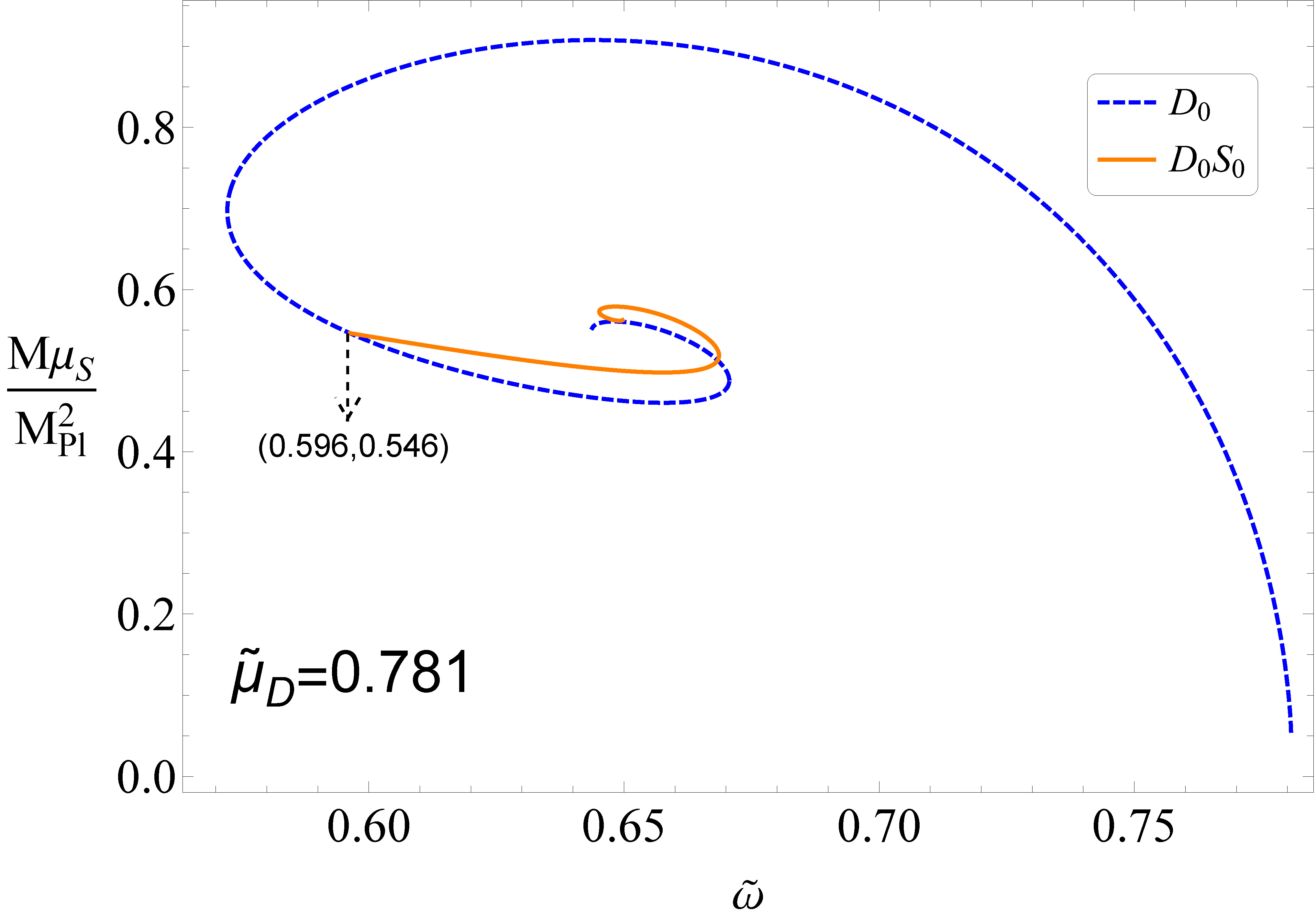}
\includegraphics[height=.24\textheight]{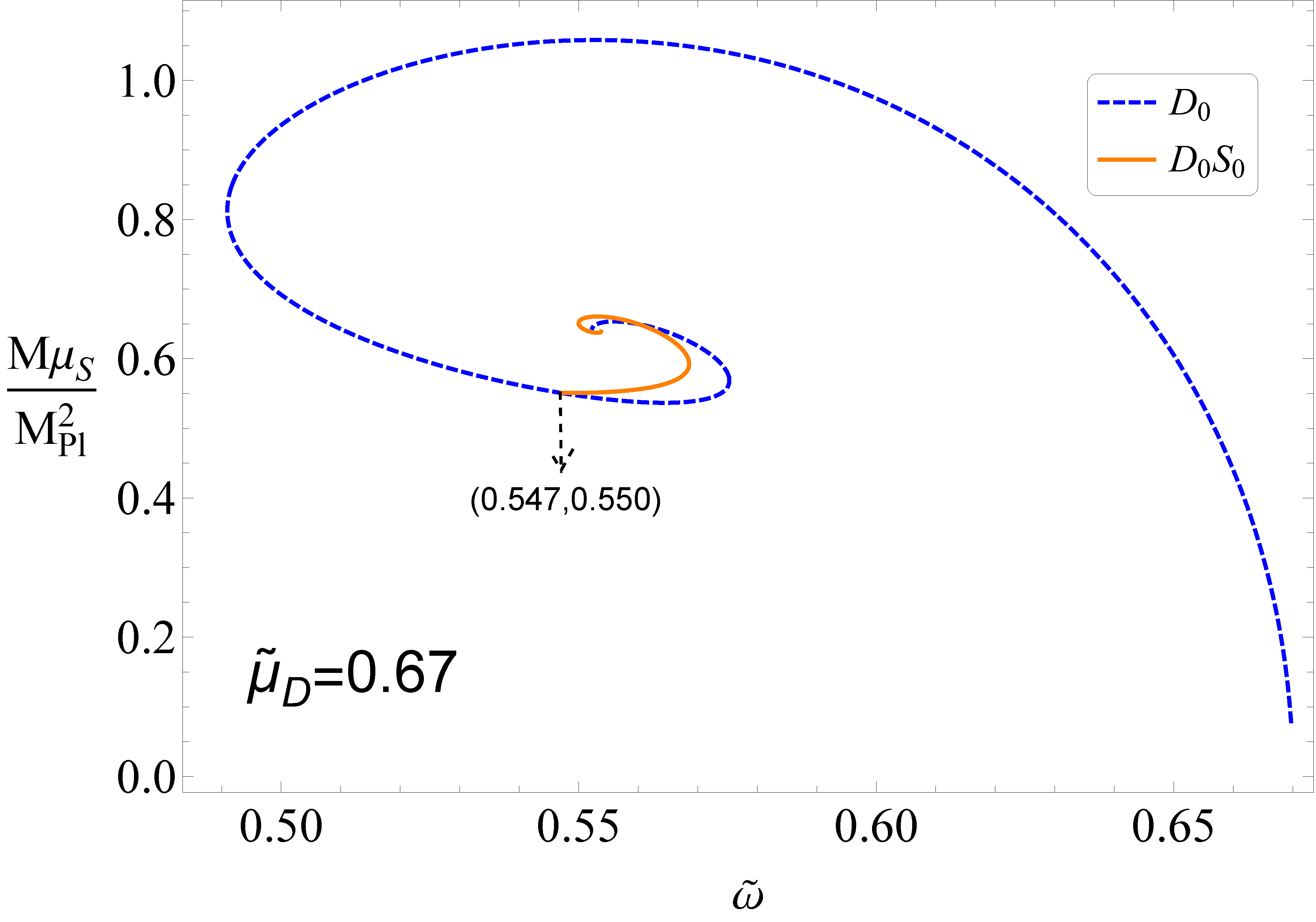}
\includegraphics[height=.24\textheight]{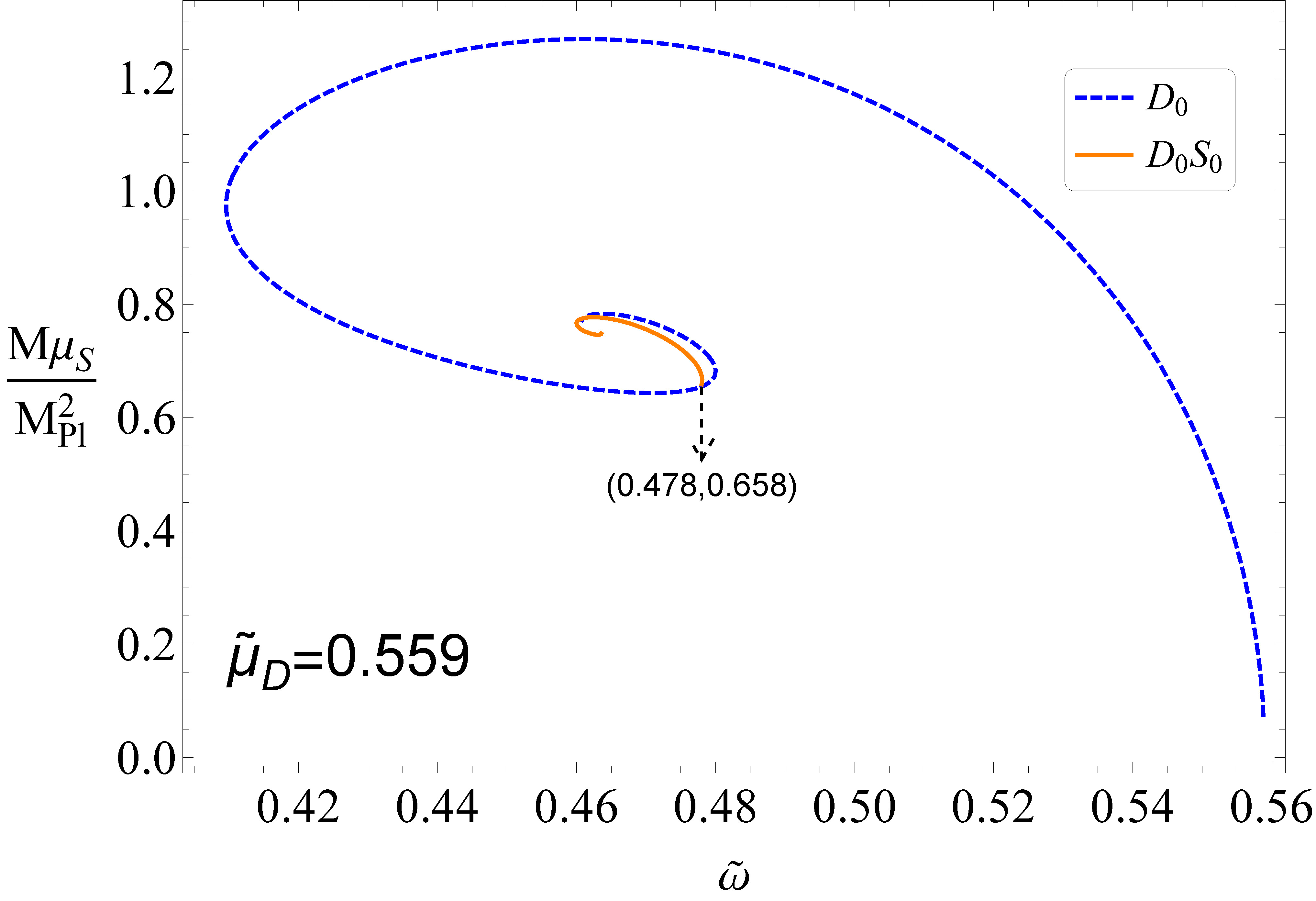}
\end{center}
\caption{The ADM mass $M$ of the DBSs as a function of the synchronized frequency $\tilde{\omega}$ for $\tilde{\mu}_D=0.905, 0.9, 0.894, 0.781, 0.67, 0.559$. The black dashed line represents the $S_0$ state solutions with $\tilde{\mu}_S = 1$, the blue dashed line represents the $D_0$ state solutions, and the orange line denote the coexisting state $D_0S_0$. All solutions have $\tilde{\mu}_S=1$.}
\label{mu_k-adm2}
\end{figure}

It can be seen from Fig.~\ref{mu_k-adm} that when $\tilde{\mu}_D = 0.9051$, the orange line is tangent to the second branch of the black dashed line. If we continue to reduce the mass $\tilde{\mu}_D$, we can get the solutions of DBSs with multiple branches. As shown in Fig.~\ref{mu_k-adm2}, we show the ADM mass $M$ of the DBSs versus the synchronized frequency $\tilde\omega$ for several values of the mass $\tilde{\mu}_D$ less than $0.9051$. The black dashed line is the same as in Fig.~\ref{mu_k-adm}, and the blue dashed line represents the solutions of the $D_0$ state when $\tilde{\mu}_D$ takes the value marked in the figure. Each orange line has multiple branches, and the whole line is a spiral, the variation of the mass $M$ of the DBSs with the synchronized frequency is more complicated than that in Fig.~\ref{mu_k-adm}. For any solution on the orange line, the amplitude of the Dirac field in the corresponding solution is always not zero. In other words, there is no solution similar to the intersection of the orange line and the black dashed line in Fig.~\ref{mu_k-adm}. When $\tilde{\mu}_D > 0.894$, the intersection of the orange line and the blue dashed line (the point with the coordinates marked in Fig. \ref{mu_k-adm2}) is located on the first branch of the blue dashed line; when $\tilde{\mu}_D = 0.894$, the intersection is precisely at the inflection point of the blue dashed line; when $0.559 \le \tilde{\mu}_D < 0.894$, the intersection is on the second branch of the blue dashed line.

In Table \ref{table1}, we show the existence domain of the synchronized frequency $\tilde{\omega}$ for several values of the Dirac field mass $\tilde{\mu}_D$. As the Dirac field mass $\tilde{\mu}_D$ decreases, the intervals of $B_1$, $B_2$ and $B_3$ are gradually narrowed. When $\tilde{\mu}_D$ decreases to $0.559$, the interval of $B_1$ becomes very narrow; while the interval of $B_2$ and $B_3$ change relatively little. Moreover, as the Dirac field mass $\tilde{\mu}_D$ decreases, $M_{max}$ first decreases and then increases, while $M_{min}$ keeps increasing.

\begin{table}[!htbp]
  \centering
    \begin{tabular}{|c|c|c|c|c|c|}
    \hline
     $\tilde{\mu}_D$  & $B_1$    & $B_2$    & $B_3$    & $M_{max}$  & $M_{min}$ \\
    \hline
    $0.905$  & $0.664\sim0.819$ & $0.787\sim0.819$ & $0.787\sim0.793$ & $0.620$ & $0.359$ \\
    \hline
    $0.900$  & $0.660\sim0.811$ & $0.779\sim0.811$ & $0.779\sim0.785$ & $0.615$ & $0.370$ \\
    \hline
    $0.894$  & $0.655\sim0.802$ & $0.770\sim0.802$ & $0.770\sim0.776$ & $0.609$ & $0.382$ \\
    \hline
    $0.781$  & $0.596\sim0.669$ & $0.645\sim0.669$ & $0.645\sim0.650$ & $0.579$ & $0.498$ \\
    \hline
    $0.670$  & $0.547\sim0.569$ & $0.550\sim0.569$ & $0.550\sim0.554$ & $0.661$ & $0.551$ \\
    \hline
    $0.559$  & $0.4780\sim0.4781$ & $0.460\sim0.478$ & $0.460\sim0.464$ & $0.777$ & $0.658$ \\
    \hline
    \end{tabular}
\caption{Maximum and minimum ADM masses($M_{max}$ and $M_{min}$) of the DBSs, and the existence domain of the synchronized frequency $\tilde{\omega}$ for several values of the Dirac field mass $\tilde{\mu}_D$. $B_1$, $B_2$ and $B_3$ represent the first, second and third branches of the orange line in Fig.~\ref{mu_k-adm2}, respectively. All solutions have $\tilde{\mu}_S=1$.}
\label{table1}
\end{table}

\subsection{ Nonsynchronized frequency }

Similar to the case of synchronized frequency, we divide the families of nonsynchronized frequency solutions of DBSs into three categories: the \textit{one-branch-A} solution family, \textit{multi-branch} solution family, and \textit{one-branch-B} solution family. The value ranges of $\tilde{\omega}_S$ corresponding to these three solution families are $0.7676 \le \tilde{\omega}_S \le 0.9082$, $0.7128 \le \tilde{\omega}_S \le 0.7675$ and $0.6794 \le \tilde{\omega}_S \le 0.7127$, respectively. The remainder of this section discusses these three families of solutions in detail.

\subsubsection{One-Branch-A}

\begin{figure}[!htbp]
\begin{center}
\includegraphics[height=.16\textheight]{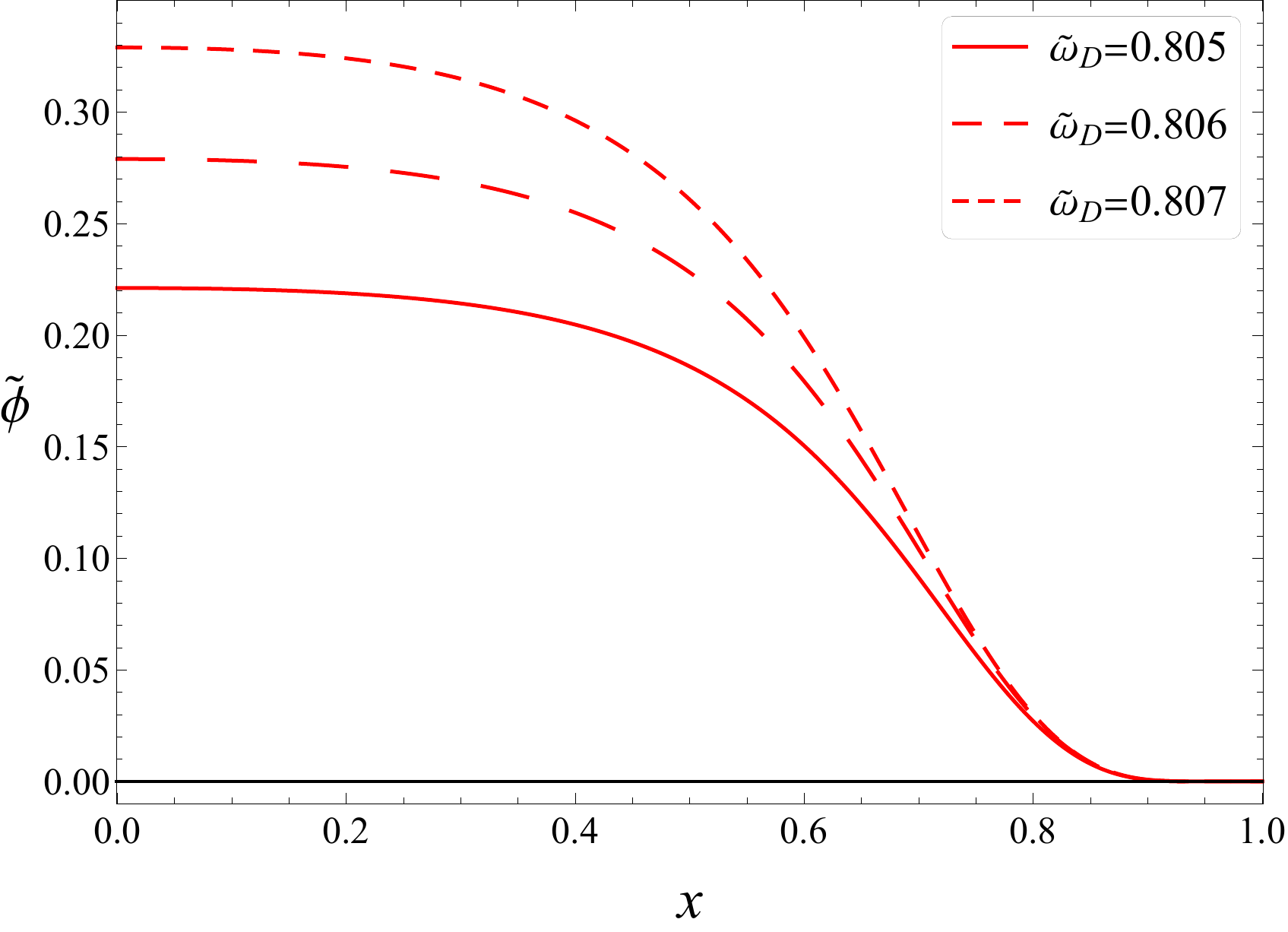}
\includegraphics[height=.16\textheight]{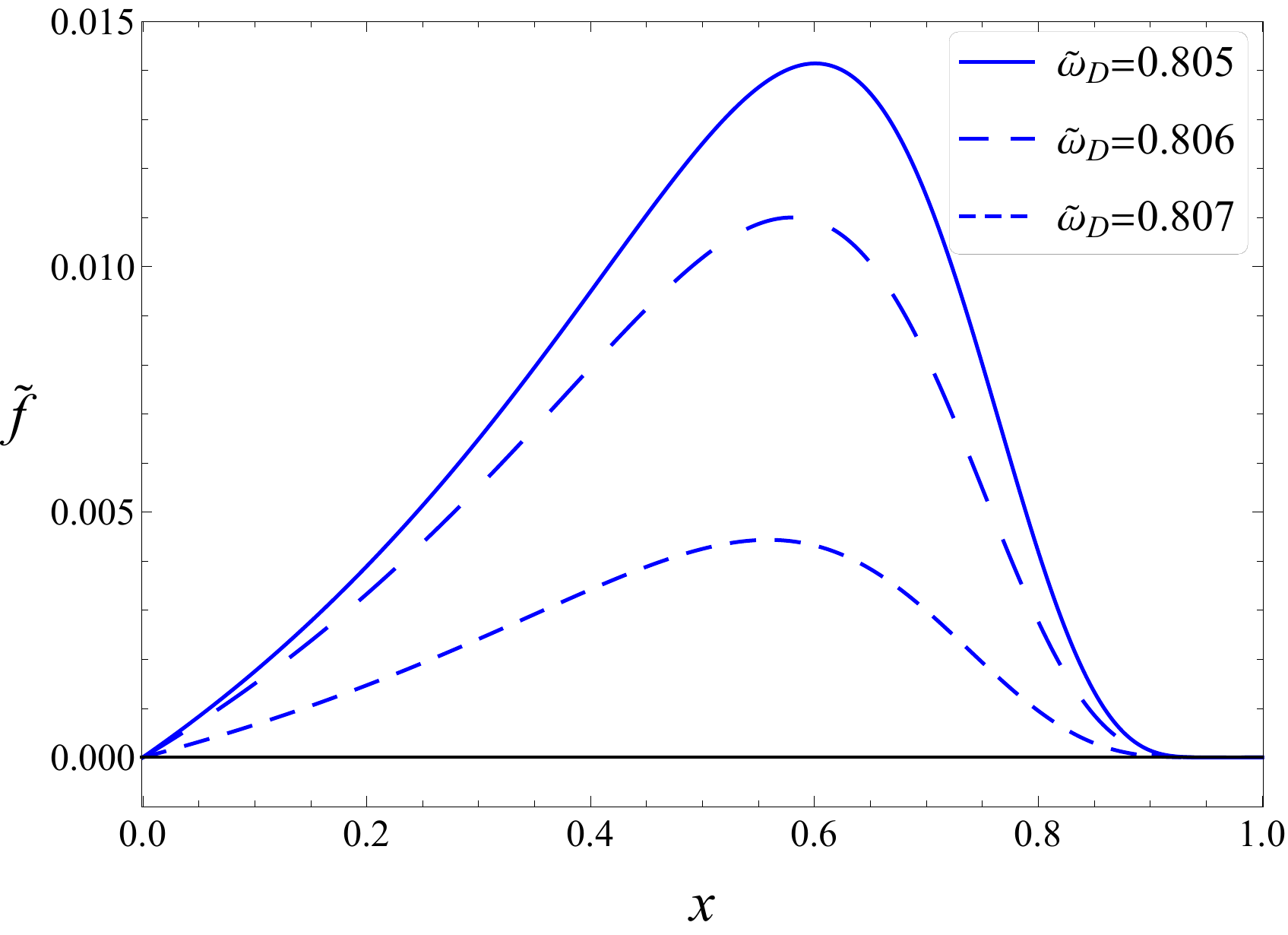}
\includegraphics[height=.16\textheight]{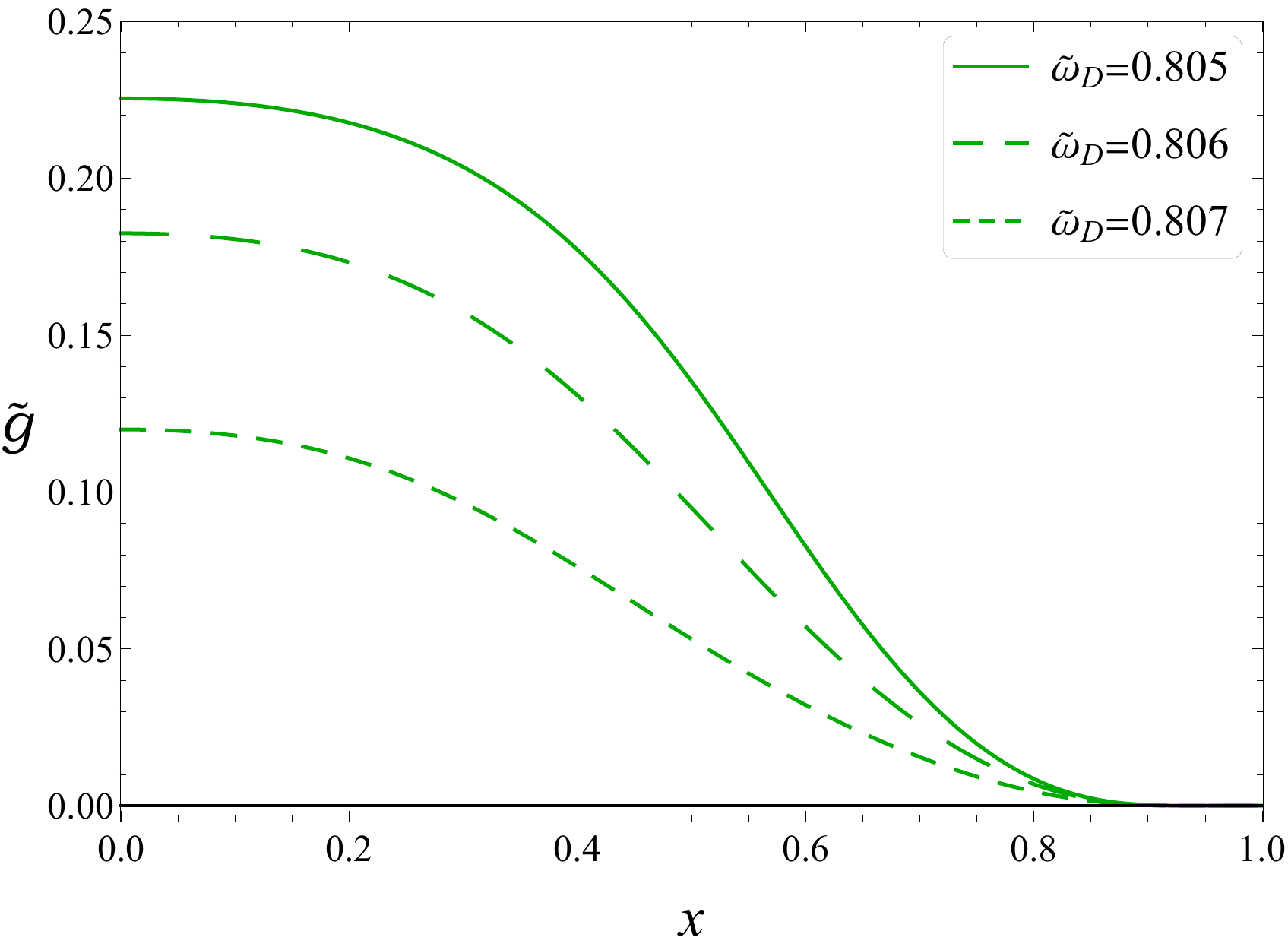}
\end{center}
\caption{Scalar field function $\tilde{\phi}$(left panel) and Dirac field functions $\tilde{f}$(middle panel) and $\tilde{g}$(right panel) as functions of $x$ with $\tilde{\omega}_D = 0.805, 0.806, 0.807$. All solutions have $\tilde{\omega}_S= 0.79$ and $\tilde{\mu}_S = \tilde{\mu}_D = 1$.}
\label{psi_f_g_ob-a}
\end{figure}

For \textit{one-branch-A} solution families, the profiles of the field functions $\tilde{\phi}$, $\tilde{f}$ and $\tilde{g}$ with several values of the nonsynchronized frequency $\tilde{\omega}_D$ in the solutions of the $D_0S_0$ state are shown in Fig.~\ref{psi_f_g_ob-a}. Similar to the case in Fig.~\ref{psi_f_g}, as the nonsynchronized frequency $\tilde{\omega}_D$ increases, $\tilde{\phi}_{max}$ increases and $\tilde{f}_{max}$ and $\tilde{g}_{max}$ decreases.

\begin{figure}[!htbp]
\begin{center}
\includegraphics[height=.24\textheight]{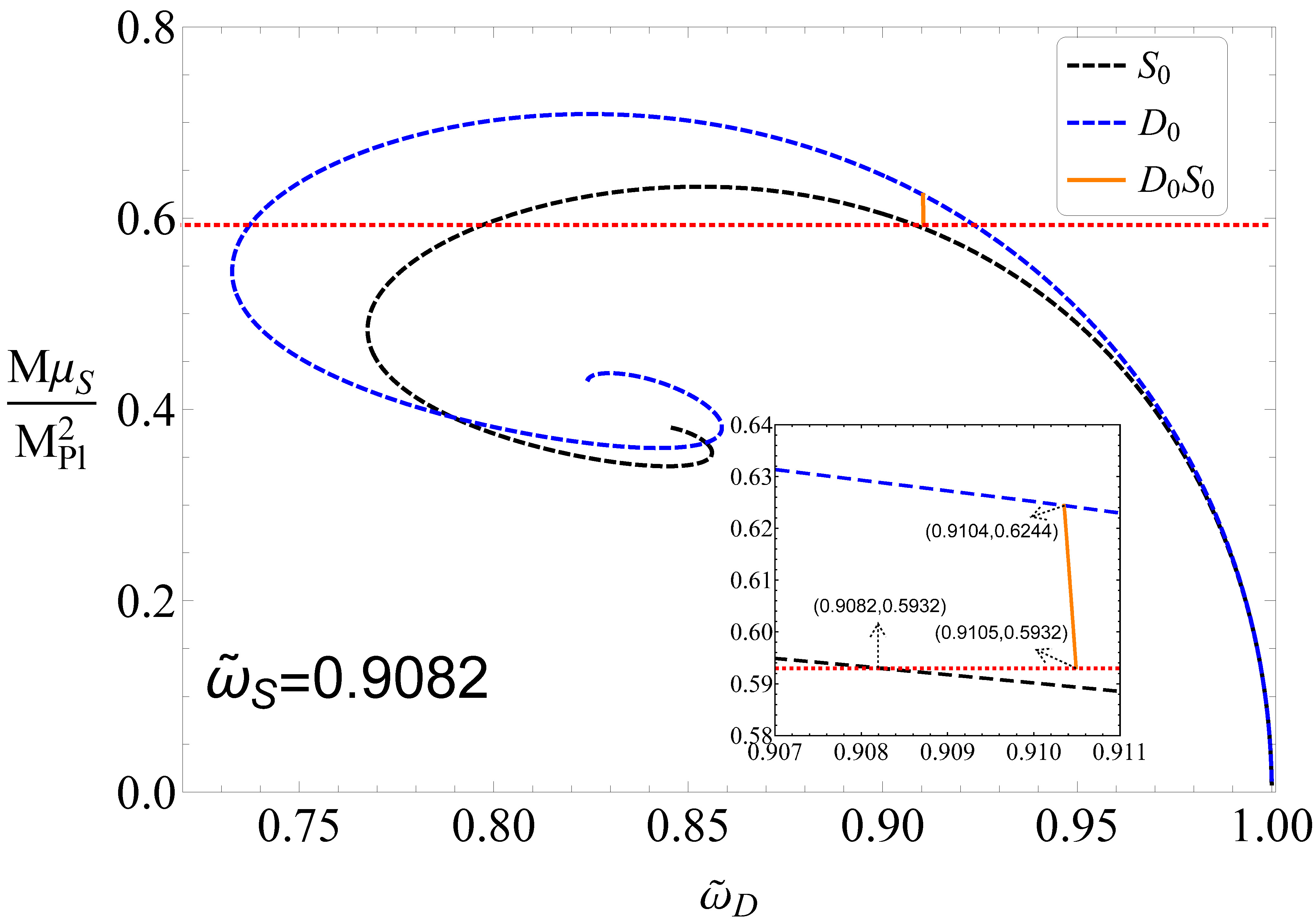}
\includegraphics[height=.24\textheight]{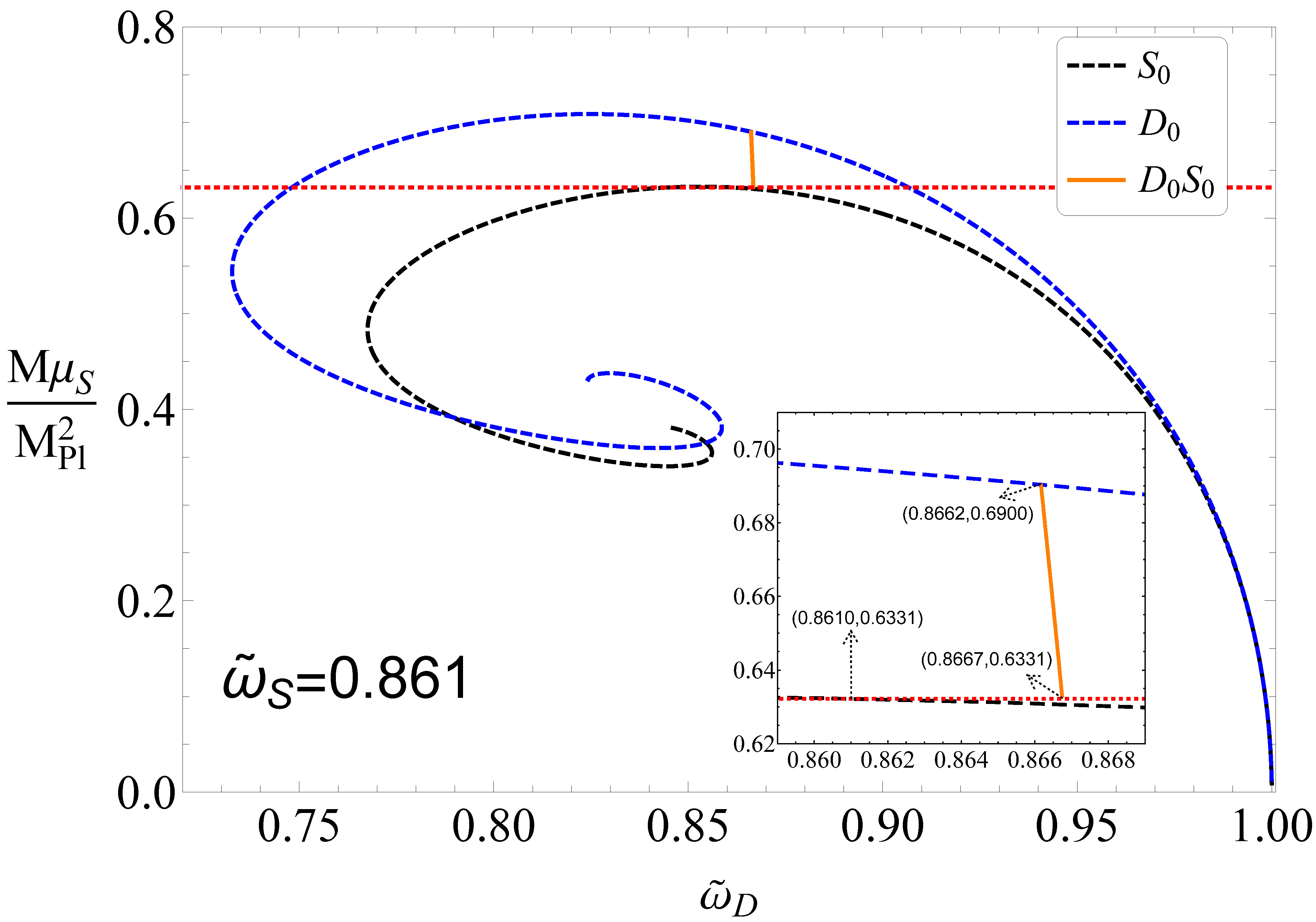}
\includegraphics[height=.24\textheight]{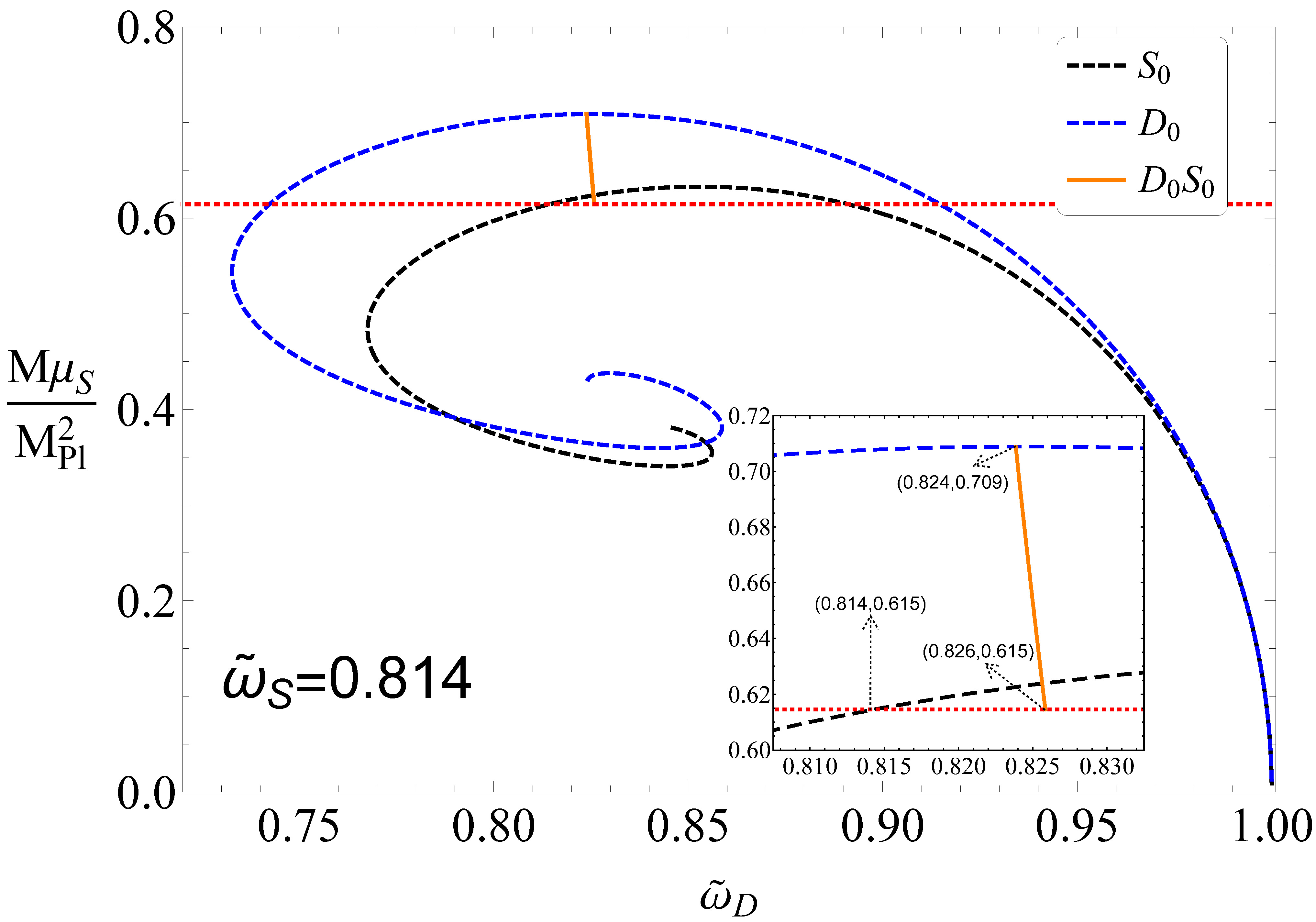}
\includegraphics[height=.24\textheight]{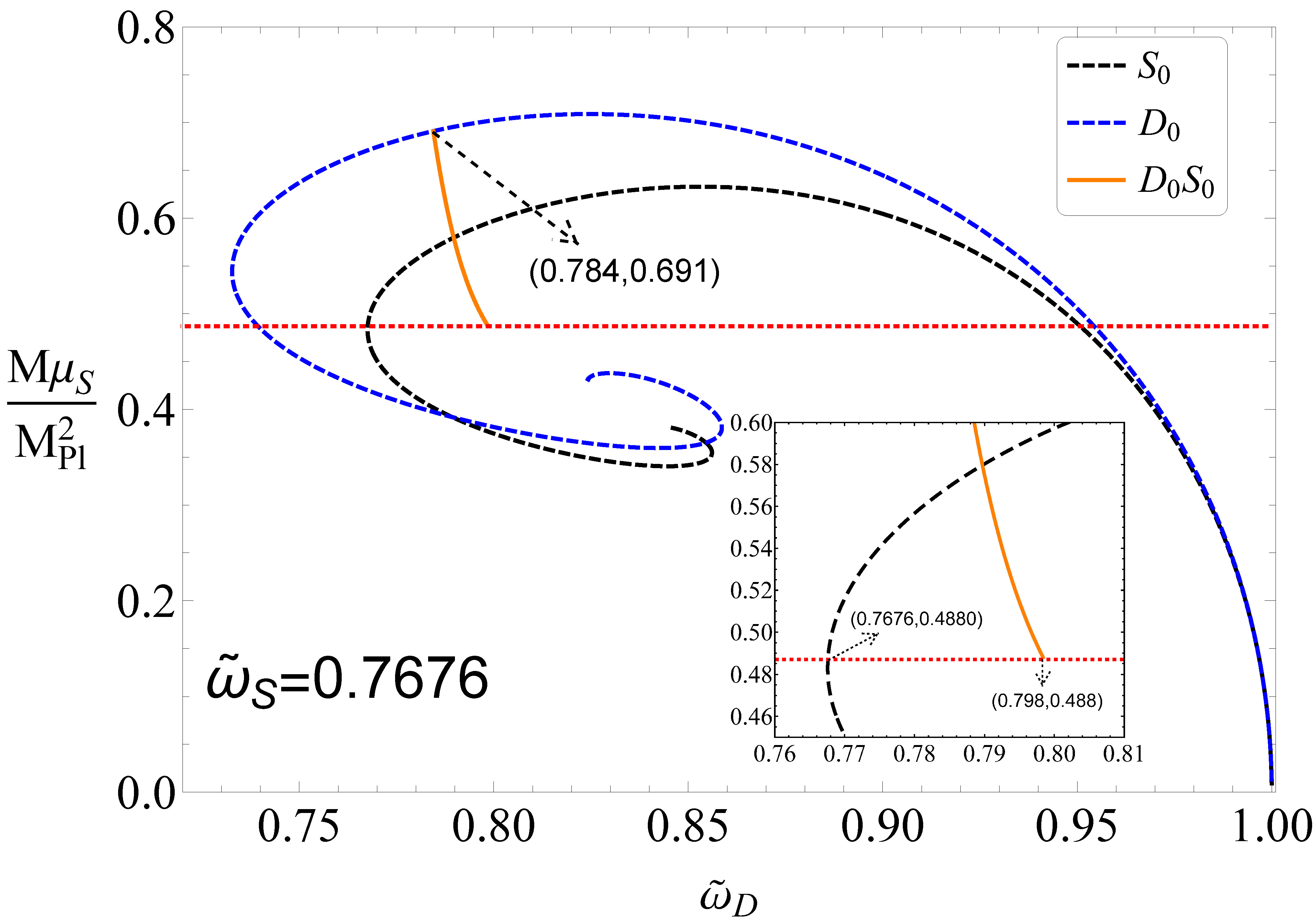}
\end{center}
\caption{The ADM mass $M$ of the DBSs as a function of the nonsynchronized frequency $\tilde{\omega}_D$ for $\tilde{\omega}_S=0.9082, 0.861, 0.814, 0.7676$. The black dashed line represents the $S_0$ state solutions, the blue dashed line represents the $D_0$ state solutions, the red horizontal dashed line represents the ADM mass of the BSs with $S_0$ state when $\tilde{\omega}_S$ takes the value marked in the figure, and the orange line denote the coexisting state $D_0S_0$. All solutions have $\tilde{\mu}_D = \tilde{\mu}_S=1$.}
\label{mu_k-adm-k0-2}
\end{figure}

In Fig.~\ref{mu_k-adm-k0-2}, we show the ADM mass $M$ of the DBSs versus the nonsynchronized frequency $\tilde{\omega}_D$ for several values of the scalar field frequency $\tilde{\mu}_S$, where $0.7676 \le \tilde{\mu}_S \le 0.9082$. The black and blue dashed lines represent the solutions of $S_0$ and $D_0$, and the orange line represents the DBSs. As can be seen from the insets in each panel, as the scalar field frequency $\tilde{\omega}_S$ decreases, the existence domain of the nonsynchronized frequency $\tilde{\omega}_D$ increases. In addition, the ADM mass $M$ of the DBSs decreased with increasing $\tilde{\omega}_D$. Similar to Fig.~\ref{mu_k-adm}, when the ADM mass $M$ of the DBSs reaches a maximum, the scalar field disappears, and the DBSs transform into Dirac stars, while when the ADM mass $M$ of the DBSs comes to a minimum, the Dirac fields disappear, and the DBSs transform into BSs. The ADM mass of the BSs is equal to the minimum of the ADM mass of the DBSs when the scalar field frequency $\tilde{\mu}_S$ is taken at the value marked in each panel. In other words, for the \textit{one-branch-A} solution families, the minimum value of the ADM mass $M$ of the DBSs depends on the scalar field frequency $\tilde{\omega}_S$.

\subsubsection{Multi-Branch}

\begin{figure}[!htbp]
\begin{center}
\includegraphics[height=.16\textheight]{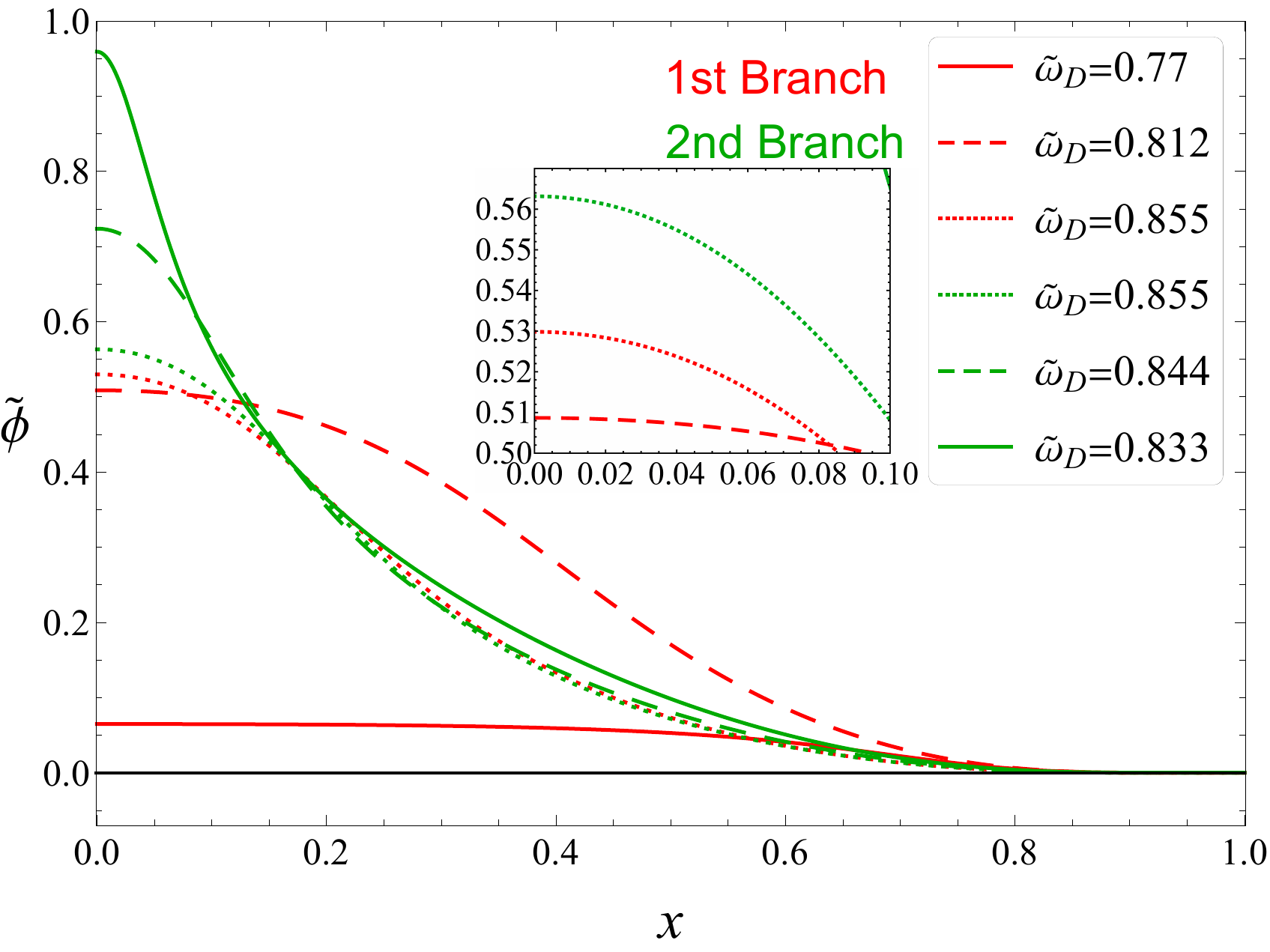}
\includegraphics[height=.16\textheight]{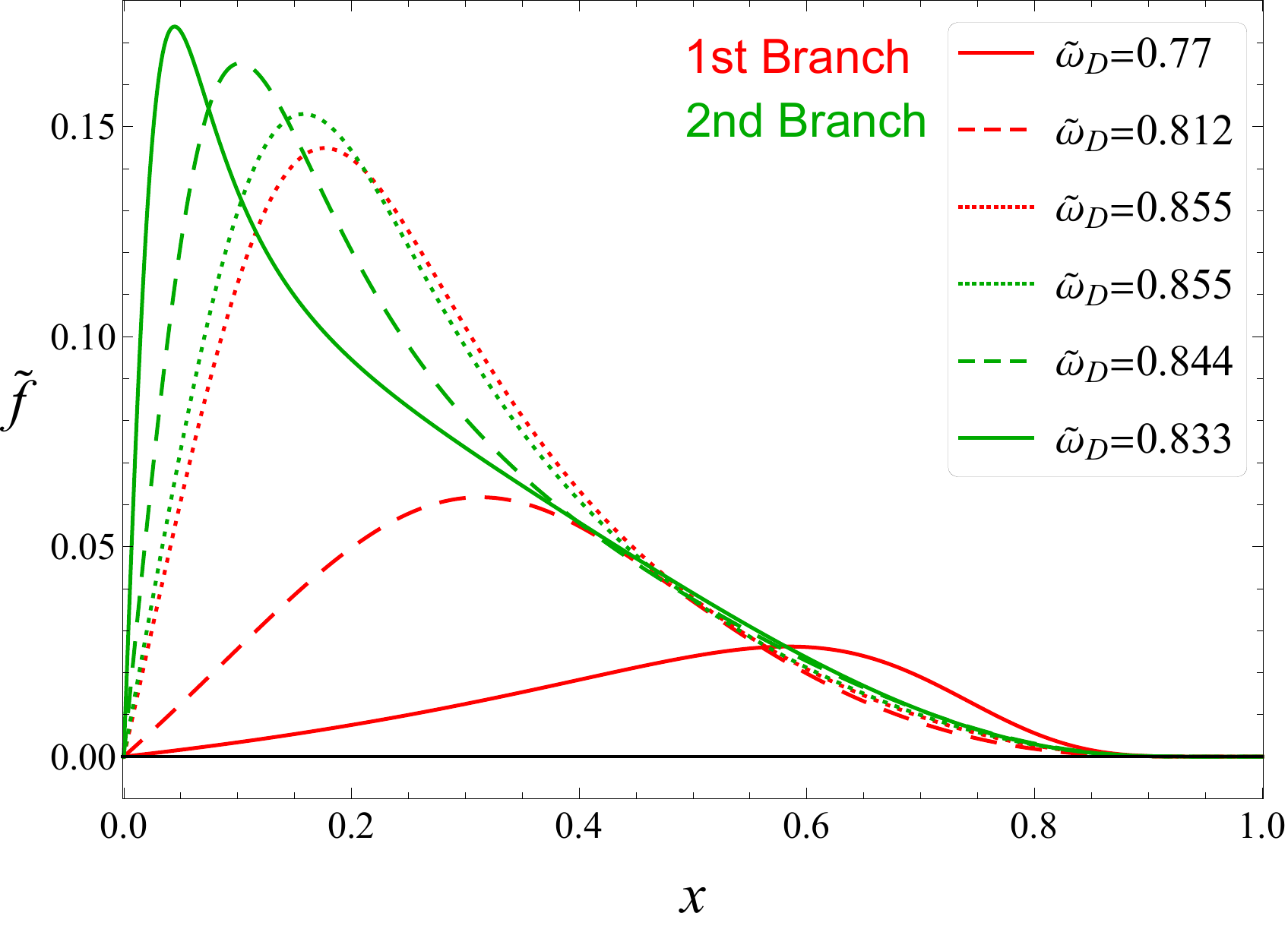}
\includegraphics[height=.16\textheight]{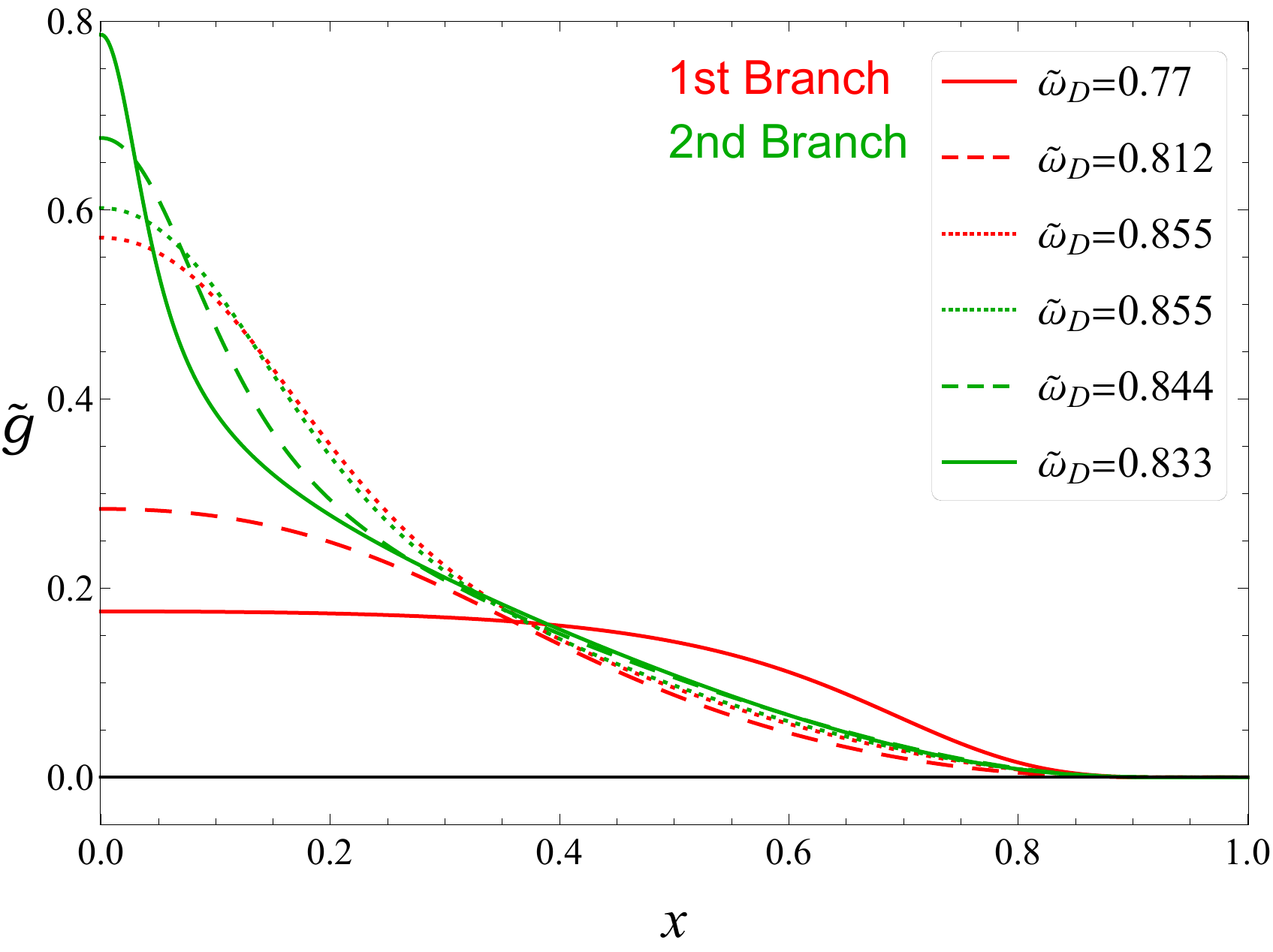}
\includegraphics[height=.16\textheight]{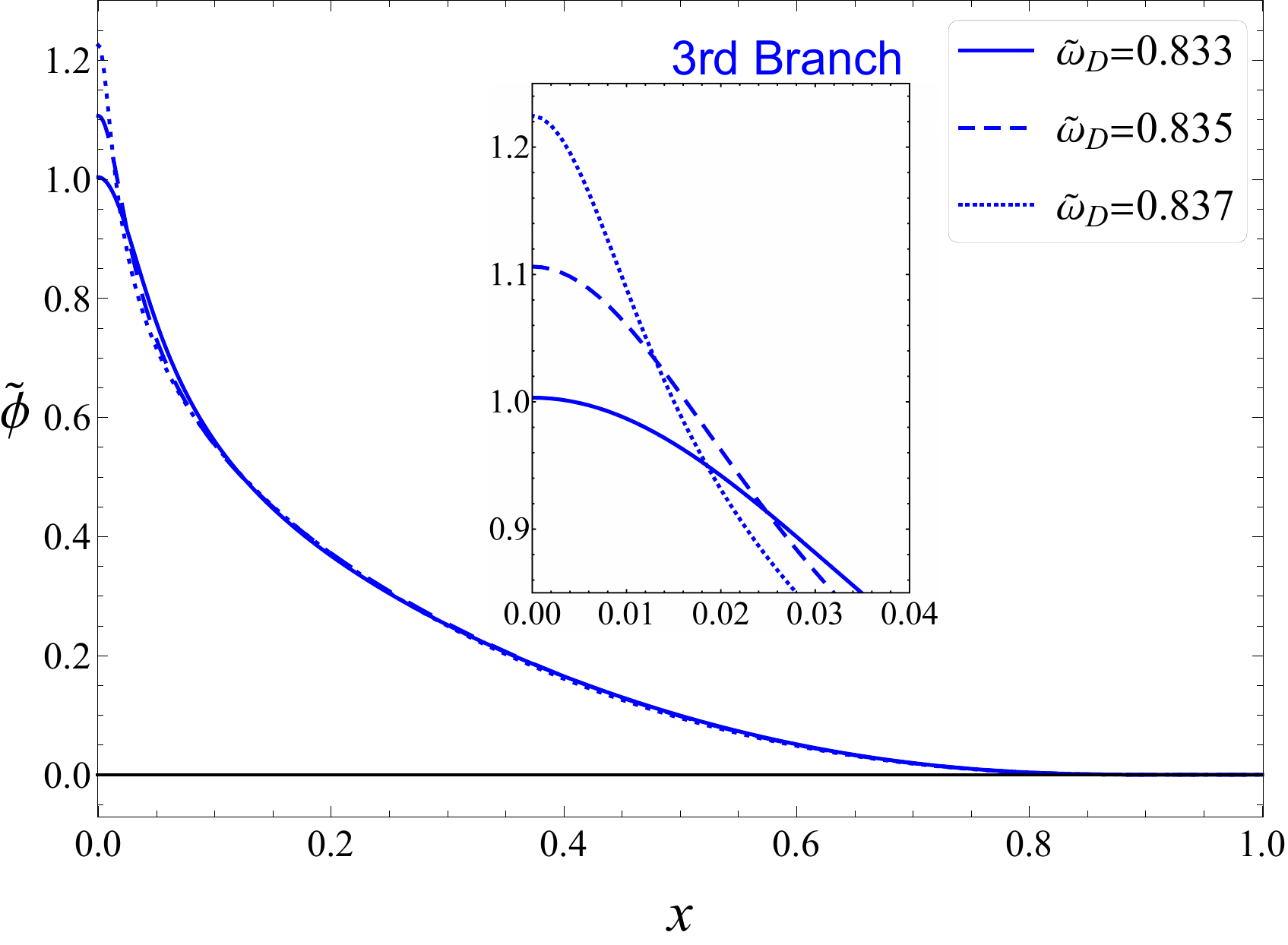}
\includegraphics[height=.16\textheight]{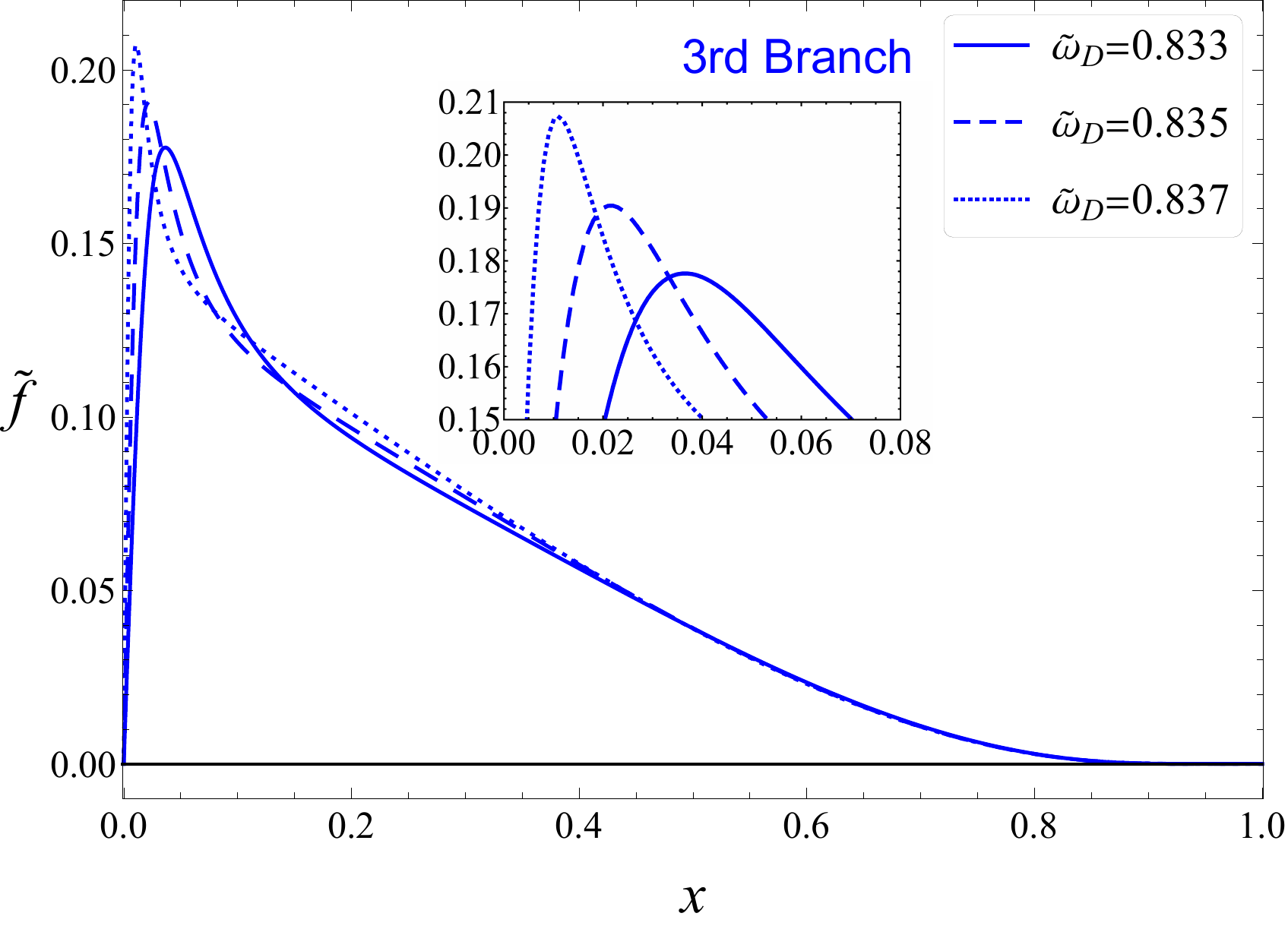}
\includegraphics[height=.16\textheight]{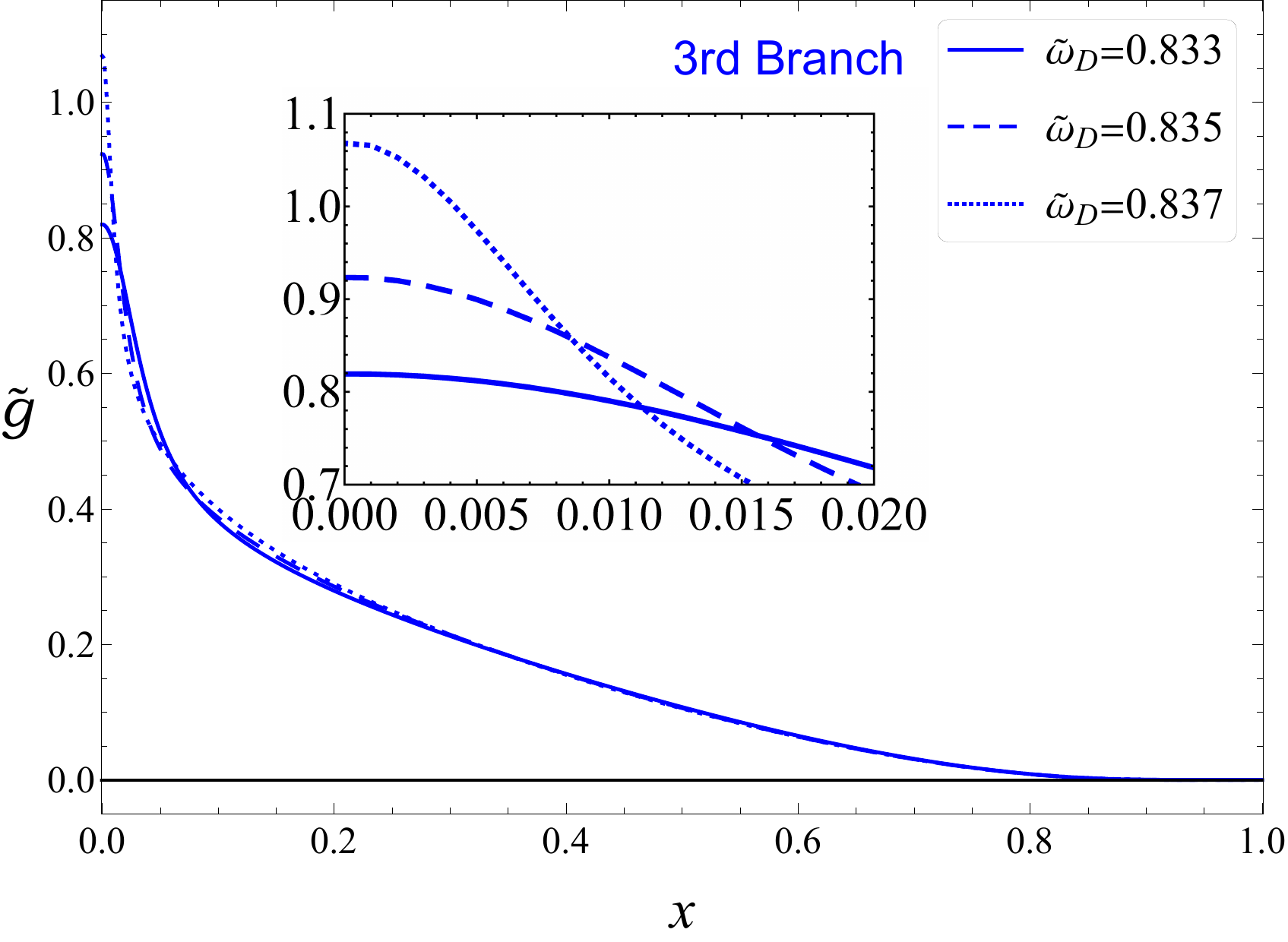}
\end{center}
\caption{Scalar field function $\tilde{\phi}$(left panel) and Dirac field functions $\tilde{f}$(middle panel) and $\tilde{g}$(right panel) as functions of $x$ with several values of nonsynchronized frequency $\tilde{\omega}_D$, where the field functions on the first, second and third branches are represented by the red, orange and blue lines. All solutions have $\tilde{\omega}_S= 0.749$ and $\tilde{\mu}_S = \tilde{\mu}_D = 1$.}
\label{psi_f_g_mb2}
\end{figure}

For \textit{multi-branch} solution families, the change of the field functions $\tilde{\psi}$, $\tilde{f}$ and $\tilde{g}$ on each branch with the nonsynchronized frequency $\tilde{\omega}_D$ is shown in Fig.~\ref{psi_f_g_mb2}. The field functions on the first, second and third branches are represented by the red, orange and blue lines, respectively. For the first and third branch solutions, $\tilde{\psi}_{max}$, $\tilde{f}_{max}$ and $\tilde{g}_{max}$ gradually increase as the nonsynchronized frequency $\tilde{\omega}_D$ increases. For the second branch solutions, $\tilde{\psi}_{max}$, $\tilde{f}_{max}$ and $\tilde{g}_{max}$ increase as the nonsynchronized frequency $\tilde{\omega}_D$ decreases.

\begin{figure}[!htbp]
\begin{center}
\includegraphics[height=.24\textheight]{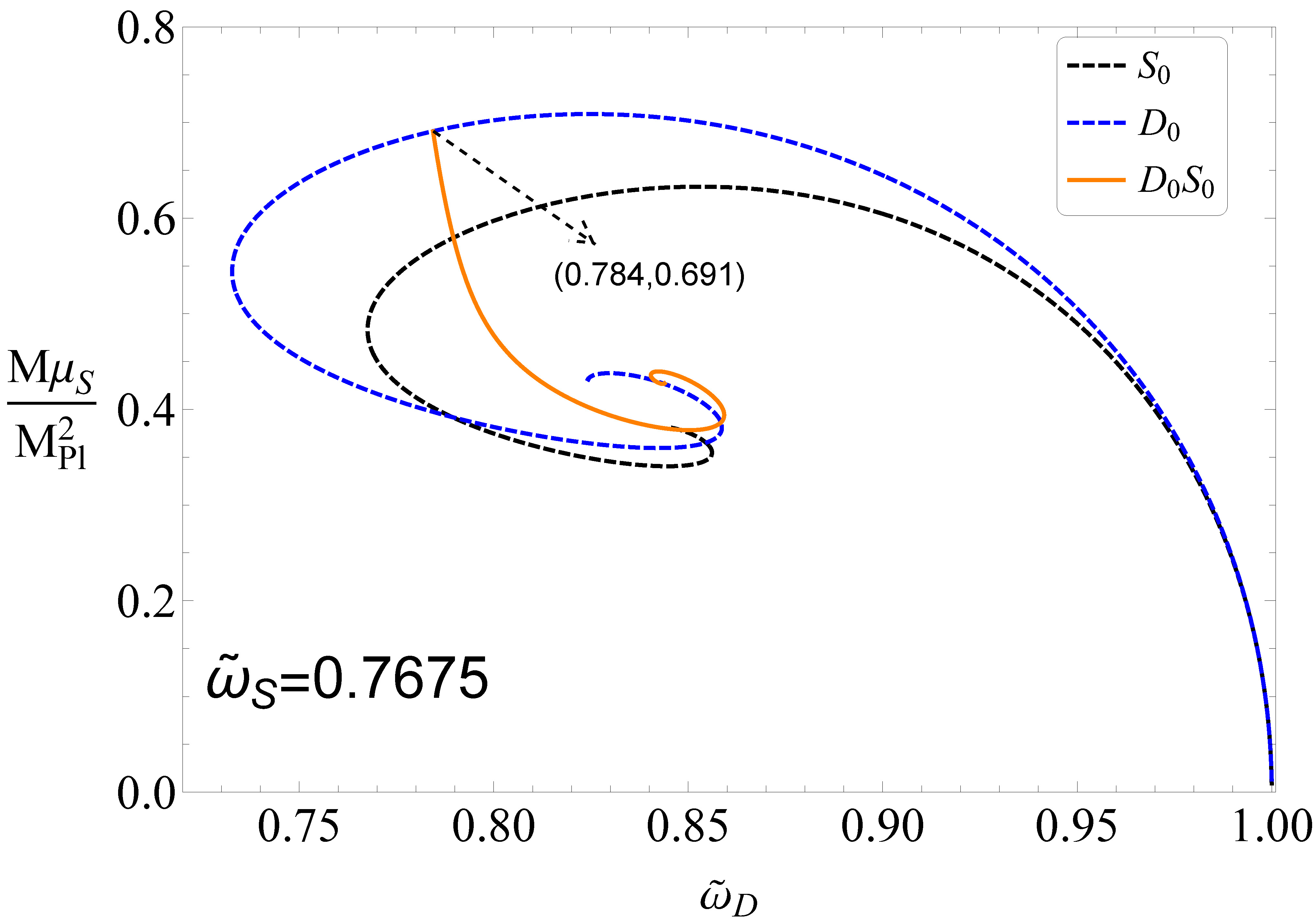}
\includegraphics[height=.24\textheight]{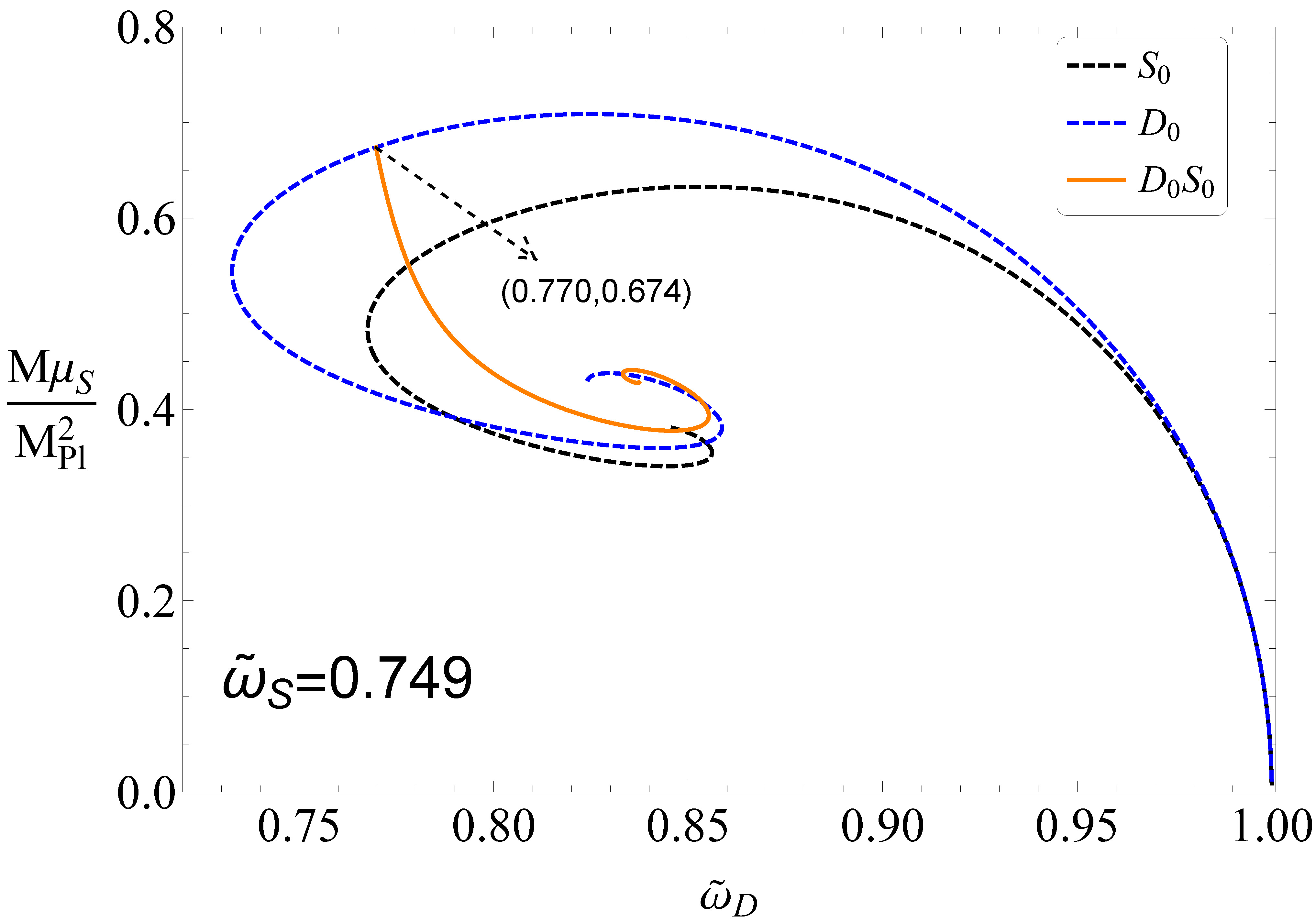}
\includegraphics[height=.24\textheight]{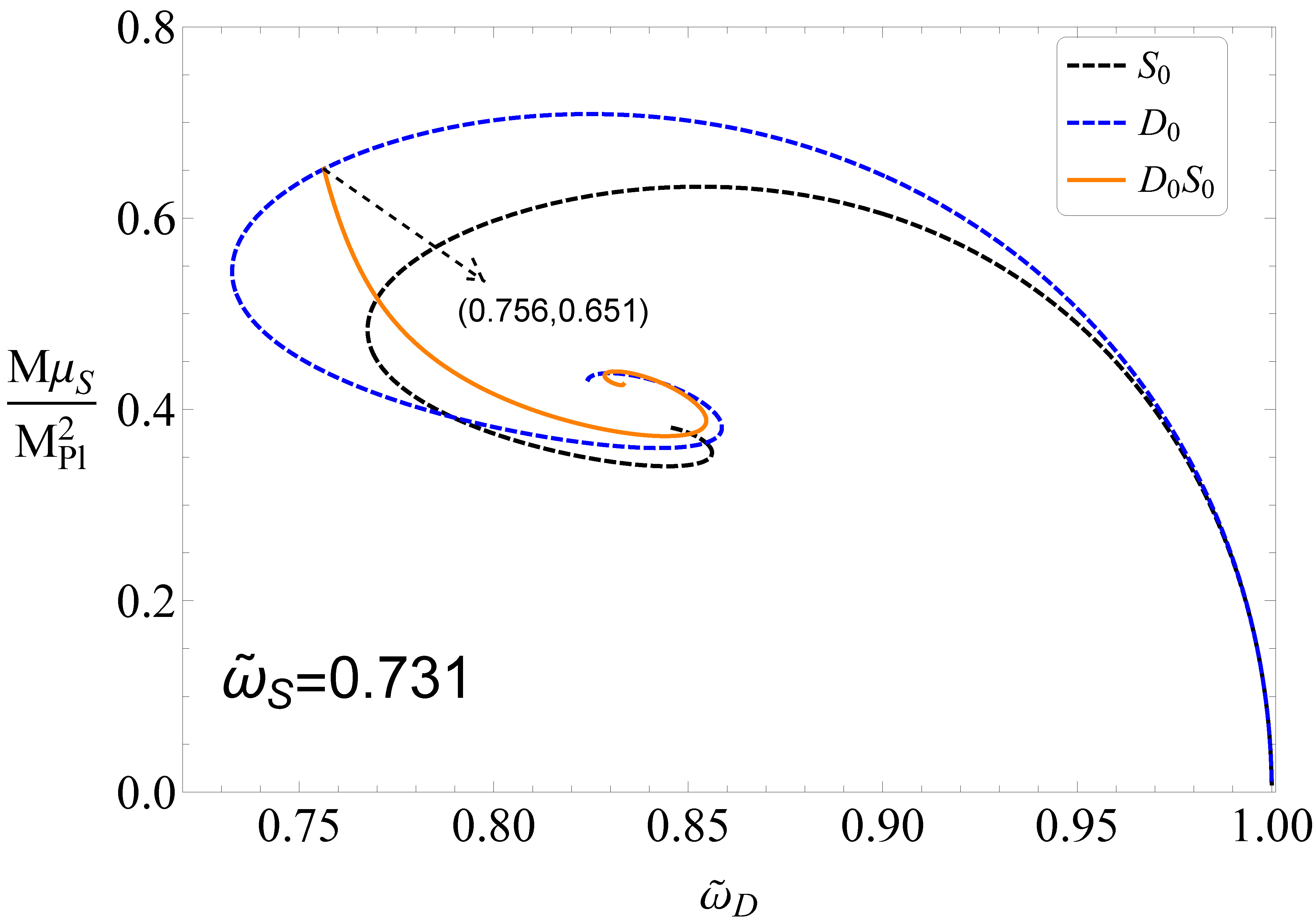}
\includegraphics[height=.24\textheight]{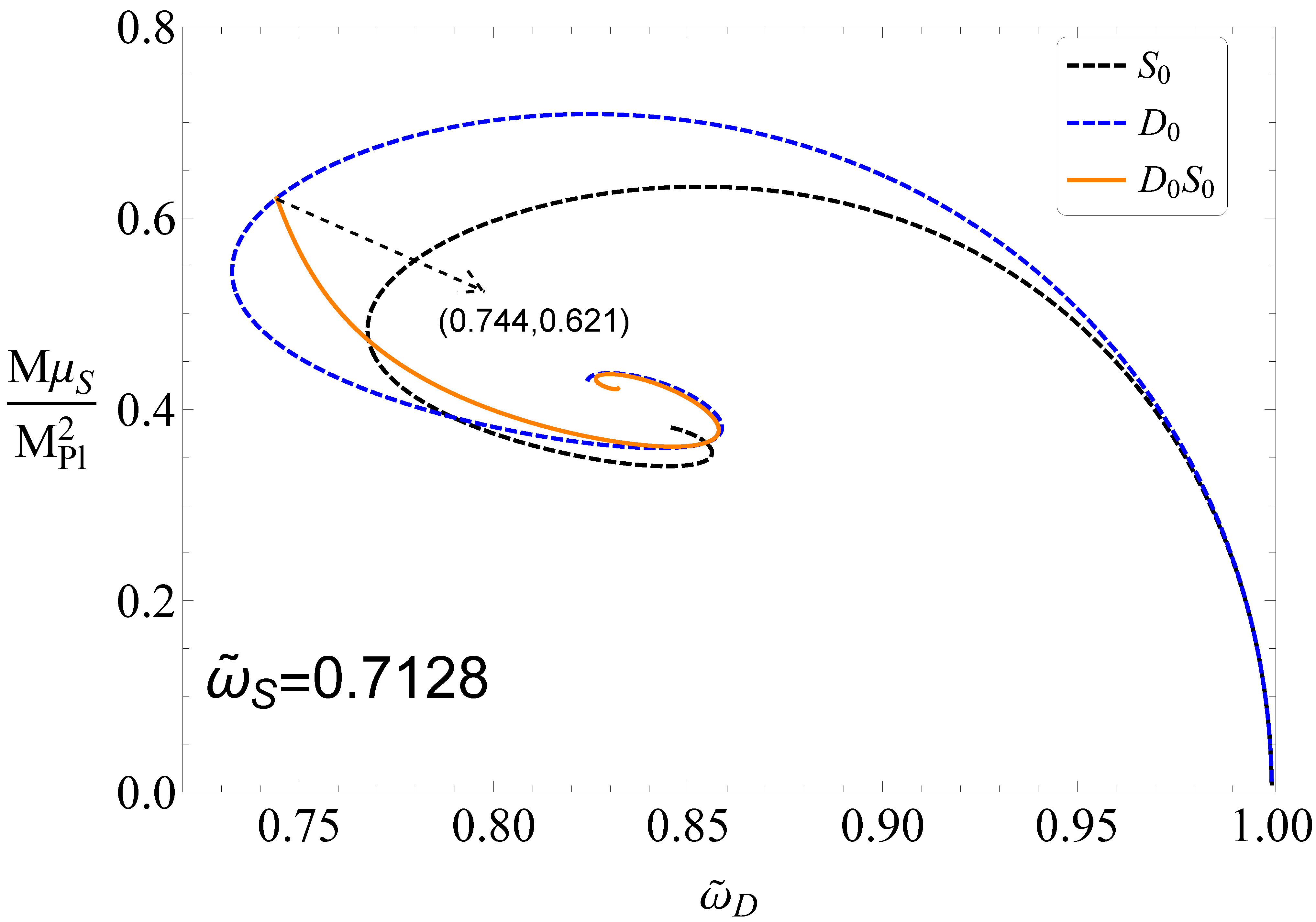}
\end{center}
\caption{The ADM mass $M$ of the DBSs as a function of the nonsynchronized frequency $\tilde{\omega}_D$ for $\tilde{\omega}_S=0.7675, 0.749, 0.731, 0.7128$. The black dashed line represents the $S_0$ state solutions, the blue dashed line represents the $D_0$ state solutions, and the orange line denote the coexisting state $D_0S_0$. All solutions have $\tilde{\mu}_D = \tilde{\mu}_S=1$.}
\label{mu_k-adm-k0-1}
\end{figure}

\begin{table}[!htbp]
  \centering
    \begin{tabular}{|c|c|c|c|c|c|}
    \hline
       $\tilde{\omega}_S$   & $B_1$    & $B_2$    & $B_3$    & $M_{max}$  & $M_{min}$ \\
    \hline
    $0.7675$  & $0.784\sim0.859$ & $0.840\sim0.859$ & $0.840\sim0.844$ & $0.691$ & $0.378$ \\
    \hline
    $0.749$  & $0.770\sim0.855$ & $0.833\sim0.855$ & $0.833\sim0.837$ & $0.674$ & $0.378$ \\
    \hline
    $0.731$  & $0.756\sim0.855$ & $0.829\sim0.855$ & $0.829\sim0.833$ & $0.651$ & $0.372$ \\
    \hline
    $0.7128$  & $0.744\sim0.858$ & $0.826\sim0.858$ & $0.826\sim0.832$ & $0.621$ & $0.361$ \\
    \hline
    \end{tabular}
\caption{Maximum and minimum ADM masses($M_{max}$ and $M_{min}$) of the DBSs, and the existence domain of the nonsynchronized frequency $\tilde{\omega}_D$ for $\tilde{\omega}_S=0.7675, 0.749, 0.731, 0.7128$. $B_1$, $B_2$ and $B_3$ represent the first, second and third branches of the orange line in Fig.~\ref{mu_k-adm-k0-1}, respectively. All solutions have $\tilde{\mu}_S = \tilde{\mu}_D =1$.}
\label{table2}
\end{table}

In Fig.~\ref{mu_k-adm-k0-1}, we show the ADM mass $M$ of the DBSs versus the nonsynchronized frequency $\tilde{\omega}_D$ for several values of the scalar field frequency $\tilde{\mu}_S$, where $0.7128 \le \tilde{\mu}_S \le 0.7675$. The black dashed and blue dashed lines are the same as in Fig.~\ref{mu_k-adm-k0-2}. Like Fig.~\ref{mu_k-adm2}, each orange line has multiple branches, and the whole line is a spiral. At the intersection of the orange line and the blue dashed line, the DBSs transform into Dirac stars. For the other solutions on the orange line, both the scalar and Dirac fields coexist. Moreover, there is no solution on the orange line where the Dirac field functions vanish, i.e., there is no solution similar to the intersection of the black dashed line and the orange line in Fig.~\ref{mu_k-adm}). In addition, unlike Fig.~\ref{mu_k-adm2}, the intersection of the orange line and the blue dashed line is always in the first branch of the blue dashed line.

In Table \ref{table2}, we show the existence domain of the nonsynchronized frequency $\tilde{\omega}_D$ for several values of the scalar field frequency $\tilde{\omega}_S$. As the frequency $\tilde{\omega}_S$ decreases, the intervals of $B_1$, $B_2$ and $B_3$ are gradually narrowed, where the change in the intervals of $B_3$ is relatively small. Moreover, as the frequency $\tilde{\omega}_S$ decreases, both $M_{max}$ and $M_{min}$ gradually decrease.

\subsubsection{One-Branch-B}

\begin{figure}[!htbp]
\begin{center}
\includegraphics[height=.16\textheight]{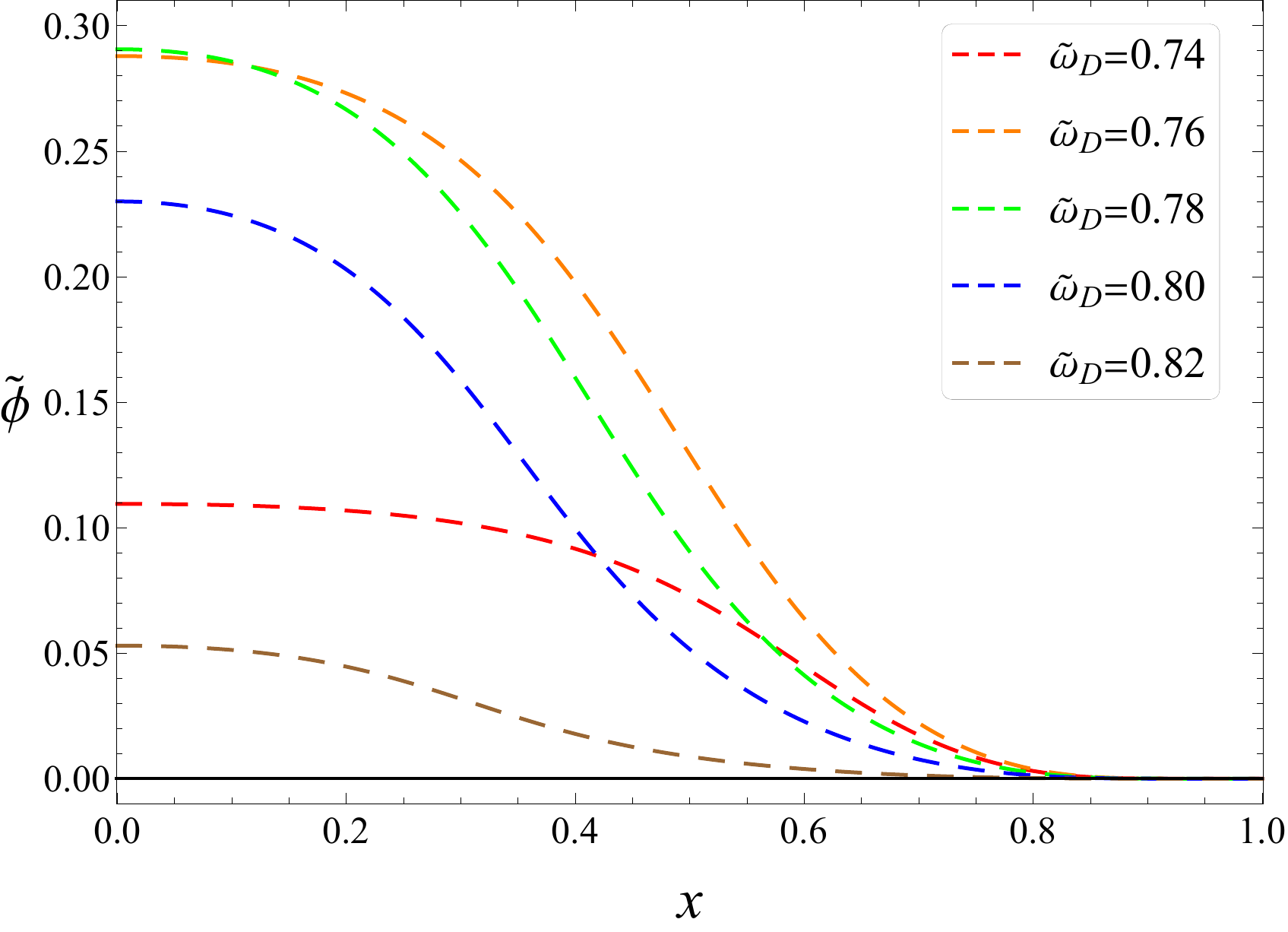}
\includegraphics[height=.16\textheight]{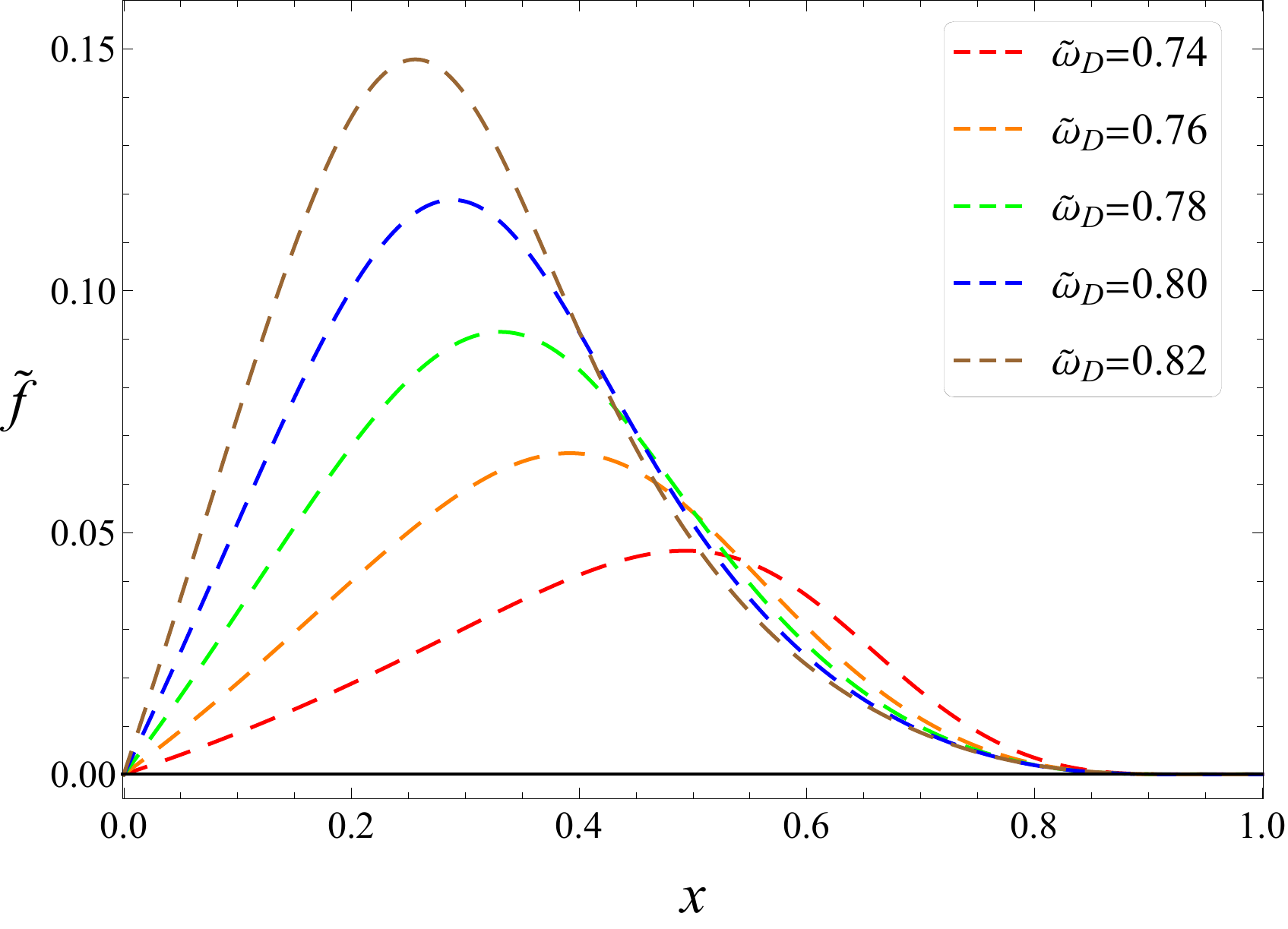}
\includegraphics[height=.16\textheight]{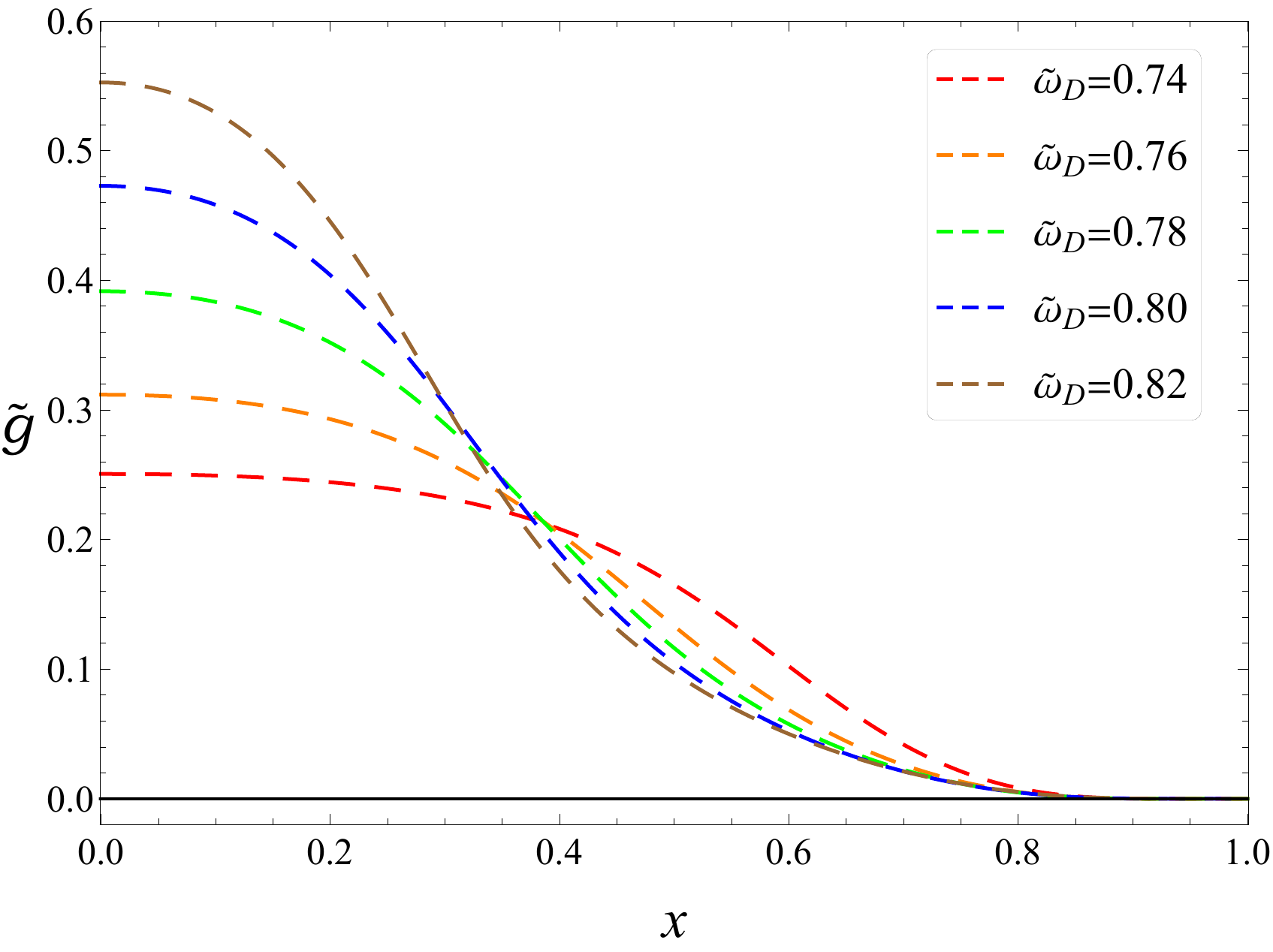}
\end{center}
\caption{Scalar field function $\tilde{\phi}$(left panel) and Dirac field functions $\tilde{f}$(middle panel) and $\tilde{g}$(right panel) as functions of $x$ with $\tilde{\omega}_D = 0.74, 0.76, 0.78, 0.80, 0.82$. All solutions have $\tilde{\omega}_S= 0.702$ and $\tilde{\mu}_S = \tilde{\mu}_D = 1$.}
\label{psi_f_g_ob-b}
\end{figure}

For \textit{one-branch-B} solution families, the profiles of the field functions $\tilde{\phi}$, $\tilde{f}$ and $\tilde{g}$ with several values of nonsynchronized frequency $\tilde{\omega}_D$ in the solutions of the $D_0S_0$ state are shown in Fig.~\ref{psi_f_g_ob-b}. For the scalar field, $\tilde{\phi}_{max}$ first increases and then decreases with the increase of the nonsynchronized frequency $\tilde{\omega}_D$. Whereas for Dirac field, $\tilde{f}_{max}$ and $\tilde{g}_{max}$ increase with increasing the frequency $\tilde{\omega}_D$.

\begin{figure}[!htbp]
\begin{center}
\includegraphics[height=.24\textheight]{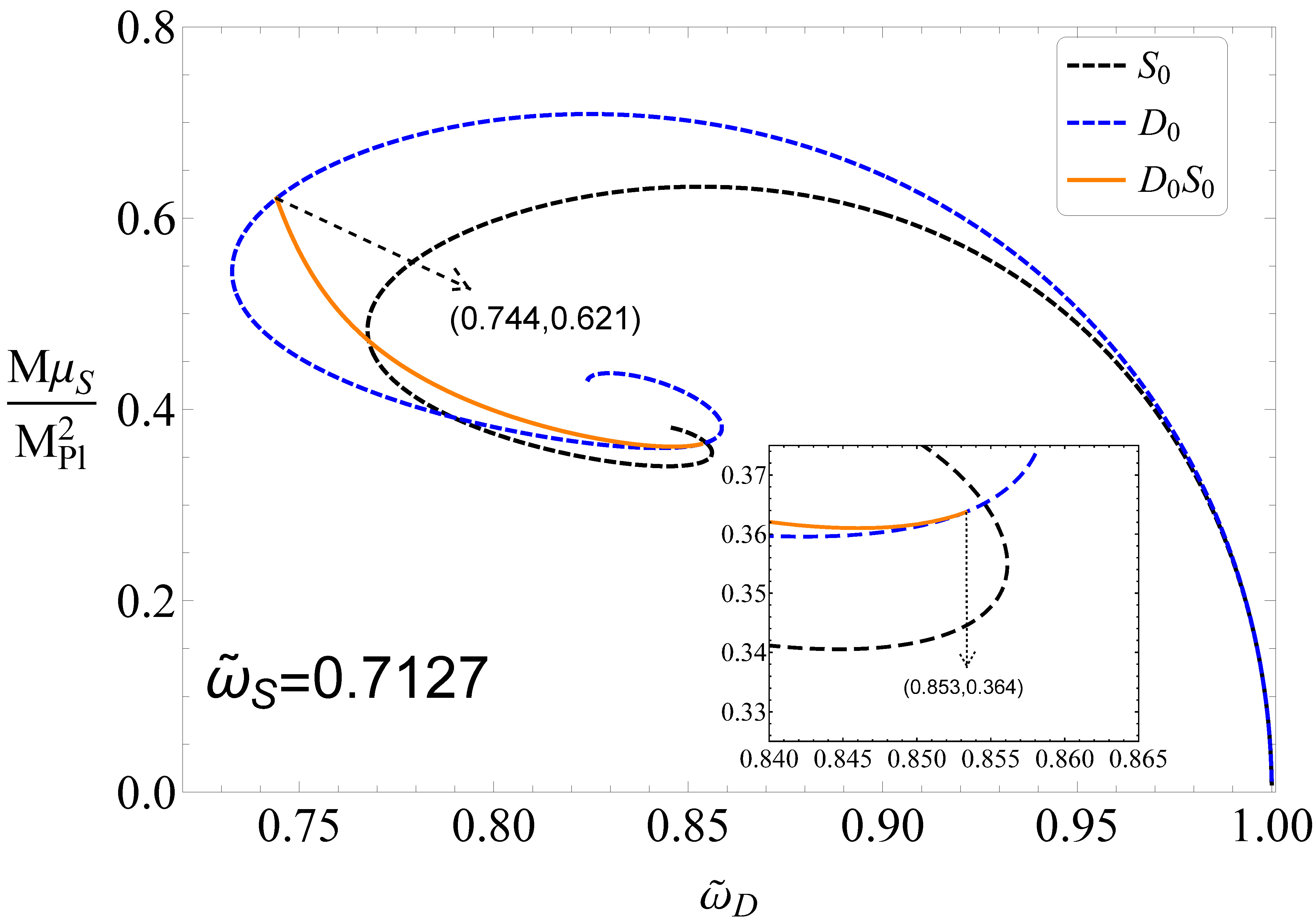}
\includegraphics[height=.24\textheight]{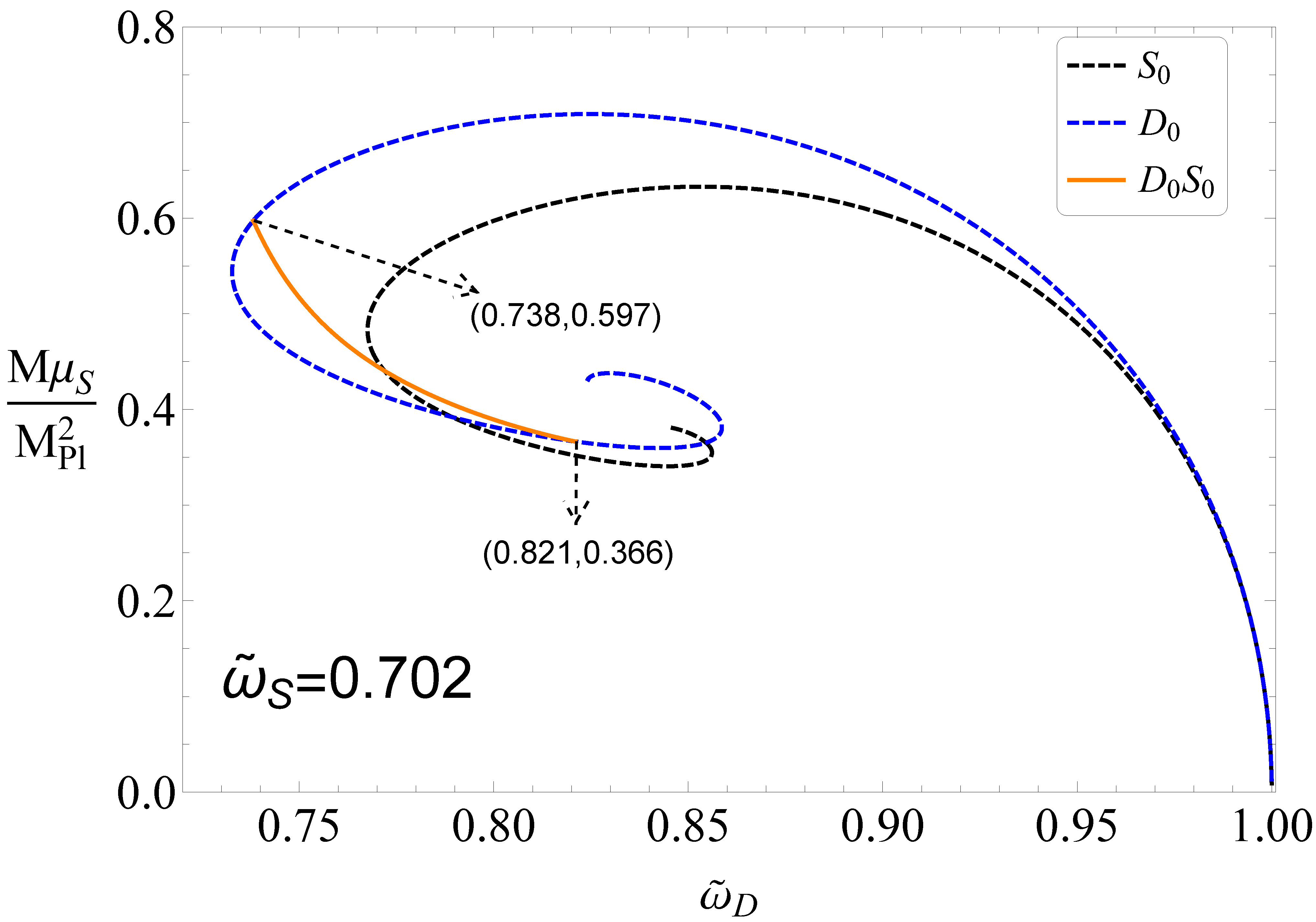}
\includegraphics[height=.24\textheight]{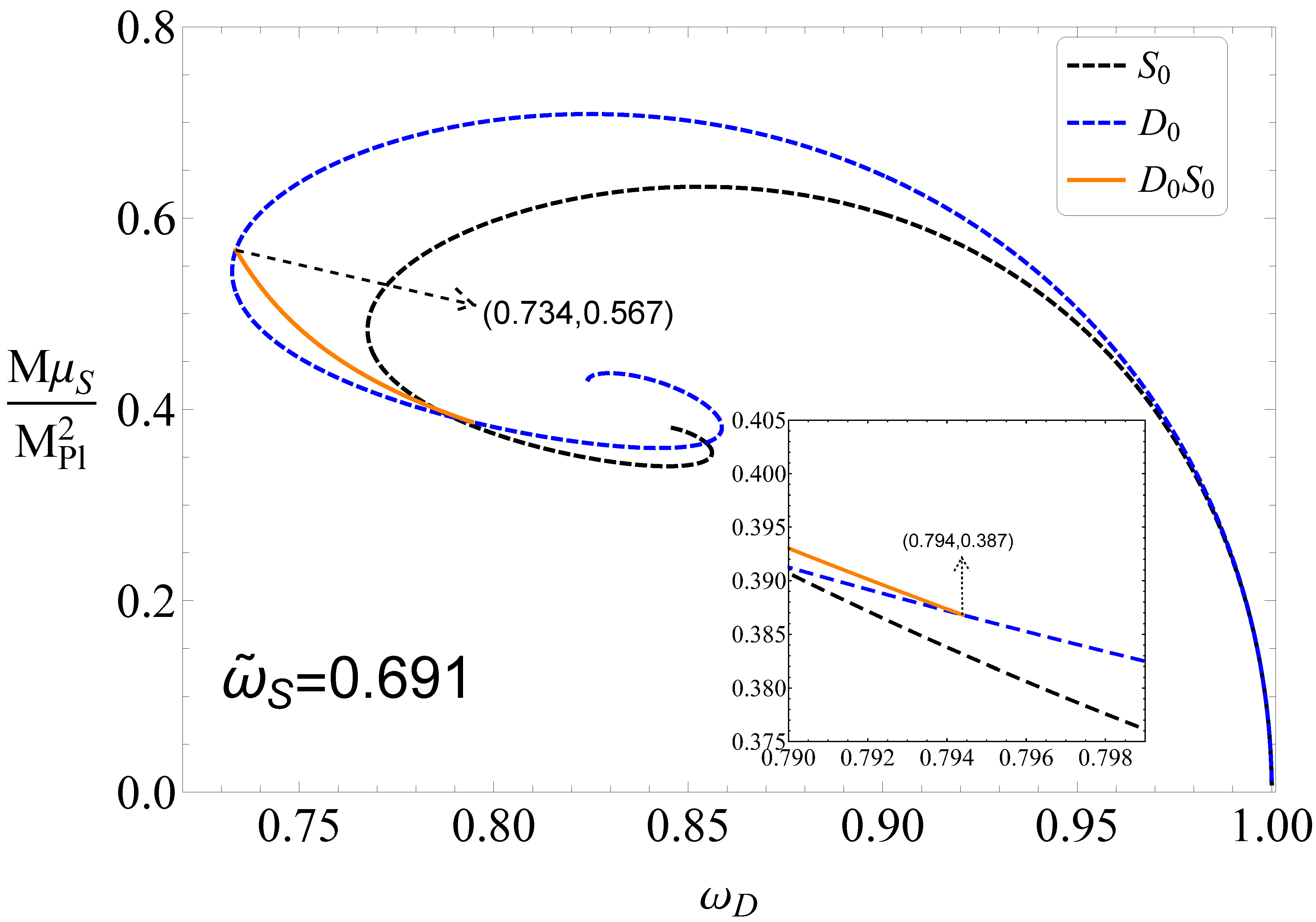}
\includegraphics[height=.24\textheight]{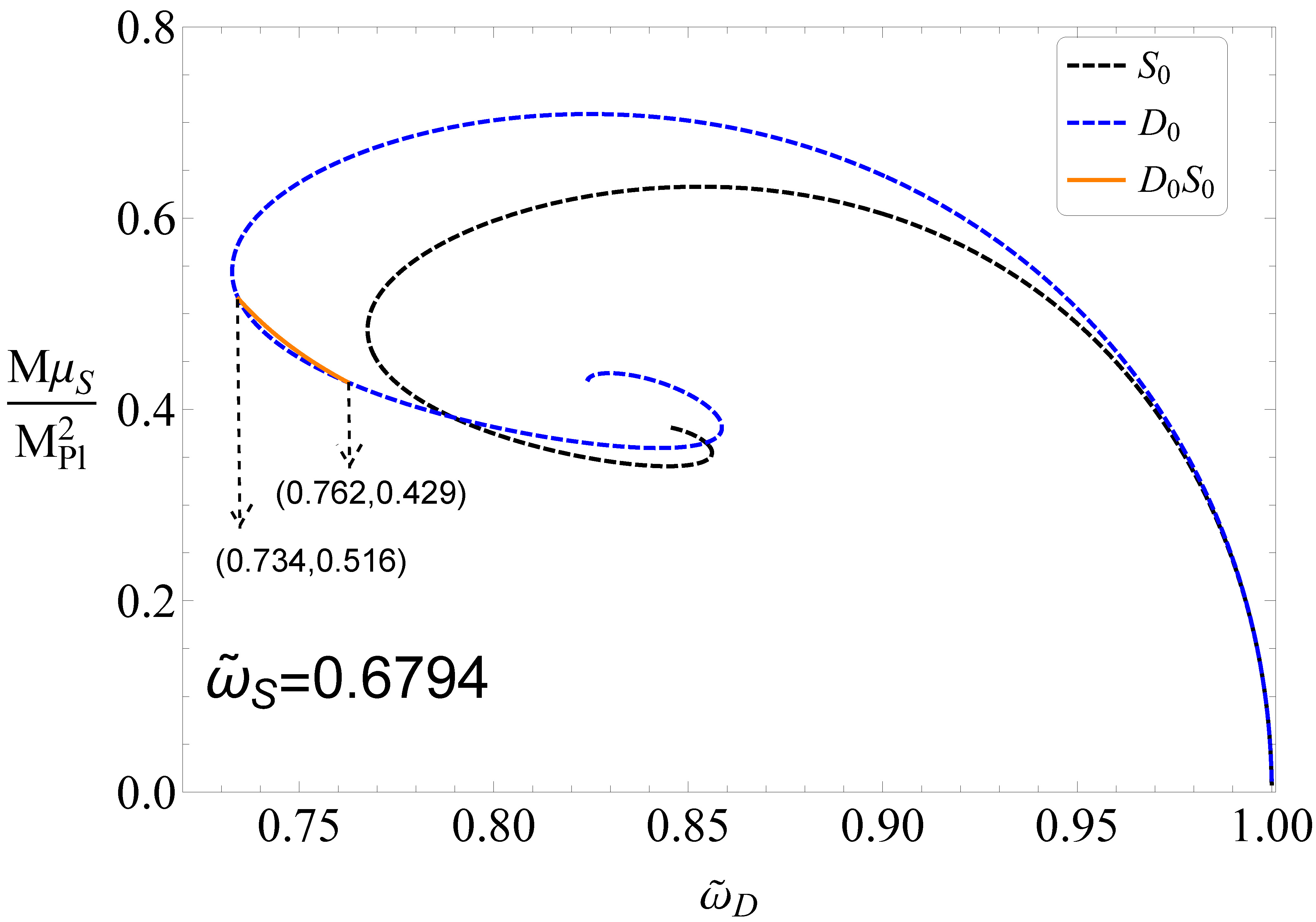}
\end{center}
\caption{The ADM mass $M$ of the DBSs as a function of the nonsynchronized frequency $\tilde{\omega}_D$ for $\tilde{\omega}_S=0.7127, 0.702, 0.691, 0.6794$. The black dashed line represents the $S_0$ state solutions, the blue dashed line represents the $D_0$ state solutions, and the orange line denote the coexisting state $D_0S_0$. All solutions have $\tilde{\mu}_D = \tilde{\mu}_S=1$.}
\label{mu_k-adm-k0-3}
\end{figure}

In Fig.~\ref{mu_k-adm-k0-3}, we show the ADM mass $M$ of the DBSs versus the nonsynchronized frequency $\tilde{\omega}_D$ for several values of the scalar field frequency $\tilde{\mu}_S$, where $0.6794 \le \tilde{\mu}_S \le 0.7127$. The black and blue dashed lines are the same as in Fig.~\ref{mu_k-adm-k0-1} and Fig.~\ref{mu_k-adm-k0-2}, the orange line represents the DBSs. As seen from the inset in the top left panel, the ADM mass $M$ of DBSs has a slight upward trend after reaching a minimum value. However, if the scalar field frequency $\tilde{\mu}_S$ is small enough, the mass $M$ of DBSs decreases with increasing the nonsynchronized frequency $\tilde{\omega}_D$ (there is no tendency for the mass $M$ of DBSs to increase). As the scalar field frequency $\tilde{\omega}_S$ decreases, the existence domain of the nonsynchronized frequency $\tilde{\omega}_D$ decreases. The one-branch solutions shown in Fig.~\ref{mu_k-adm-k0-3} and Fig.~\ref{mu_k-adm} are different. In Fig.~\ref{mu_k-adm-k0-3}, the Dirac fields are always in existence as the nonsynchronized frequency $\tilde{\omega}_D$ increases, while the scalar field appears and then disappears. At the two intersections of the orange line and the blue dashed line, the DBSs transform into Dirac stars. 

\begin{figure}[!htbp]
\begin{center}
\includegraphics[height=.23\textheight]{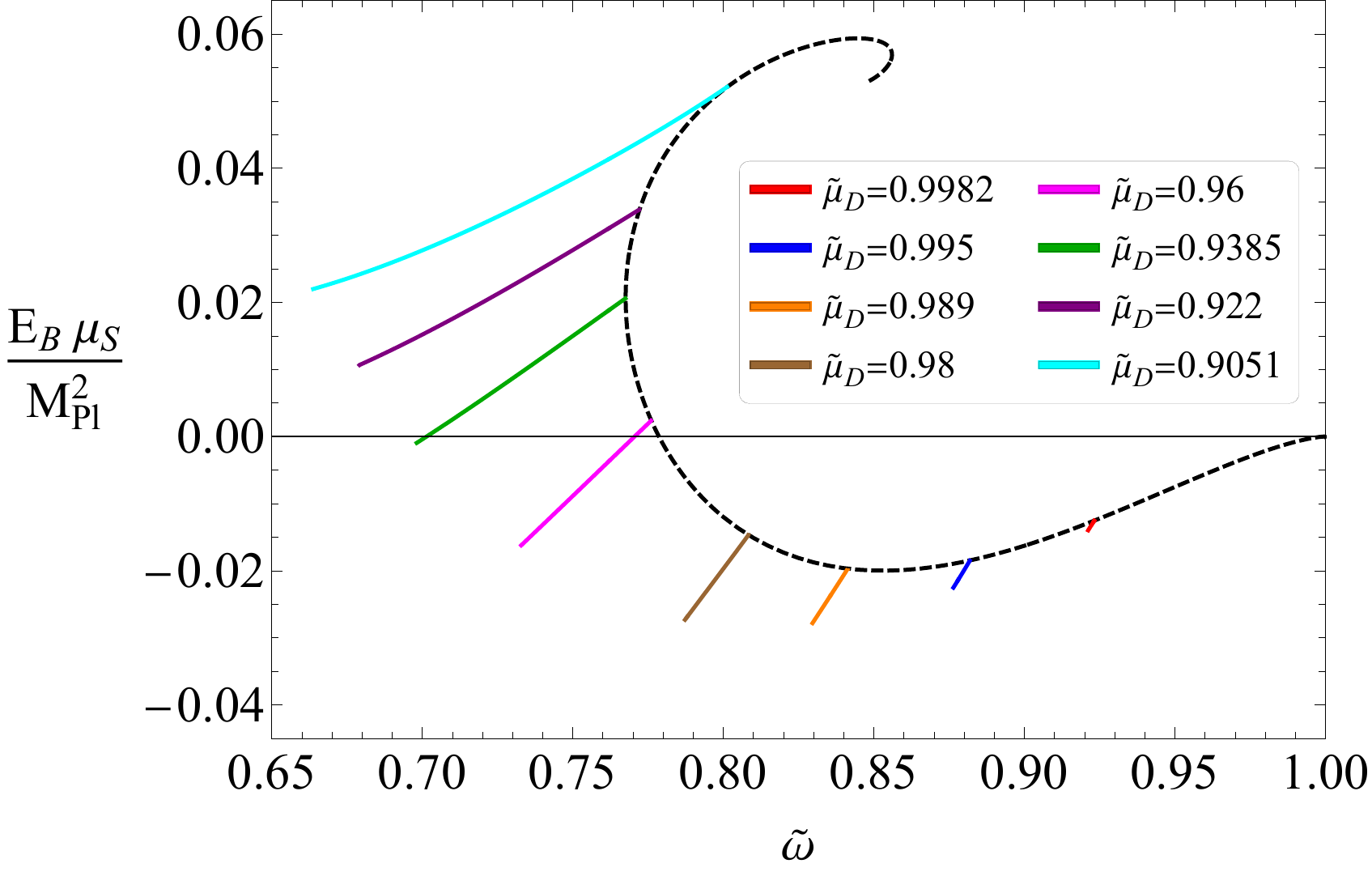}
\includegraphics[height=.23\textheight]{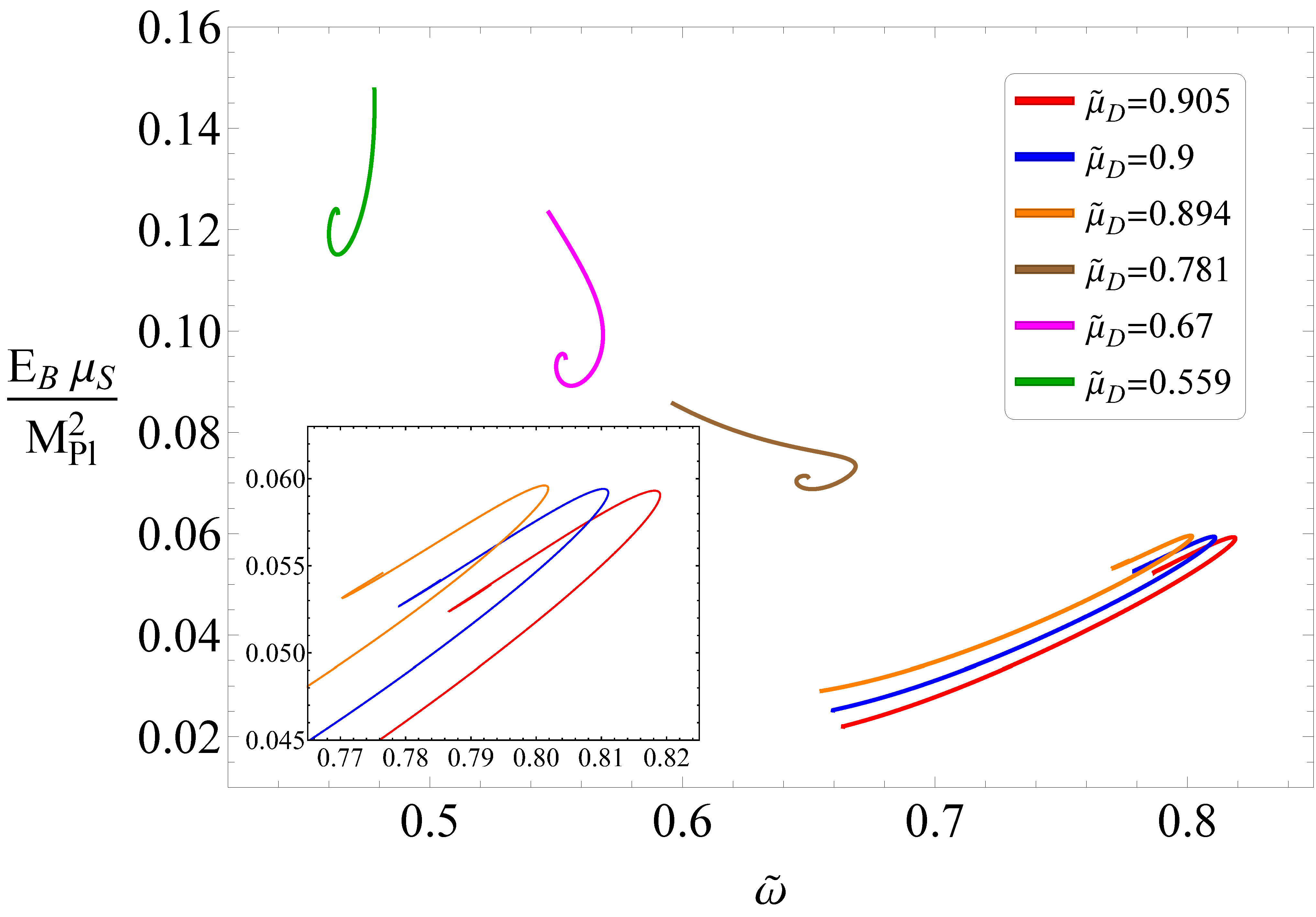}
\end{center}
\caption{\textit{Left}: The binding energy $E_B$ of the \textit{one-branch} solution family as a function of the synchronized frequency $\tilde{\omega}$ for several values of the Dirac field mass $\tilde{\mu}_D$. \textit{Right}: Same as left panel for the \textit{multi-branch} solution family. The black dashed line represents the $S_0$ state solutions. All solutions have $\tilde{\mu}_S=1$.}
\label{BE_sf}
\end{figure}

\begin{table*}[!htbp]
\caption{For \textit{one-branch} (left panel) and \textit{multi-branch} (right panel) solution families, the maximum value $E_{max}$ and the minimum value $E_{min}$ of the binding energy $E_{B}$ of the DBSs, the synchronized frequency $\tilde{\omega}$ corresponding to $E_{max}$ and $E_{min}$ ($\omega_{max}$ and $\omega_{min}$). $\omega_{0}$ is the value of the synchronized frequency $\tilde{\omega}$ when $E_B = 0$.}
\centering
\subtable[
]{
\scalebox{1}{
       \begin{tabular}{|c|c|c|c|c|c|}
    \hline
       $\tilde{\mu}_D$  & $E_{max}$    & $E_{min}$    & $\omega_{max}$    & $\omega_{min}$  & $\omega_{0}$ \\
    \hline
    $0.9982$  & $-0.012$ & $-0.014$ & $0.923$ & $0.921$ & \diagbox[width=3em,height=1.5em]{}{} \\
    \hline
    $0.98$  & $-0.015$ & $-0.027$ & $0.808$ & $0.787$ & \diagbox[width=3em,height=1.5em]{}{} \\
    \hline
    $0.96$  & $0.003$ & $-0.016$ & $0.776$ & $0.733$ & $0.770$ \\
    \hline
    $0.9385$  & $0.021$ & $-0.001$ & $0.768$ & $0.698$ & $0.701$ \\
    \hline
    $0.922$  & $0.034$ & $0.011$ & $0.772$ & $0.679$ & \diagbox[width=3em,height=1.5em]{}{} \\
    \hline
    $0.9051$  & $0.052$ & $0.022$ & $0.801$ & $0.664$ & \diagbox[width=3em,height=1.5em]{}{} \\
    \hline
    \end{tabular}}
       \label{tab:firsttable}
}
\subtable[]{
\scalebox{1}{
       \begin{tabular}{|c|c|c|c|c|c|}
    \hline
       $\tilde{\mu}_D$   & $E_{max}$    & $E_{min}$    & $\omega_{max}$    & $\omega_{min}$  & $\omega_{0}$ \\
    \hline
    $0.905$  & $0.059$ & $0.022$ & $0.818$ & $0.664$ & \diagbox[width=3em,height=1.5em]{}{} \\
    \hline
    $0.9$  & $0.059$ & $0.025$ & $0.810$ & $0.659$ & \diagbox[width=3em,height=1.5em]{}{} \\
    \hline
    $0.894$  & $0.060$ & $0.030$ & $0.801$ & $0.655$ & \diagbox[width=3em,height=1.5em]{}{} \\
    \hline
    $0.781$  & $0.086$ & $0.069$ & $0.596$ & $0.651$ & \diagbox[width=3em,height=1.5em]{}{} \\
    \hline
    $0.67$  & $0.124$ & $0.089$ & $0.547$ & $0.556$ & \diagbox[width=3em,height=1.5em]{}{} \\
    \hline
    $0.559$  & $0.148$ & $0.115$ & $0.478$ & $0.464$ & \diagbox[width=3em,height=1.5em]{}{} \\
    \hline
    \end{tabular}}
       \label{tab:secondtable}
}
\label{table3}
\end{table*}

At the end of this section, we will analyse the binding energy $E_B = M - \mu_SQ_S - 2\mu_DQ_D$ of the five solution families mentioned above. The binding energy $E_B$ of the DBSs versus the synchronized frequency $\tilde{\omega}$ for several values of the mass $\tilde{\mu}_D$ are presented in Fig.~\ref{BE_sf}. In the left panel, fix the value of the Dirac field mass $\tilde{\mu}_D$, the binding energy $E_B$ of the DBSs increases with increasing synchronized frequency $\tilde{\omega}$. When the value of $\tilde{\mu}_D$ is sufficiently small, some unstable solutions ($E_B > 0$) will occur. For the \textit{multi-branch} solution family (right panel), the relationship between the binding energy $E_B$ of DBSs and the synchronized frequency $\tilde{\omega}$ is complicated, and there is no stable DBSs ($E_B > 0$).

In Table \ref{table3}, we show the maximum and minimum values of the binding energy $E_{B}$ of the DBSs in Fig.~\ref{BE_sf}, the synchronized frequency $\tilde{\omega}$ corresponding to the maximum and minimum values of $E_{B}$, and the value of the synchronized frequency $\tilde{\omega}$ when $E_B = 0$. For the \textit{one-branch} solution family, both $E_{max}$ and $E_{min}$ increase as the Dirac field mass $\tilde{\mu}_D$ decreases, and there is no $\omega_{0}$ for the DBSs with higher or lower Dirac field mass $\tilde{\mu}_D$. For the \textit{multi-branch} solution family, both $E_{max}$ and $E_{min}$ increase as the mass $\tilde{\mu}_D$ decreases, and there is no $\omega_{0}$ for the DBSs.

\begin{figure}[!htbp]
\begin{center}
\includegraphics[height=.22\textheight]{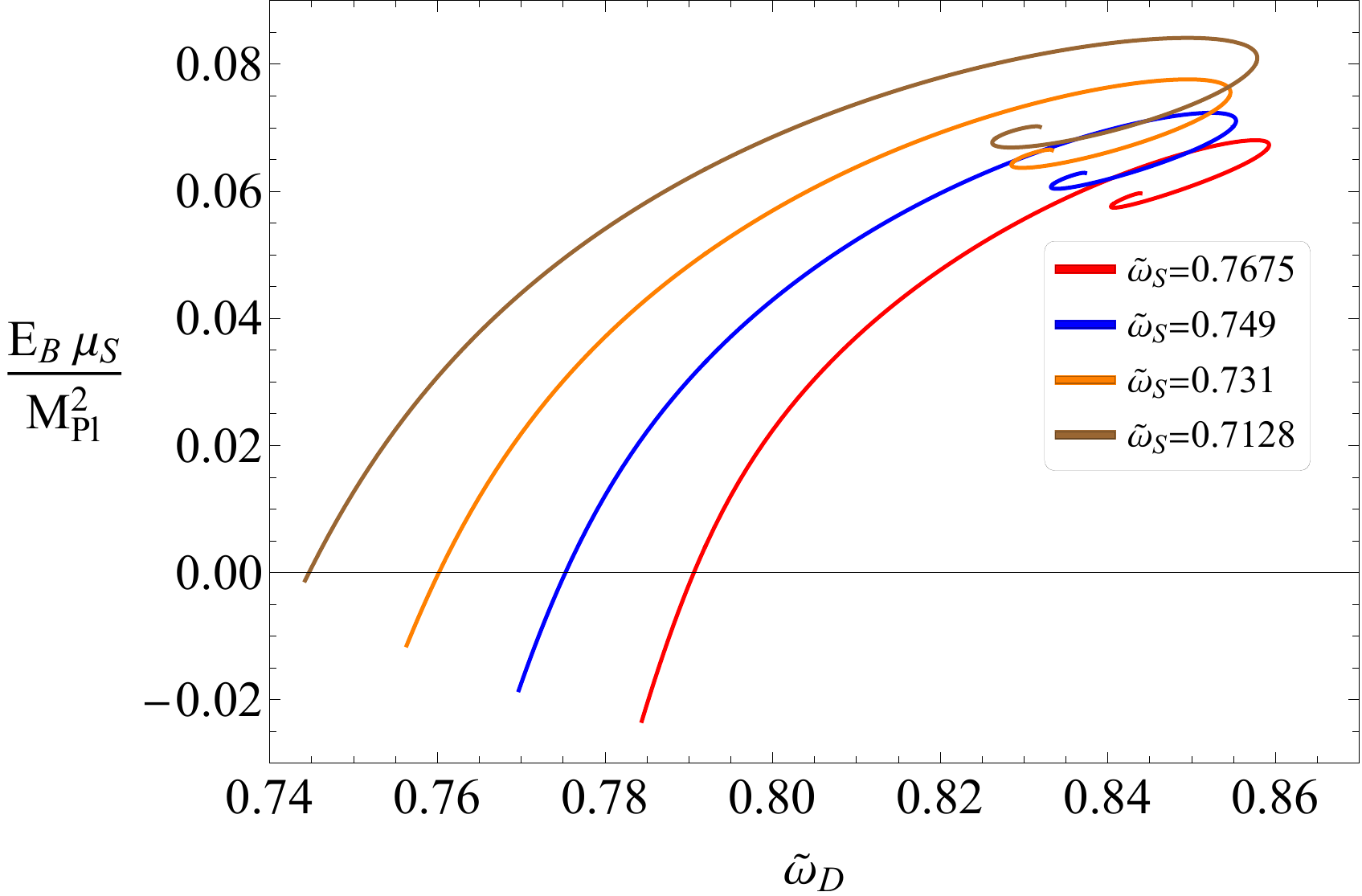}
\includegraphics[height=.22\textheight]{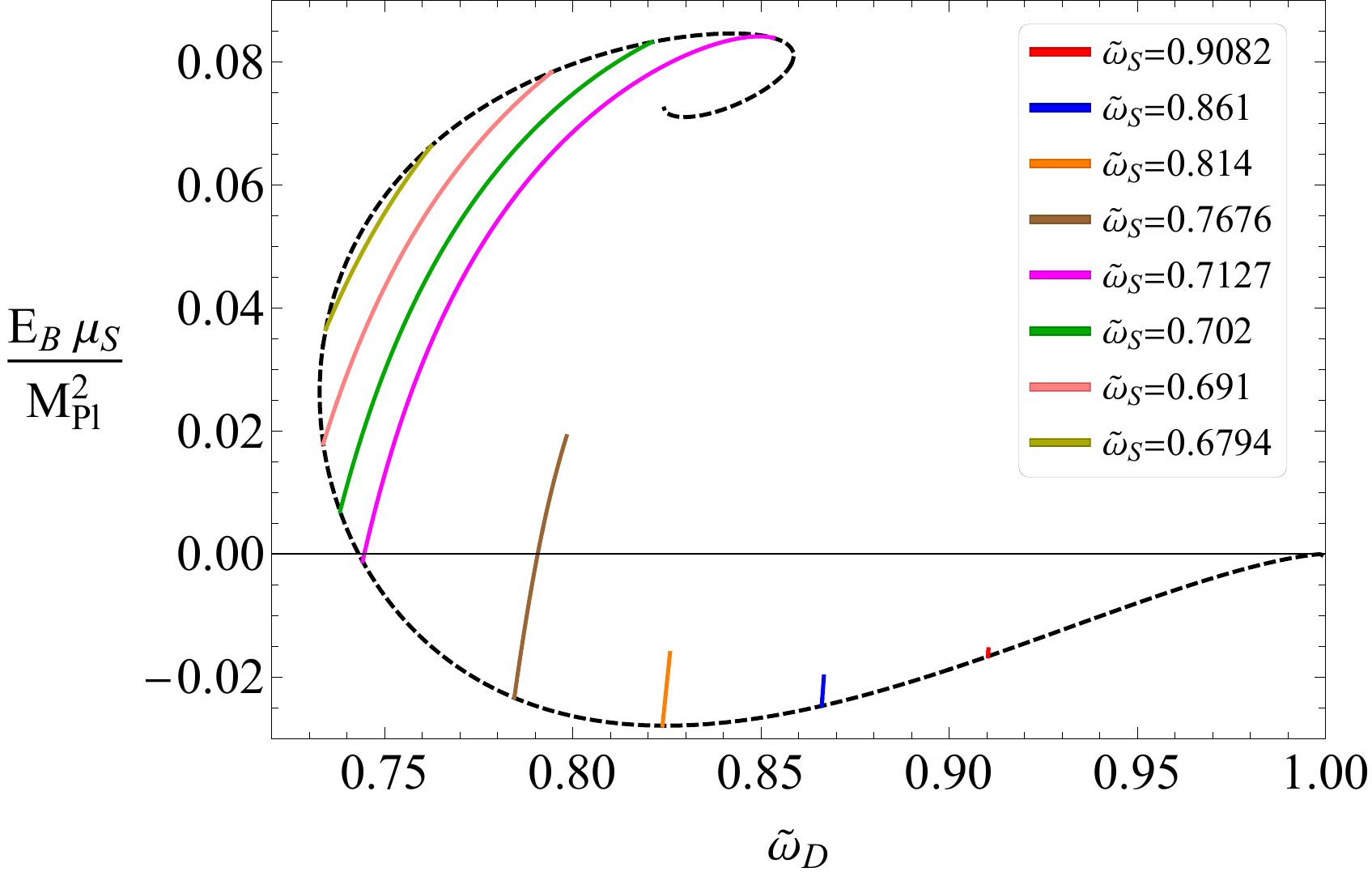}
\end{center}
\caption{\textit{Left}: The binding energy $E_B$ of the \textit{multi-branch} solution family as a function of the nonsynchronized frequency $\tilde{\omega}_D$ for several values of the scalar field frequency $\tilde{\omega}_S$. \textit{Right}: Same as left panel for the \textit{one-branch-A} and \textit{one-branch-B} solution families. The black dashed line represents the $D_0$ state solutions. All solutions have $\tilde{\mu}_S = \tilde{\mu}_D = 1$.}
\label{BE_nsf}
\end{figure}

\begin{table*}[!htbp]
\caption{For \textit{multi-branch}(left panel), \textit{one-branch-A}(middle panel) and \textit{one-branch-B}(right panel) solution families, the maximum value $E_{max}$ and the minimum value $E_{min}$ of the binding energy $E_{B}$ of the DBSs, the nonsynchronized frequency $\tilde{\omega}_D$ corresponding to $E_{max}$ and $E_{min}$($\omega_{Dmax}$ and $\omega_{Dmin}$). $\omega_{D0}$ is the value of the nonsynchronized frequency $\tilde{\omega}_D$ when $E_B = 0$.}
\centering
\subtable[
]{
\scalebox{0.7}{
       \begin{tabular}{|c|c|c|c|c|c|}
    \hline
        $\tilde{\omega}_S$  & $E_{max}$    & $E_{min}$    & $\omega_{max}$    & $\omega_{min}$  & $\omega_{D0}$ \\
    \hline
    $0.7675$  & $0.068$ & $-0.023$ & $0.857$ & $0.784$ & $0.791$ \\
    \hline
    $0.749$  & $0.072$ & $-0.018$ & $0.852$ & $0.770$ & $0.775$ \\
    \hline
    $0.731$  & $0.078$ & $-0.011$ & $0.849$ & $0.756$ & $0.760$ \\
    \hline
    $0.7128$  & $0.084$ & $-0.001$ & $0.849$ & $0.744$ & $0.745$ \\
    \hline
    \end{tabular}}
       \label{tab:firsttable}
}
\subtable[]{
\scalebox{0.7}{
       \begin{tabular}{|c|c|c|c|c|c|}
    \hline
       $\tilde{\omega}_S$   & $E_{max}$    & $E_{min}$    & $\omega_{max}$    & $\omega_{min}$  & $\omega_{D0}$ \\
    \hline
    $0.9082$  & $-0.015$ & $-0.017$ & $0.9105$ & $0.9104$ & \diagbox[width=3em,height=1.5em]{}{} \\
    \hline
    $0.861$  & $-0.020$ & $-0.025$ & $0.8667$ & $0.8662$ & \diagbox[width=3em,height=1.5em]{}{} \\
    \hline
    $0.814$  & $-0.016$ & $-0.028$ & $0.826$ & $0.824$ & \diagbox[width=3em,height=1.5em]{}{} \\
    \hline
    $0.7676$  & $0.019$ & $-0.023$ & $0.596$ & $0.651$ & $0.791$ \\
    \hline
    \end{tabular}}
       \label{tab:secondtable}
}
\subtable[]{
\scalebox{0.7}{
       \begin{tabular}{|c|c|c|c|c|c|}
    \hline
        $\tilde{\omega}_S$  & $E_{max}$    & $E_{min}$    & $\omega_{max}$    & $\omega_{min}$  & $\omega_{D0}$ \\
    \hline
    $0.7127$  & $0.084$ & $-0.001$ & $0.849$ & $0.744$ & $0.745$ \\
    \hline
    $0.702$  & $0.083$ & $0.007$ & $0.821$ & $0.738$ & \diagbox[width=3em,height=1.5em]{}{} \\
    \hline
    $0.691$  & $0.078$ & $0.018$ & $0.794$ & $0.734$ & \diagbox[width=3em,height=1.5em]{}{} \\
    \hline
    $0.6794$  & $0.066$ & $0.037$ & $0.762$ & $0.734$ & \diagbox[width=3em,height=1.5em]{}{} \\
    \hline
    \end{tabular}}
       \label{tab:thirdtable}
}
\label{table4}
\end{table*}

The binding energy $E_B$ of the DBSs versus the nonsynchronized frequency $\tilde{\omega}_D$ for several values of the scalar field frequency $\tilde{\omega}_S$ are presented in Fig.~\ref{BE_nsf}. For the \textit{multi-branch} solution family, with the increase of nonsynchronized frequency $\tilde{\omega}_D$, the binding energy $E_B$ of DBSs first increases, then turns to the second branch, and then changes in a spiral shape. For the \textit{one-branch-A} solution family, the DBSs are stable ($E_B < 0$) when the scalar field frequency $\tilde{\omega}_S$ is sufficiently large, and unstable DBSs ($E_B > 0$) appear as the scalar field frequency $\tilde{\omega}_S$ decreases. For the \textit{one-branch-B} solution family, stable DBSs hardly exist.

In Table \ref{table4}, we show the maximum and minimum values of the binding energy $E_{B}$ of the DBSs in Fig.~\ref{BE_nsf}, the nonsynchronized frequency $\tilde{\omega}_D$ corresponding to the maximum and minimum values of $E_{B}$, and the value of the nonsynchronized frequency $\tilde{\omega}_D$ when $E_B = 0$. For the \textit{multi-branch} solution family, as the scalar field frequency $\tilde{\omega}_S$ decreases, both $E_{max}$ and $E_{min}$ increase and $\omega_{D0}$ decreases.  For the \textit{one-branch-A} solution family, as the scalar field frequency $\tilde{\omega}_S$ decreases, $E_{max}$ and $E_{min}$ first decrease and then increase, and $\omega_{D0}$ occurs. For the \textit{one-branch-B} solution family, as the scalar field frequency $\tilde{\omega}_S$ decreases, $E_{max}$ decreases and $E_{min}$ increases, $\omega_{D0}$ disappears.

\section{Conclusions}\label{sec5}
In this paper, we construct spherically symmetric multistate Dirac-boson stars consisting of a scalar field and two spin fields, both in the ground state. Furthermore, the characteristics of different types of solution families of DBSs are discussed.

First, for the case of synchronized frequency, we divide the solution families of DBSs into \textit{one-branch} and \textit{multi-branch} based on the number of branches of the solutions. For the \textit{one-branch} solution family, as the synchronized frequency $\tilde{\omega}$ increases, $\phi_{max}$ increases and $f_{max}$ and $g_{max}$ decrease. Moreover, the mass $M$ of DBSs decreases as $\tilde{\omega}$ decreases, while the binding energy $E_B$ changes in the opposite direction. In addition, the existence domain of the synchronous frequency $\tilde{\omega}$ increases as the mass $\tilde{\mu}_D$ of the Dirac field decreases. For the multi-branch solution family, the curve of the mass $M$ versus the synchronized frequency $\tilde{\omega}$ is spiral, and no similar family of solutions was found in Ref.~\cite{Li:2019mlk}. For the first branch of the DBSs, $\phi_{max}$ increases as the synchronized frequency $\tilde{\omega}$ increases, $f_{max}$ and $g_{max}$ first decrease and then increase. For the second and third branches of the DBSs, as the synchronized frequency $\tilde{\omega}$ increases, $\phi_{max}$, $f_{max}$ and $g_{max}$ all have monotonic behaviour. For each branch, the existence domain of the synchronized frequency $\tilde{\omega}$ increases as the mass $\tilde{\mu}_D$ of the Dirac fields decreases. Furthermore, as the mass $\tilde{\mu}_D$ decreases, $M_{min}$ increases, and $M_{max}$ first decreases and then increases. It is worth noting that for the family of multi-branch solutions, all solutions are unstable ($E_B > 0$).

Then, for the case of nonsynchronized frequency, we divide the solution families of DBSs into \textit{multi-branch}, \textit{one-branch-A} and \textit{one-branch-B}. For the \textit{multi-branch} solution family, in contrast to the case of synchronized frequency, $\phi_{max}$, $f_{max}$ and $g_{max}$ have monotonic behaviour on all three branches as the nonsynchronized frequency $\tilde{\omega}_D$ increases. For each branch, the existence domain of the nonsynchronized frequency $\tilde{\omega}_D$ decreases as the frequency $\tilde{\mu}_S$ of the scalar field decreases. And as the frequency $\tilde{\mu}_S$ decreases,  both $M_{min}$ and $M_{max}$ decrease. Furthermore, the stable solution ($E_B < 0$) exists only on the first branch. For the \textit{one-branch-A} solution family, as the nonsynchronized frequency $\tilde{\omega}_D$ increases, $\phi_{max}$, $f_{max}$, $g_{max}$, the mass $M$ and the existence domain of the nonsynchronized frequency $\tilde{\omega}_D$ vary in the same way as in the \textit{one-branch} solution family for the case of synchronized frequency. However, for the case of nonsynchronized frequency, $M_{min}$ in the \textit{one-branch-A} solution family is determined by the scalar field frequency $\tilde{\mu}_S$. Moreover, no matter what value of the scalar field frequency $\tilde{\mu}_S$ is taken, there is always a stable solution (provided that this solution belongs to the \textit{one-branch-A} solution family). For the \textit{one-branch-B} solution family, as the nonsynchronized frequency $\tilde{\omega}_D$ increases, $f_{max}$ and $g_{max}$ increases, $\phi_{max}$ increases and then decreases, and the mass $M$ of the DBSs decreases. The existence domain of the nonsynchronized frequency $\tilde{\omega}_D$ decreases as the scalar field frequency $\tilde{\mu}_S$ increases. In addition, there are almost no stable solutions in the \textit{one-branch-B} solution family.

The \textit{multi-branch} solution family for the cases of the synchronized/nonsynchronized frequencies and the \textit{one-branch-B} solution family for the cases of the nonsynchronized frequency are unique, as similar families of solutions have not been found in past studies of multistate boson stars~\cite{Li:2019mlk,Li:2020ffy}. Moreover, these newly discovered solution families are relatively unstable (most of them satisfy $E_B > 0$). We will continue to study DBSs. The scalar and Dirac fields in this work are in the ground state, and later we will consider the case where the scalar and Dirac fields are in different energy levels, respectively. Furthermore, we believe that spherically symmetric multistate Dirac stars (SMSDSs) should exist where the matter fields are several Dirac fields. The construction of the model for SMSDSs is also part of our future work.

\section*{Acknowledgements}
This work is supported by National Key Research and Development Program of China (Grant No. 2020YFC2201503) and  the National Natural Science Foundation of China (Grant No.~12047501). Parts of computations were performed on the shared memory system at institute of computational physics and complex systems in Lanzhou university. 

\section*{Appendix}\label{appendix1}
In order to study the Dirac equation in curved spacetime, we need to introduce the vierbein $e_a^{\alpha}$, defined by
\begin{equation}
g_{\alpha\beta}e_a^{\alpha}e_b^{\beta} = \eta_{ab}\,,
\end{equation}
where $g_{\alpha\beta}$ is the curved space metric and $\eta_{ab} = (-1,+1,+1,+1)$. We use Greek letters $\alpha,\beta,\cdots$ for curved spacetime indices and the Latin letters $a,b,\cdots$ for flat spacetime indices. Greek indices are raised and lowered with $g_{\alpha\beta}$ and Latin indices are raised and lowered with $\eta_{ab}$. 

$\gamma^a$ are flat spacetime $\gamma$-matrices and $\gamma^{\alpha}$ are curved spacetime $\gamma$-matrices. $\gamma^a$ are related to $\gamma^{\alpha}$ through the vierbein:

\begin{equation}\label{equ27}
\gamma^{\alpha} = e_a^{\alpha}\gamma^a,\qquad \gamma^a = e^a_{\alpha}\gamma^{\alpha}.
\end{equation}
$\gamma^a$ and $\gamma^{\alpha}$ satisfying the anticommutation relations:
\begin{equation}
\{\gamma^{\alpha},\gamma^{\beta}\} = 2g^{\alpha \beta},\qquad \{\gamma^a,\gamma^b\} = 2\eta^{ab}.
\end{equation}
$\gamma$-matrices indices are raised and lowered in the following ways:
\begin{equation}
\gamma_a = \eta_{ab} \gamma^b,\qquad \gamma_{\alpha} = g_{\alpha\beta}\gamma^{\beta}.
\end{equation}
The non-zero component of the vierbein we choose in this paper is
\begin{equation}\label{equ30}
e_0^t = \frac{1}{\sqrt{N}\sigma},\quad e_1^\theta = \frac{1}{r},\quad e_2^\varphi = \frac{\csc\theta}{r},\quad e_3^r = \sqrt{N}.
\end{equation}

For $\gamma^a$ we use the Weyl representation:
\begin{equation}\label{equ31}
\setlength{\arraycolsep}{8pt}
\gamma^0 = i\begin{pmatrix} 0&1\\1&0 \end{pmatrix},\qquad \gamma^k = i\begin{pmatrix} 0&\sigma_k\\-\sigma_k &0 \end{pmatrix},\qquad k = 1,2,3,
\end{equation}
where $\sigma_k$ are the Pauli matrices:
\begin{equation}
\setlength{\arraycolsep}{8pt}
\sigma_1 = \begin{pmatrix} 0&1\\1&0 \end{pmatrix},\qquad \sigma_2 = \begin{pmatrix} 0&-i\\i&0 \end{pmatrix},\qquad \sigma_3 = \begin{pmatrix} 1&0\\0&-1 \end{pmatrix}.
\end{equation}
It can be obtained from Eq.~(\ref{equ27}), Eq.~(\ref{equ30}) and Eq.~(\ref{equ31}) that $\gamma^{\alpha}$ in this paper are
\begin{equation}
\gamma^t = \frac{\gamma^0}{\sqrt{N}\sigma},\qquad \gamma^r = \sqrt{N}\gamma^3,\qquad \gamma^{\theta} = \frac{\gamma^1}{r}, \qquad \gamma^{\varphi} = \frac{\csc\theta \gamma^2}{r}.
\end{equation}

The covariant derivatives of $\overline{\Psi}^{(k)}$ and $\Psi^{(k)}$ are
\begin{equation}
\hat{D}_{\mu}\Psi^{(k)} = \partial_\mu \Psi^{(k)} + \Gamma_\mu \Psi^{(k)},\quad \hat{D}_{\mu}\overline{\Psi}^{(k)} = \partial_\mu \overline{\Psi}^{(k)} - \Gamma_\mu \overline{\Psi}^{(k)},
\end{equation}
where $\Gamma_\mu = \omega_{\alpha bc}\gamma^b\gamma^c/4$, with $\omega_{\alpha bc}$ is the spin connection:
\begin{equation}
\omega_{\mu ab} = e_{a\nu}e_b^{\lambda}\Gamma_{\mu \lambda}^{\nu} - e_b^{\lambda}\partial_\mu e_{a\lambda}\,,
\end{equation}
where $\Gamma_{\mu \lambda}^{\nu}$ is the affine connection.

\providecommand{\href}[2]{#2}\begingroup\raggedright
\endgroup

\end{document}